\documentclass[iop]{emulateapj-rtx4}
\usepackage{graphicx, natbib,amsmath, lscape, afterpage}

\def\nmin{$N_{\rm min}(\dot{\rm O}$)}

\def\lgwv{0.13 \pm 0.44}
\def\alphwv{-0.84 \pm 0.24}
\def\vlgnr{14.1 \pm 0.2}
\def\valphn{-1.11 \pm 0.29}

\def\N#1{N({\rm #1})}
\def\Ntild#1{\tilde N({\rm #1})}
\def\ltp{\left ( \,}
\def\rtp{\, \right  ) }
\def\sci#1{{\; \times \; 10^{#1}}}
\def\mkms{{\rm \; km\,s\,^{-1}}}
\newcommand{\lgnr}{\log N_{(10\,\rm kpc)}}

\newcommand{\nsit}{$N({\rm Si^{++}})$}
\newcommand{\nsiw}{$N({\rm Si^{+}})$}
\newcommand{\ncit}{$N({\rm C^{++}})$}
\newcommand{\nciw}{$N({\rm C^{+}})$}
\newcommand{\mCf}{C_{\rm f}}

\newcommand{\ntot}{44}
\newcommand{\nhires}{35}

\newcommand{\npair}{7}

\newcommand{\nbonus}{21}
\newcommand{\ndlow}{22}  % Detections of a Low-ion
\newcommand{\ndet}{33}  % Detections of a Low-ion
\newcommand{\exsys}{J1233+4758 94\_38}
\newcommand{\exqso}{J1233+4758}
\newcommand{\cosha}{COS-Halos}
\newcommand{\mRperp}{R}
\newcommand{\Rperp}{$R$}
\newcommand{\zphot}{z_{\rm phot}}
\newcommand{\ntarg}{38}
\newcommand{\mmsun}{M_\odot}

\newcommand{\etal}{et~al.\/}

\newcommand{\mwlya}{W_{\rm Ly\alpha}}
\newcommand{\ew}{\mbox{$W_{\rm r}$}}
\newcommand{\mew}{EW}

\newcommand{\kms}{\mbox{km\thinspace s$^{-1}$}}

\newcommand{\mnhi}{N_{\rm HI}}
\newcommand{\nhi}{\mbox{$N_{\rm HI}$}}
\newcommand{\mnlow}{N_{\rm Low}}
\newcommand{\nlow}{\mbox{$N_{\rm Low}$}}

\newcommand{\lya}{\mbox{Ly$\alpha$}}
\def\cm#1{\, {\rm cm^{#1}}}

\newcommand{\avg}[1]{\left< #1 \right>} % for average

\shortauthors{Werk \etal}
\shorttitle{COS-Halos LowIons}

\begin{document} 
%\nocite{*}
\slugcomment{Re-submitted to ApJS}

\title{The COS-Halos Survey: An Empirical Description of Metal-Line Absorption in the Low-Redshift Circumgalactic Medium}

\author{Jessica K.\ Werk\altaffilmark{1},
J. Xavier Prochaska\altaffilmark{1},
 Christopher Thom\altaffilmark{2},
  Jason Tumlinson \altaffilmark{2}, 
  Todd M. Tripp \altaffilmark{3}, 
 John M. O'Meara \altaffilmark{4}, 
 \& Molly S. Peeples \altaffilmark{5}
  }

\altaffiltext{1}{UCO/Lick Observatory; University of California, Santa Cruz, CA $jwerk@ucolick.org$}
\altaffiltext{2}{ Space Telescope Science Institute, 3700 San Martin Drive,  Baltimore, MD}
\altaffiltext{3}{Department of Astronomy, University of Massachusetts, Amherst, MA}
\altaffiltext{4}{Department of Chemistry and Physics, Saint Michael's College, Colchester, VT}
\altaffiltext{5}{Department of Physics and Astronomy, University of California, Los Angles, CA}

\begin{abstract}
We present the equivalent width and column density measurements for
low and intermediate ionization states of the circumgalactic medium
(CGM) surrounding \ntot\ low-$z$, $L\approx L^*$ galaxies drawn from
the \cosha\ survey.  These measurements are derived from far-UV
transitions observed in HST/COS and Keck/HIRES spectra of background
quasars within an impact parameter $\mRperp < 160$\,kpc to the
targeted galaxies.  The data show significant metal-line absorption
for \ndet\ of the \ntot\ galaxies, including quiescent systems,
revealing the common occurance of a cool ($ T \approx ~ $10$^{4 - 5}$ K), 
metal-enriched CGM.  The detection rates and column densities derived for these metal lines decrease with increasing impact parameter, a trend we interpret as a declining metal surface density profile for the CGM.
A comparison of the relative column densities of adjacent ionization
states indicates the gas is predominantly ionized.  The large surface
density in metals demands a large reservoir of metals and gas in the
cool CGM (very conservatively, $M_{\rm CGM}^{\rm cool} > 10^{9}
\mmsun$), which likely traces a distinct density and/or temperature regime from the highly-ionized CGM traced by O${+5}$ absorption.  The large dispersion in absorption strengths (including
non-detections) suggests the cool CGM traces a wide range of densities or a mix of local ionizing conditions.  Lastly, the
kinematics inferred from the metal-line profiles are consistent with the
cool CGM being bound to the dark matters halos hosting the galaxies;
this gas may serve as fuel for future star-formation.
Future work will leverage this dataset to provide estimates on the
mass, metallicity, dynamics, and origin of the cool CGM in low-$z$,
$L^*$ galaxies.
\end{abstract}


\section{Introduction}
\label{sec:intro}

Galaxies are traditionally discovered and characterized by the emission
from their stellar components. 
%and the dark matter halos that surround them. 
 The majority of a galaxy's stars are localized at one to a
few tens of kiloparsecs from the center of its potential well, depending on the
angular momentum and total mass of the system.  
Over the course of a Hubble time, the stellar system dominates the
energetic output of the galaxy, drives its chemical enrichment, and
ultimately defines its standing in the pantheon of modern galaxies. 

Between the stars, at least in star-forming galaxies, lies
a diffuse medium of gas, dust, and metals that comprises the
interstellar medium (ISM).  This predominantly neutral gas-phase fuels
star formation, receives energetic and kinetic feedback, and collects
the material by-products from the stars that form and die. The ISM is a crucial
component of young and growing galaxies.  It further serves as a
conduit for the light radiated by massive stars (and their explosions)
in the form of \ion{H}{2} regions and the photo-dissociation regions
that surround them, and dust that converts UV/optical photons into
infrared light.

It is now well-recognized that a third, major baryonic component
exists for galaxies in the form of a
diffuse and ionized medium that extends to many tens, and even
several hundred kpc from the galaxy
\citep[e.g.][]{mwd+93, lbt+95,stocke+95,clw+98,tripp+98,pss02,wakker09,pwc+11}. 
This reservoir, referred to as halo gas or the circumgalactic
medium (CGM), permeates the dark matter halo and serves
as both a supply of material for future star formation and as the gutter
for gas, metals, and dust expelled from the central galaxy and/or its
satellites and progenitors.  
Because it can mediate galaxy accretion and feedback, developing a complete picture of the CGM has clear and vital importance for understanding the formation of galaxies and their stellar
systems. Furthermore, it is the fundamental intermediary of the baryonic cycle connecting galaxies to
the intergalactic medium (IGM).

Although predicted many decades ago \citep{bs69},
the discovery of the CGM awaited the pioneering efforts
of J. Bergeron, who associated strong \ion{Mg}{2} absorption in quasar
spectra to a handful of $L \approx L^*$ galaxies at close impact
parameters to the sightlines \citep{bergeron86}.  The launch of the {\it Hubble Space
  Telescope} and successful operation of UV spectrographs have
extended the
study of the CGM to a multitude of far-UV transitions including the
\ion{H}{1} Lyman series \citep{lbt+95,pss02, bowen+02}.  The first few
generations of UV spectrographs (FOS, GHRS, STIS), however, had only
sufficient throughput for observations of the brightest AGN on the
sky. Consequently, while a few pioneering HST programs studied the gaseous envelopes of specific galaxies (e.g. Bowen et al. 1995)\nocite{bowen+95}, the majority of UV absorber surveys selected targets based on QSO properties such as brightness or redshift and not based on the properties of foreground galaxies \citep[e.g.][]{ssp96,pss00,dt01,tripp08, tc08b, tc08a, ctp+10, tilton+12}.
This has essentially limited studies of the low-$z$ CGM to small samples
and/or galaxies at large impact parameters. 

Our understanding of the CGM for $L \approx L^*$ galaxies is especially
limited. This limitation arises from the observation that a random QSO sightline will pass within 300 kpc of an L$^*$ galaxy only once for every $\Delta$z = 0.25 interval, and only one in 4 such sightlines will have a QSO pairing within 100 kpc separation \citep{tf05}.    Most previous studies traced the inner CGM ($\mRperp < 100$\,kpc) using the
\ion{Mg}{2} doublet \citep[e.g.][]{bowen+95, bc09,chg+10, bc11}.
The results indicate a high incidence of strong \ion{Mg}{2} absorption
(70\%\ covering fraction for $\mRperp < 75$\,kpc and $W_{\rm r}  >
0.3$\AA), revealing the presence of a cool and metal-enriched CGM on
these scales.  % [Mention Bordoloi?]
Regarding far-UV
transitions, only a handful of systems have been studied at \lya,
O${+5}$, or C${+3}$ transitions with previous generations of UV
spectrographs \citep{stl98, ts00, clw01, trippetal01, pks0405_uv, araciletal06, stockeetal06,mc09, wakker09,pwc+11}.  These analysis
have suggested a high incidence of \ion{H}{1} gas around $L^*$ galaxies
and the occasional detection of highly-ionized metals within a few
hundred kpc.    The samples have been too small, however, to properly
assess the ionization state, to characterize its basic properties, or to explore
trends with impact parameter, stellar mass, galaxy color, etc.  
It is worth noting
that $L^*$ galaxies are too rare to dominate the cosmic census of gas,
dust, or metals.  Analyses of their CGM, therefore, offers greater
insight on the processes of galaxy formation than cosmology.
Nevertheless, a detailed assessment may provide crucial
insight into the formation of galaxies of all shapes and sizes. 

With the explicit goal of assessing the multiphase nature of halo
gas in $L \approx L^*$, low-redshift galaxies, we have designed  and executed a large program
with the {\emph{Cosmic Origins Spectrograph}}
(COS; Froning \& Green 2009\nocite{froning09}, Green et al. 2012\nocite{green+12}) on the {\emph {Hubble Space Telescope (HST)}}.
Specifically, we targeted the halo gas of 38 galaxies drawn from the
imaging dataset of the Sloan
Digital Sky Survey (SDSS) whose angular offsets from quasar sightlines
and photometric redshifts implied impact parameters (\Rperp) well inside
their virial radii. The \cosha\
survey provides sensitive absorption-line measurements for a
comprehensive suite of multiphase ions from the spectra of 38
z$<$ 1 QSOs lying behind the `target' galaxies.
Our follow-up survey of these fields has revealed an additional
\nbonus\ 
`bonus' galaxies included in our analysis that also lie at $\mRperp < 160$\,kpc to the
sightlines. In aggregate, these data comprise a
carefully-selected statistically-sampled map of the physical state and
metallicity of the CGM for $L \approx L^*$ galaxies.

In the first survey paper by \cosha, we examined the incidence of
highly ionized, metal-enriched gas traced by the O${+5}$ doublet
\citep{tumlinson11}.  To our surprise, the data revealed a very high
incidence of strong O${+5}$ absorption, $\N{O^{+5}} > 10^{14.3}
\cm{-2}$, in the CGM of star-forming $L^*$ galaxies yet a relative lack of
such ions in the CGM of non-SF galaxies.  The mass in metals (and
presumably baryons) of this highly ionized and possibly warm/hot
($T>10^5$\,K) gas is substantial;  we estimate a reservoir comparable to and likely far
exceeding the mass of metals and gas in the ISM of $L^*$ galaxies.
The reasons(s) for the absence of O${+5}$ around red galaxies
remains an open question.

In  companion papers, we have surveyed the incidence and distribution
of \ion{H}{1} gas of the CGM for these galaxies
\citep{thom12,tumlinson12}.  The data reveal a very high incidence of
strong \ion{H}{1} \lya\ absorption ($W_{\rm r} \gtrsim 1$\AA) at nearly all
impact parameters to the survey edge ($R=160$\,kpc), for both SF and
non-SF $L^*$ galaxies.  These results establish the presence of a
cool-phase CGM around essentially all $L^*$ galaxies.  In this paper,
we extend the analysis to the metal lines that may be associated with
this \ion{H}{1} gas.  Such data are critical to assessing the
ionization state, metallicity, and mass of the CGM.
  
This specific paper focuses on the absorption-line measurements for
all ions except  \ion{H}{1} \citep{tumlinson12} and O${+5}$ \citep{tumlinson11}.    We present our
Keck/HIRES spectroscopy for the first time and a comprehensive
analysis of the COS spectra.  We provide tables of the measured
equivalent widths and column densities for all important transitions.
We then examine empirical relations between the CGM and galaxy properties.  
%Weintentionally limit the analysis to have minimal modeling (e.g.\
%avoiding ionization modeling).  Future papers will focus on the
%science derived from such efforts.
%outline
In $\S$~2 we present the sample definition and the multi-wavelength dataset; $\S$~3 details the metal-line identification and subsequent measurements of equivalent widths and column densities; $\S$~4 details the results of this empirical analysis, providing an examination of trends in CGM metal line absorption with galaxy properties;  $\S$~5 argues, based on these results, that the cool CGM (  10$^4$ K $\lesssim$ T $<$ 10$^5$ K) is predominantly ionized, metal enriched, bound to the galaxy's dark matter halo, and contains a significant reservoir of baryons for $L^*$ galaxies; and finally, $\S$~6 offers a concise, bulleted list of our key results and conclusions.

  Throughout this work we assume the 5-year WMAP cosmology with
  $\Omega_{\Lambda}$ = 0.74, $\Omega_{m}$ = 0.26, and H$_{0}$ = 72 km
  s$^{-1}$ Mpc$^{-1}$ \citep{wmap05}. For the purpose of this work, we define cool CGM gas in the temperature range 10$^4$ K $\le$ T $<$ 10$^5$ K, and warm CGM gas in the temperature range  10$^5$ K $\le$ T $<$ 10$^6$ K. 
    
%%%%%%%%%%%%%%%%%%%%%%%%%%%%%%%
%%%%%%%%%%%%%%%%%%%%%%%%%%%%%%%
\section{The Data}

%This section summarizes the diverse, multi-wavelength datasets that
%comprise the \cosha\ survey.  

 \subsection{Sample Definition}
\label{sec:sample}

  The \cosha\ program targeted \ntarg\ galaxies with photometric redshifts
  $\zphot \approx 0.15 - 0.4$ having a range of galaxy color
  $(u-r)$ and stellar mass $M_* = 10^{9.5} - 10^{11.5} \mmsun$,
  and distributed at impact parameters $\mRperp \approx 15-160$\,kpc from a
 background, UV-bright quasar \citep[see][for a complete description
 of the sample definition]{tumlinson12}.  
 The survey selection criteria implied 
 no explicit biases regarding the galaxy inclination or local
 environment.  Previous studies of  Ly$\alpha$ absorption in galaxy halos have indicated that the CGM of  $L \approx L^*$ galaxies extends to at least 300 kpc \citep{rudie12, pwc2+11}. We emphasize that our survey probes the inner 160 kpc, and therefore  may not probe the entire,
metal-bearing CGM of $L^*$ galaxies. Because of the occasional semi-catastrophic error in the photometric
redshift, a small fraction of the galaxies have $z \lesssim 0.1$ and
correspondingly lower stellar mass and luminosity than expected.  In
the following, we restrict the study to galaxies with $L> 0.1L^*$ to
isolate $L \approx L^*$ galaxies, where we use the R-band absolute magnitude of $-21.2$ for an $L^*$ galaxy \citep{blantonetal03}.  Future works will study the CGM of the fainter, dwarf galaxy population.

\begin{figure*}[t!]
\begin{centering}
\hspace{0.5in}
\includegraphics[height=0.80\linewidth,angle=90]{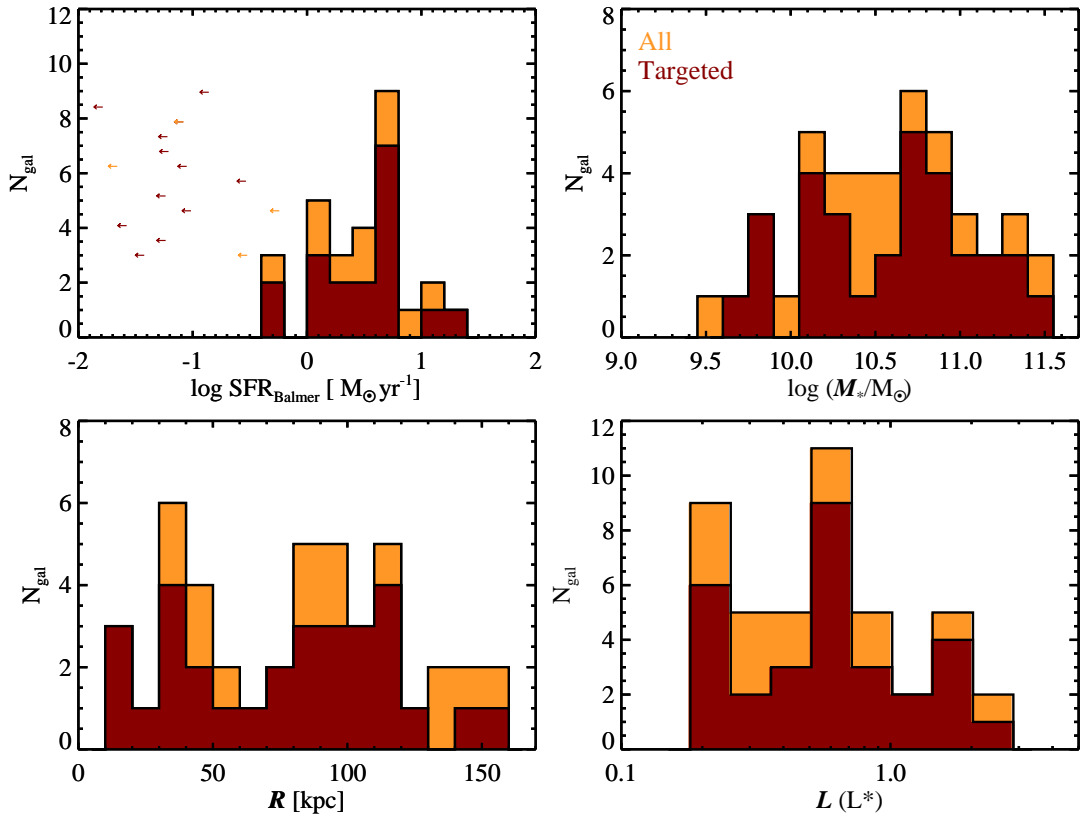}
\end{centering}
\caption{Histograms of the properties for the \cosha\ galaxies
  analyzed in this manuscript \citep[see also][]{werk12}.  
  At a given redshift, the sample comprises the closest,
  luminous ($L>0.1L^*$) galaxy with an impact parameter $\mRperp <
  160$\,kpc of each sightline.  The panels show the explicitly targeted galaxies
  (red) and the full sample (orange) which includes `bonus'
  galaxies discovered in the same quasar fields and which satisfy our
  sample criteria.
}
\label{fig:hist}
\end{figure*}

\begin{figure*}[t!]
\begin{centering}
\includegraphics[height=0.90\linewidth,angle=90]{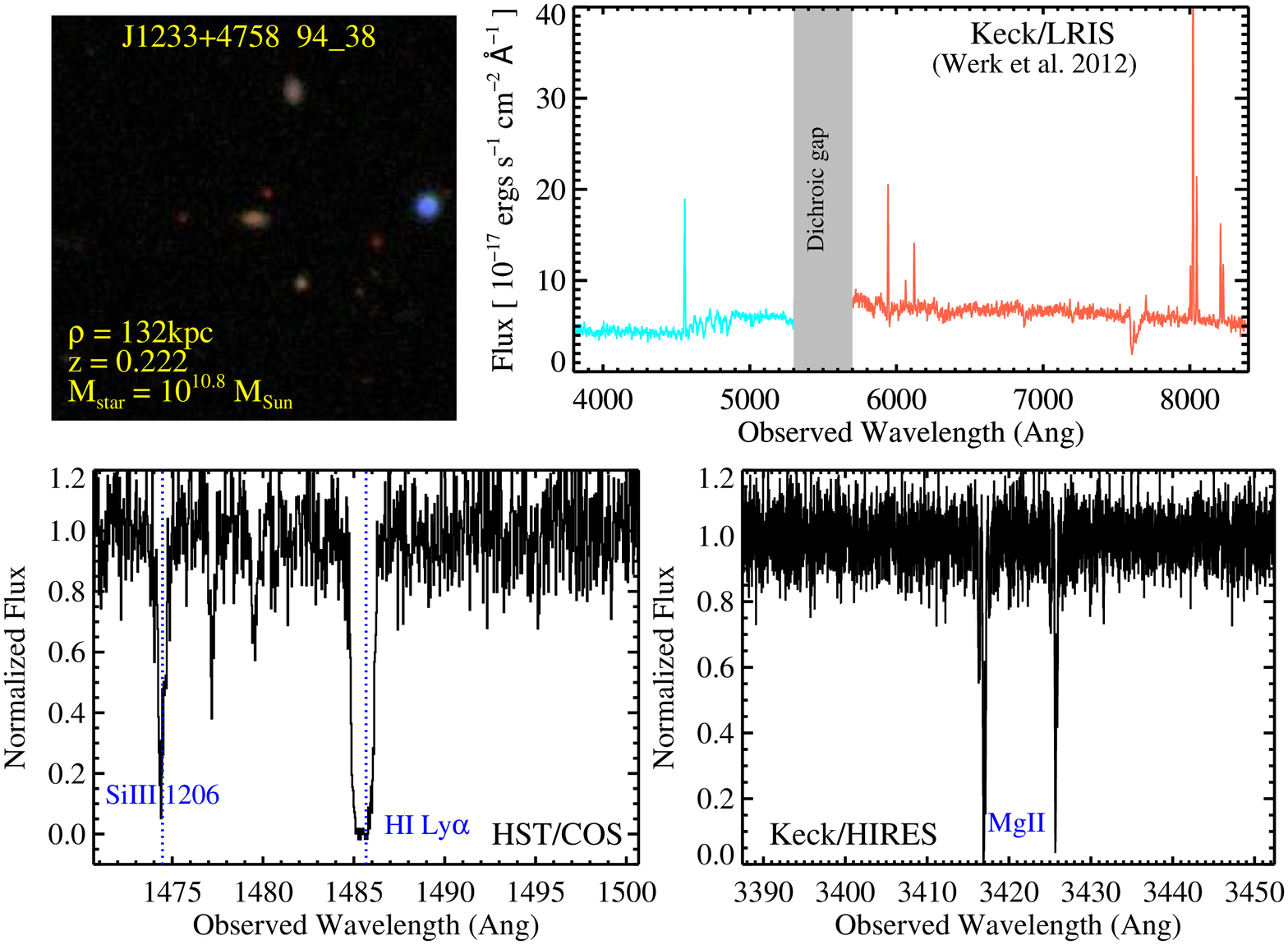}
\end{centering}
\caption{
Montage of the SDSS image (upper left panel; $1.5'$ on a side) 
and Keck/LRIS spectrum
(upper right) of an example target galaxy (J1233+4758\_94\_38),
together with the HST/COS spectrum (lower left) zoomed-in on
\ion{H}{1} \lya\ and the Keck/HIRES spectrum (lower right) zoomed-in
on the \ion{Mg}{2} doublet. These data are characteristic of the
dataset that has been gathered for nearly all of the galaxies in the sample.
}
\label{fig:example}
\end{figure*}

 In the course of acquiring spectra for these targeted
 galaxies, we discovered an additional \nbonus\ galaxies at
 close impact parameters to the quasar and also with $z < z_{\rm
   qso}$ and $L>0.1L^*$.  Although not our primary targets, these galaxies were
 selected without bias to CGM absorption and therefore are included
 in the following analyses, in identical fashion as the targeted sample.  
A full description of the galaxy spectra, photometry, and the
 inferred galaxy properties (e.g.\ SFR, stellar mass) is given in
 \cite{werk12}.  For convenience, Table~\ref{tab:galprops} provides a
 summary of the galaxies studied here and Figure~\ref{fig:hist} shows
 histograms of a few key properties.  % M*, L*, rho, SFR  
Throughout the paper, we identify each galaxy
 according to the quasar field and its angular offset from the quasar,
 e.g., galaxy J0042--1037\_358\_9 is located $9''$ from the quasar 
 J004222.29--103743.8 oriented 358$^\circ$ east of North.  
 Figure~\ref{fig:example} shows the SDSS imaging and our Keck/LRIS
 spectrum of the representative galaxy \exsys. Once we trim the sample to exclude galaxies with $L < 0.1L^*$ and duplicate absorbers, defined as galaxies meeting all criteria outlined above with a more massive companion $L^*$ galaxy at the same redshift (see discussion below), we are left with 44 galaxy absorbers in total. 

%%%%%%%%%%%%%%%%%%%%%%%%%%%%%%%%%
%%%%%%%%%%%%%%%%%%%%%%%%%%%%%%%%%%%%%%%%%%%%%%%%%%%%%%%%%%%%%%%%%%%%%

%As described in depth in \cite{tumlinson12}, the experimental design of
%\cosha\ was to identify fields where one or more galaxies with a
%photometric redshift $\zphot \approx 0.2$ and whose angular separation
%from a far-UV bright quasar was consistent with $\Rperp < 150$\,kpc.
%[Comment on success]

Galaxies, owing to the processes of hierarchical structure formation,
are known to cluster, form groups, and even merge with one another.
For these reasons, `random' $L^*$
galaxies often reside within a virialized galaxy cluster or group, and a smaller
subset are currently experiencing a merger with another galaxy. In the former cases, standard
treatment is to refer to the cluster (group) member as a satellite of
the far more massive, underlying dark matter halo.  At z$\sim$0, models predict that tidal stripping significantly diminishes the size of the individual satellites' halos \citep{okamoto99}. By extension, we infer that a large fraction of the gas at $\Rperp < 160$\,kpc (pertaining to
a galaxy's CGM) would likely be distributed throughout the intracluster or intragroup medium.
In clusters, then,  the localized intracluster or intragroup
medium may dominate any observed absorption.  This
may be especially true for highly ionized species like O${+5}$.  

%%%%%%%%%%%%%%%%%%

%%%%%%%%%%%%%%%

For the \cosha\ sample, \cite{werk12} carefully considered the environments
of our target and bonus samples, as assessed 
from previous analyses of SDSS
spectroscopy and photometry (maxBCG cluster catalog; Koester et al. 2007;\nocite{maxbcg}), and reported that only 5 of the
galaxies (within 3 QSO fields) are probable members of a massive group or galaxy
cluster: J0928+6025: 110\_35 (0.766 Mpc projected from cluster center;
620 km s$^{-1}$ velocity offset), 129\_19 (0.731 Mpc projected from
cluster center; 570 km s$^{-1}$ velocity offset), and 187\_15 (0.701
Mpc projected from cluster center; 700 km s$^{-1}$ velocity offset);
J1016+4706: 359\_16 (0.573 Mpc projected from cluster center; 985 km
s$^{-1}$ velocity offset); and J1514+3620: 287\_14 (19.3 Mpc from the
cluster center; 1550 km s$^{-1}$ velocity offset). 
For reference, \cite{maxbcg} consider a galaxy to be a cluster member if the velocity offset is less than 2000 km s$^{-1}$. In addition to those galaxies that match a maxBCG cluster, we find three galaxies in the QSO field J2257$+$1340 (270\_40, 238\_31, and 230\_25) that have confirmed spectroscopic redshifts z$\sim$0.177, all of which are probable members of a previously unidentified group of galaxies.  In our analysis, we treat these systems in identical fashion to the others. 

While the complications related to group/cluster environments is
predicted (and observed) to be a rare concern for our program,
projected pairs
of $L \approx L^*$ galaxies are more common. Deep  sky surveys reveal
that  the fraction of $\sim L^*$ galaxies  at z$\sim$0.2  that lie in
close projected pairs  is 2 $-$ $3 \%$ for $\mRperp \le 30$\,kpc
(e.g. Zepf \& Koo 1989; Kartaltepe et
al. 2007\nocite{zepf89,kartaltepe07}) and $\sim$20 \% for $\mRperp \le 100$\,kpc \citep{lin04}. 
From our follow-up spectroscopy on the fields, there are \npair\ cases
where two or more $L>0.1L^*$ galaxies with a very similar spectroscopic redshift
($\Delta z < 0.004$) 
lay within $\mRperp = 160$\, kpc of the same quasar sightline. This
number includes two cases of galaxy cluster/groups identified in the
previous paragraph (QSO sightlines J2257+1340 and J0928+6025). Based
on SDSS galaxy catalogs and less-accurate photometric redshifts, at
least 10 additional galaxies included in our study could have an $L >
0.1L^*$ galaxy (in two cases,  multiple galaxies) within a 300 kpc projected radius.  In these cases, it is possible if not probable that the CGM of each galaxy contributes to the observed absorption.  For
example, the very high incidence of \lya\ absorption from the CGM
%at all galaxy luminosities for $\Rperp < 300$\,kpc
\citep[e.g.][]{pwc+11,tumlinson12}  implies that any galaxy close to a quasar
sightline will contribute \ion{H}{1} absorption. In
the case of galaxy pairs, assigning all of the observed absorption to a single galaxy would
overestimate the CGM of that galaxy.  
It will be difficult if not impossible, however, to separate the
contributions from each galaxy: the gas need not have identical
velocity and line-blending further complicates the analysis.

For these reasons, we have taken the following approach to CGM
analysis in fields where multiple galaxies lie close to the sightline
and have velocity offset $|\delta v| < 500\kms$ based on spectroscopic redshifts.  We associate all
observed absorption with the CGM of the most massive galaxy (gauged by
the stellar mass) with virial radius $r_{\rm vir} \ge \mRperp$
\citep[see][]{tumlinson12}. This approach is partly justified by the 
expectation that lower mass galaxies near the sightline will 
lie within the dark matter halo of the most massive galaxy.   In this case,
lower mass galaxies are considered satellites.
In short, our procedure yields constraints on the CGM
for a population of 44 $\approx L^*$ galaxies with a diverse set of
environments.  Future analysis will focus on the importance of
environment for the nature of the CGM from such galaxies.

There are several systematic errors associated
with this procedure.  First, the effects of projection.  On occasion, placing the
smaller galaxy within the halo of the larger will be incorrect because the
two are truly well separated. 
Second, stellar mass is not a perfect proxy for dark matter halo mass.
Third, interactions amongst galaxies and their satellites may be a
crucial aspect of the formation and evolution of the CGM.  It may
inappropriate or at least non-representative, to focus on the CGM of
individual, isolated galaxies.
Nevertheless, a proper treatment of modeling the CGM
must take into account projection effects, the matching of stellar
mass to dark matter halos, etc.
Such analysis will be presented in a future paper of our own on this
topic. 
This paper provides an empirical assessment of the CGM
subject to the uncertainties described above.

\subsection{COS Spectroscopy}

\cite{tumlinson12} describe fully the acquisition, reduction and calibration of the data from the Large Cycle~17 {\emph {HST}} program (PID:
11598) known as \cosha\ whose COS spectroscopy provides all of the far-UV spectra analyzed here.
For every sightline, these observations yielded a continuous spectrum
spanning $\lambda \approx 1150-1800$\AA.  The exposure times were
chosen to achieve a signal-to-noise (S/N) of $7-10$ per resolution
element (FWHM~$\approx 15 \kms$) at $\lambda \approx 1300$\AA.
The analysis that follows was performed on the data binned by 3 native
spectral `pixels' to a dispersion of $\Delta \lambda \approx
0.0367$\AA.  

Because the COS optics do not correct for the mid-frequency wavefront
errors arising from zonal irregularities in the {\emph{HST}} primary, the
 true COS line-spread-function (LSF) 
is not characterized by the single Gaussian.  Instead, it is well-described by
a Gaussian convolved with a power-law that extends to many tens of
pixels beyond the line-center \citep{cos_lsf}.
These broad wings affect both the
precision of our equivalent width measurements and complicate
assessments of line-saturation. We mediate these effects when we fit absorption lines (described in $\S$ 3.3) by using the nearest wavelength grid point and convolving with the real LSF. 
%We explore these effects in greater
%detail in Appendix~\ref{appx:LSF}.

Although our coadding procedures properly preserve the Poisson
counting statistics of the data, we find that we usually have enough counts in each pixel so that we can use standard propagation of
error in the Gaussian regime.  All of the spectra were continuum
normalized with automated routines in $\approx 20$\AA\ chunks centered
on transitions of interest.  These continuum models were visually
inspected and manually adjusted as necessary.
Figure~\ref{fig:example} shows a representative slice of the COS
spectra centered at the \lya\ transition associated with the galaxy
\exsys.  One also notes strong \ion{Si}{3}~ $\lambda1206$ absorption
consistent with this redshift.

\subsection{HIRES data}
 To supplement the far-UV spectra from {\emph {HST}}/COS, we obtained
 Keck/HIRES echelle spectra for \nhires\ quasars
 (Table~\ref{tab:hires}).   For galaxies at $z>0.1$, these data provide coverage
 of the \ion{Mg}{2}~$\lambda\lambda$2796,2803 doublet,  an
 excellent diagnostic of cool ($T \le 10^4$\,K), metal-enriched gas.
 Because these data were taken at substantially higher spectral
 resolution and (generally) higher S/N than the COS data, they also
 offer additional constraints on line saturation and the kinematics of
 the CGM.  

 The HIRES spectra were obtained over four runs spanning the nights of UT
 2008 October 6, 2010 March 26, 2010 September 2, 
 and 2012 April 12-13.
On each night, the instrument was configued with the blue
cross-disperser and collimator, i.e.\ HIRESb.  We employed an echelle
angle ECH=$0^\circ$ and a cross-disperser angle XDANGL$\approx 1.0$ to
give nearly continuous wavelength coverage from $\lambda =
3050-5880$\AA\ with $\approx 20$\AA\ gaps at $\lambda \approx
3970$\AA\ 
and 4965\AA\
owing to the CCD mosaic configuration.  We used the C1 decker for all observations
giving a FWHM~$\approx 6 \kms$ spectral resolution\footnote{Unlike the
  COS spectra, the LSF for HIRES is well approximated by a single
  Gaussian.}.  The data were
reduced and calibrated using standard techniques with the
HIRedux\footnote{http://www.ucolick.org/$\sim$xavier/HIRedux/index.html}
software package.  The optimally coadded
exposures (generally two per target) were continuum normalized with
custom software, and a single combined spectrum sampled at $\Delta v =
1.3~ \kms$ per pixel was generated for the absorption-line analysis.  A
portion of the Keck/HIRES spectrum for \exqso, centered at the
\ion{Mg}{2} doublet of the $z=0.2221$ 94\_38 galaxy, is shown in
Figure~\ref{fig:example}.

%%%%%%%%%%%%%%%%%%%%%%%%%%%
\section{Line Measurements}

%-- describe identification, measurement (baseline integration), blends, non-detections, saturation, continuum fitting. \\
%-- Show the data in stacks. 
%-- Tabulate

This section discusses the metal-line measurements obtained from the
HST/COS and Keck/HIRES quasar absorption-line spectra.  The analysis
focuses solely on the absorption systems associated with
spectroscopically confirmed galaxies foreground to our \cosha\ quasar
sample.  We further restrict the analysis to galaxies with impact
parameters $\mRperp \le 160$\,kpc and luminosity $L > 0.1L^*$.    
In several fields, there are
multiple galaxies located within 160\,kpc and having systemic velocity
offsets $|\delta v| < 400 \kms$ (e.g.\ J0928+6025).  In
$\S$~\ref{sec:sample}, we discuss our approach to assigning properties
of the CGM in such cases.  For the following analysis, these issues
only affect the zero point for the velocities reported.

%The association of an absorption system (defined to be the set of
%absorption lines from all transitions at a common absorption redshift
%$z_{\rm abs}$) to a given galaxy is, in principle, a complex process.
%For the \cosha\ sample, however, we have demonstrated that nearly
%every galaxy exhibits strong \ion{H}{1} \lya\ absorption within $100
%\kms$ of the galaxy systemic redshift \citep{tumlinson12}.  Further, there
%is rarely any other strong \ion{H}{1} \lya\ absorption  (equivalent
%width EW~$>100$\,m\AA) within $1000 \kms$ of the galaxy redshift.  
%As such, we are highly confident in associating these strong
%absorption systems with each of our galaxies.  Nearly any set of
%quantitative criteria for establishing an association (e.g.\ the
%strongest line within $\pm 400\kms$) would connect our systems to the
%galaxy.   [wording]

We initiated the metal-line identifications by first searching for
possible \ion{H}{1} Lyman series absorption within 
$\pm 600~ \kms$ of the galaxy spectroscopic redshift.  
\cite{tumlinson12} describe in detail that effort and the results. We then visually inspected the HST/COS and Keck/HIRES spectra at the
expected locations of a large suite of far-UV transitions.  If
positive and consistent absorption was apparent, we defined a velocity
interval to perform further analysis, attempting to maintain a constant
interval for all metal-line transitions.  This process also enabled a
first-pass for flagging blends with Galactic absorption and/or
coincident absorption from unrelated systems.  
% In all but two cases, there was  \ion{H}{1} absorption within $200 \kms$. 

In parallel with this process, we 
generated velocity plots to further inspect the
putative detections, refine the velocity intervals for analysis, and
confirm probable blends.  The plots of positive detections are given
in Figure~\ref{fig:stackend} (multi-page species-stack plots shown at the end of the paper). 
Table~\ref{tab:lowlines} lists all of the transitions analyzed and
the velocity interval for the analysis. For all line measurements, we have used
atomic data from \cite{morton03}.

\subsection{Equivalent Widths}
\label{sec:EW}

For all important transitions not severely compromised by a line-blend or
sky emission, we measured the rest equivalent width
\ew\ and estimated its uncertainty with simple boxcar summation over the
analysis interval. These intervals and the \ew\ measurements 
are given in Table~\ref{tab:lowlines}.  
The errors reported are statistical only;  one may adopt an additional
$\approx 20$\,m\AA\ systematic error owing to uncertainty in continuum placement (of order 5\%). We estimate this typical error by considering an absorption line profile that spans 100 km s$^{-1}$, corresponding to 430\ m\AA\ at 1300 \AA\ (observed), where 5\% continuum placement error leaves us with an uncertainty of $\approx 20$\,m\AA\ . While 20\,m\AA\ is a characteristic value for this uncertainty based on our data, one must bear in mind that the continuum placement uncertainty will more adversely affect the weakest, broadest, and shallowest absorption lines with a low central optical depth. 

In cases when a line is significantly blended with another feature, 
we report the \ew\ measurement as an upper limit. 
We also provide upper limits ($2\sigma$ statistical) for key transtions
when the line is not detected at $3\sigma$ statistical
significance.  For the majority of the COS spectra, this corresponds
to a detection threshold of 
$\approx 60$\,m\AA.  The important exceptions are the \ion{Si}{4} and
\ion{C}{4} doublets which lie at observed wavelengths $\lambda >
1500$\AA\ where the sensitivity of COS is degraded.

\subsection{Apparent Optical Depth Method}

At a spectral resolution of $R \approx 20,000$, the COS spectra will
generally not resolve individual components of the metal-line profiles with characteristic line-widths $b \le 10 \kms$.   Nevertheless,
for weak absorption lines ($W_{\rm r} \lesssim 200$\,m\AA) the
effects of line saturation should be modest.  
%{\bf [Discuss a test from 0405]}

Our primary evaluation of the column densities was derived using the
apparent optical depth method \citep{ss96}, over the same interval
used to measure the equivalent width.  Additionally, we have derived column densities using the curve-of-growth technique,   but they rarely give converged values because of significant component structure of the observed absorption lines. While the  curve-of-growth technique can be more accurate for a single component, or profiles that are dominated by a single strong component,  that condition does not often hold in these strong absorbers.  These measurements and associated error estimates (statistical only) are presented in
Table~\ref{tab:lowlines}.  

If the normalized flux of a given transition
goes below 0.1 at any of the binned pixels, we have automatically
set the column density to be a lower limit (saturated).  Non-detections are listed as
$2\sigma$ (statistical) upper limits.  The values for ions with multiple transitions
were carefully inspected for consistency and signatures of
line saturation.  
The measurements for many transitions are reported as lower
limits due to line-saturation concerns.
In cases with mutiple measurements, we report the
weighted mean in Table~\ref{tab:lowlines}.

\subsection{Profile Fitting}

In addition to the AODM measurements of column densities, we fit Voigt
profiles, where possible,  to the detected absorption features in
order to assess kinematic component structure and improve column
density estimates of severely saturated lines. Additionally, profile fitting is essential for estimating the column densities of severely blended absorption lines, such as NIII $\lambda989$, which is often badly blended with SiII $\lambda989$. The procedure used to
perform the fits and derive the column density $N$, Doppler $b$, and
velocity offset $v$ for each component  is described in greater detail
in \cite{tumlinson12}. In short, it is an iterative fitting program
that makes use of the  \verb1MPFIT1
software\footnote{http://cow.physics.wisc.edu/$\sim$craigm/idl/fitting.html}
to optimize the fit and to generate errors near the best-fit
point. The initial parameters, including the number of components, are
set based on a by-eye examination of the data. Different transitions
of the same ionic species are required to have the same component
structure\footnote{In this process, we include a nuisance parameter
  that allows for wavelength errors (i.e.\ a shift) between
  transitions.  Owing to errors in the COS wavelength solution, these
  can be as large as 30\kms\ \citep[e.g.][]{tumlinson12}.}, 
and are therefore fit simultaneously to give a single
solution. However, we do not impose such requirements on the different
ionization states of the same element.  

We tabulate the results of
this analysis in Table \ref{tab:final_columns}, and include the
AODM-derived column density in the final column for comparison. The
profiles that are fit to the data are shown along with the data in
Figure \ref{fig:stackend}.   In the subsequent analysis, we use the AODM-derived total column densities in figures and throughout most of the analysis. As a conservative measure, we exclude badly blended lines from the analysis and treat saturated lines as lower limits. The total fit column densities are predominantly consistent with the AODM-derived columns, even in cases of complete blending (SiII $\lambda989$ affects the total column density of NIII $\lambda989$ at the 1$-$2\% level), so this choice has no impact on the results and trends that follow. The primary utility of the fits for this work is in a component analysis, which we touch on in sections 4 and 5. 
\section{Results}
In this section, we describe the principal results drawn from the
equivalent width and column density measurements of the low,
and intermediate metal-line transitions associated with
the CGM of $L \approx L^*$ galaxies. In most of this analysis, we distinguish between star-forming
(SF) galaxies and non-SF galaxies according to a strict cut in the
observed specific star-formation rate (sSFR).  The SF galaxies
(depicted with blue points) are required to have ${\rm sSFR} > 10^{-11}
\, \rm yr^{-1}$.   All of the galaxies in
the non-SF sample (${\rm sSFR} < 10^{-11} \, \rm yr^{-1}$)
have no measurable star-formation as estimated from
H$\alpha$ and [OII] emission lines
\citep[Table~\ref{tab:galprops};][]{werk12}.

Unfortunately, the redshifts of our target galaxies generally place
the \ion{C}{4} doublet beyond the wavelength coverage of the COS spectra.
Furthermore, those systems with coverage of \ion{C}{4} frequently have
poor data quality because of the lower sensitivity of COS beyond
$\approx 1500$\AA.  As such, we have only a handful of measurements,
which may suggest a high covering fraction of C$^{+3}$
for SF galaxies. While our dataset does permit the analysis of \ion{Si}{4} for the majority
of systems, the lower COS sensitivity at longer 
wavelengths implies that strong absorption ($\gtrsim 200$\,m\AA)
is required for a positive detection.   We exclude these higher ionization state lines from this analysis (though they are tabulated) since there is little to be learned from them in this specific dataset.

%-- desrcribe the fitting algoirthm. 
%-- tabulate components. 

%\subsection{Notes Individual Systems}
%\label{sec:gals}
%\subsubsection{J0042-1037: 358\_9}
%The total column densities of SiII and SiIV are well constrained by the unsaturated transitions at $\lambda$ = 1260.42 \AA ,  $\lambda$ = 1402.77, 1393.75 \AA,  respectively. Blah blah blah... describe the fits and the component structures. 
%\subsubsection{J0226+0015: 268\_22}
%Nothing useful. No HI even. Must be included in this work, nonetheless. 

%Quite redundant. 
%In the following, we restrict the study to \cosha\ galaxies with $L>
%0.1L^*$, i.e. ignoring any dwarf galaxies detected at small impact
%parameter to these sightlines.  These galaxies will be examined as
%part of future work on less luminous systems.

Because many of the sets of measurements include limits (upper and
lower), we occasionally employ the ASURV software package to perform
statistical analysis \citep{fn85}.  For the bivariate tests, we
implicitly assume that the limits are random with respect to the
galaxies.  Given the range of impact parameters sampled and that the
quasars were selected without any knowledge of the absorption, this
method should hold.  We also perform `standard' statistical tests by treating
the limits as values and discuss
whether this implies a conservative (or risky) assessment.

\subsection{Low-Ions}

As discussed in \cite{thom12} and \citep{ tumlinson12}, the \cosha\ sample almost
uniformly exhibits strong \ion{H}{1} absorption 
($W_{\rm Ly\alpha} \gtrsim 1$\AA, $\mnhi \ge 10^{14} \cm{-2}$).
This indicates a substantial, cool gas-phase for the CGM at $\mRperp <
160$\,kpc from $L^*$ galaxies, independent of galaxy properties.
If this medium
is significantly enriched in heavy elements, then we may expect to
detect at least a trace amount of dominant low-ions (e.g.\ Si$^+$, O$^0$, Mg$^+$).  
For the purposes of this work, we define `low-ions'  to be the first ion of each heavy element that has an ionization
potential exceeding 1\,Ryd. 
These should be the dominant ions in an optically thick gas where
photons with $h \nu > 1$\,Ryd are preferentially absorbed by the
surrounding hydrogen.
%%%%%%%%%%%%%%%%%%%%%%%%%%%%%%
\begin{figure*}[t]
\begin{centering}
\includegraphics[width=0.85\linewidth,angle=0]{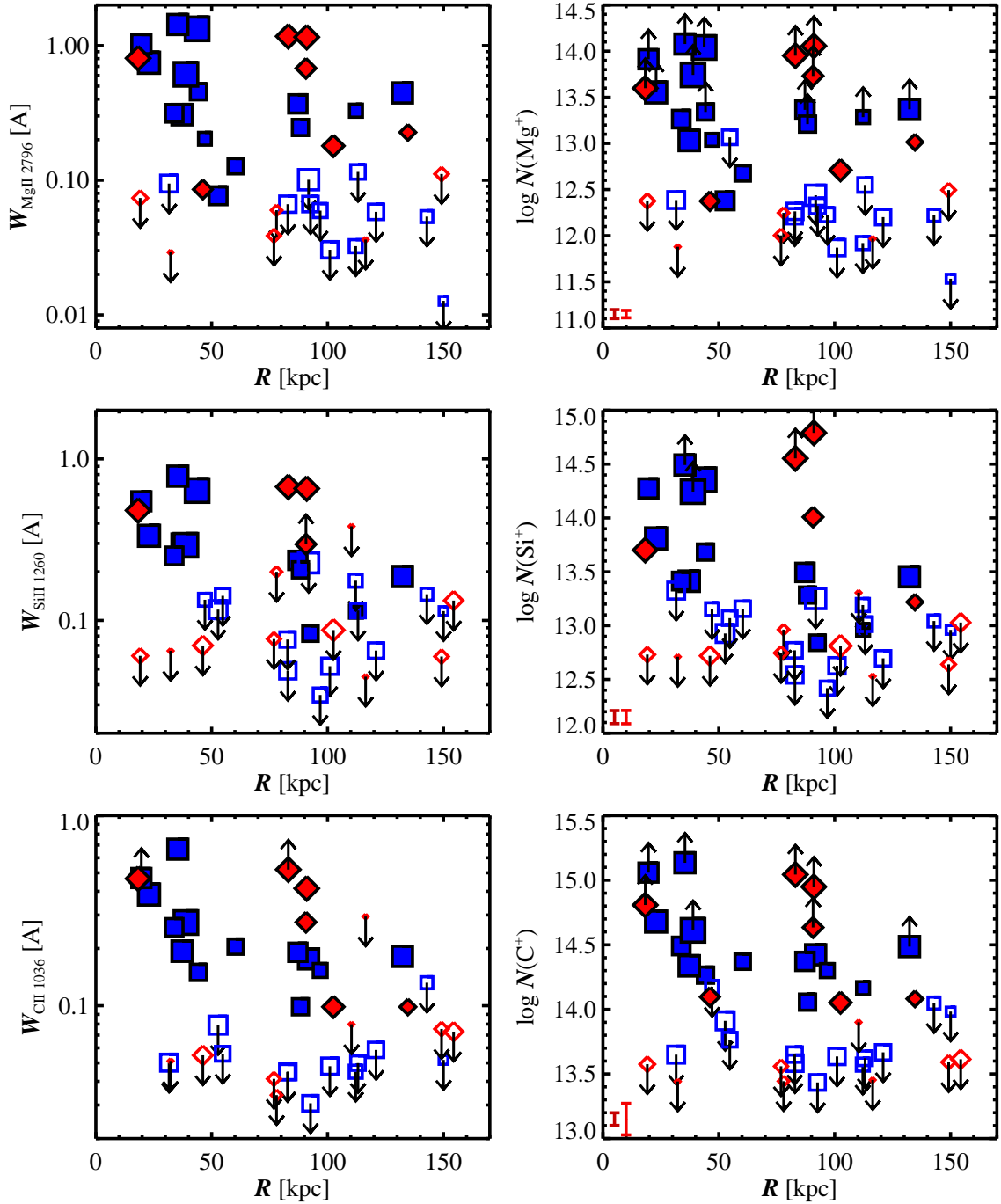}
\end{centering}
\caption{
The equivalent width (left panels) and column
density measurements (right panels) versus impact parameter \Rperp\
of the \cosha\ sample
for the set of the three most commonly detected low-ions: MgII (upper panels), SiII (middle panels), and CII (lower panels).  
In each panel, the symbol and color
distinguish between star-forming (blue squares) and non-SF galaxies
(red diamonds).   The symbol size is linearly proportional to the log\,\nhi\ value
over the range $10^{13} \cm{-2}$ to $10^{17} \cm{-2}$.  
Open symbols with errors denote $2\sigma$ statistical upper limits.
The error bars in the lower left corner of the column density plots
are representative of the measurements.  Note the significant covering
fraction to low-ion absorption at all \Rperp\ values, coupled with 
declining $W$ and $N$ values with increasing \Rperp.
It is also evident that the positive detections are dominated by
systems with larger \nhi\ values.
}
\label{fig:low_ions}
\end{figure*}
%%%%%%%%%%%%%%%%%%%%%%%%%%%%%%%%%
In Figure~\ref{fig:low_ions}, we show the equivalent width and column
density measurements for the three low-ions with a sensitive and
large sample of measurements: Mg$^+$, Si$^+$, and C$^+$.
These quantities are plotted against the physical impact parameter \Rperp\ from the
quasar sightline to the galaxy.  The SF and non-SF galaxies are
denoted by color (blue and red, respectively) and the symbol size is
proportional to the measured \ion{H}{1} column density (scaled linearly from
$\log \mnhi = 13 $ to 17).

Focusing first on the SF galaxies, which dominate the sample, each ion
shows weaker absorption 
at larger \Rperp\
(i.e.\ lower equivalent widths and column densities). 
%Table~\ref{tab:ewcolm_stats}
Similarly, the fraction of sightlines with non-detections increases
with \Rperp\ (Table~\ref{tab:cover_ew}).  For example, the median
equivalent width of \ion{C}{2}~1036 (\ion{Mg}{2}~2796)
$\tilde W_{1036}$ ($\tilde W_{2796}$) drops from 204(314)\,m\AA\ for
$\mRperp < 75$\,kpc to 59(85)\,m\AA\ for $\mRperp > 75$\,kpc while
the median low-ion column density drops from $\Ntild{C^+} = 10^{14.4}$
to $10^{14.0} \cm{-2}$ for C and
$\Ntild{Mg^{+}} = 10^{13.3}$
to $10^{12.3} \cm{-2}$ for Mg.
One draws similar results from estimates of the average values, although
such analysis is complicated by the presence of both upper and lower
limits.   A two-sample log-rank test using the ASURV package on the 
$W_{2796}$ or $\N{Mg^+}$ values\footnote{Taking the lower limits on
  $\N{Mg^+}$ as standard detections.}
rules out the null hypothesis that the $R<75$\,kpc and $R>75$\,kpc
subsets being drawn from the same population at $98\%$ confidence.

One concludes, at high confidence, that the
average surface density of low-ions decreases with increasing impact
parameter.   This must reflect a decreasing total
surface density, a decreasing metallicity, and/or an increasing ionization
state with increasing \Rperp. Since it is likely that the HI column density also decreases with increasing impact parameter \citep{tumlinson12}, an outwardly decreasing CGM metallicity gradient is the least likely to drive the main effect. %[If HI column is decreasing too
%(likely), then suggest the effect is not dominated by metallicity
%gradient] 
Despite the decreasing column of low-ion gas with increasing \Rperp, we report positive
detections to at least 130\,kpc and lack sufficient sample size to
conclude that there is negligible low-ion absorption at larger
radii.\footnote{Of course, the presence of neighboring galaxies will preclude 
zero absorption.}  Future work should assess the covering
fraction of low-ion gas to the virial radius (and beyond) for $L
\approx L^*$ galaxies.

One cannot draw strong conclusions on the
behavior of the low-ions for the smaller, non-SF sample.  There is a low detection rate at $\mRperp<75$\,kpc ($1/4$ sightlines), but 
a comparable or even higher rate of detection than the SF sample
at $\mRperp > 75$\,kpc.  The dominant factor in a positive detection appears to be
the \nhi\ value, which is uncorrelated with \Rperp\ for this small set
of galaxies.   

%Another notable result is that the distribution of
%measurements for the non-SF galaxies appears bimodal.  The positive
%detections generally occur at 0.5 to 1.0\,dex larger values than the
%non-detections.

We now consider how the low-ion absorption relates to other absorption
properties of the system and characteristics of the associated
galaxies.  We derive the quantity \nlow\  for each system, 
defined as follows:
(1) \nlow=$\N{Si^+}$, if the Si$^+$ measurement is a value or lower limit;
(2) \nlow=$\N{Mg^+}$, if the Si$^+$ measurement is an upper limit
(or there is none recorded)
  and a Mg$^+$ measurement exists.
We choose Si$^+$ as the primary ion because it has multiple
transitions in the far-UV bandpass with a range of oscillator
strengths yielding more reliable column density estimates.  The
\ion{Mg}{2} doublet, meanwhile, offers more sensitive upper limits.
Because we measure $\N{Si^+} \approx \N{Mg^+}$ in cases where
both are measured, we apply no offset when adopting one versus the
other.   In all, there are roughly half of the systems in each category.

\begin{figure}[h!]
\begin{centering}
\includegraphics[height=0.99\linewidth,angle=90]{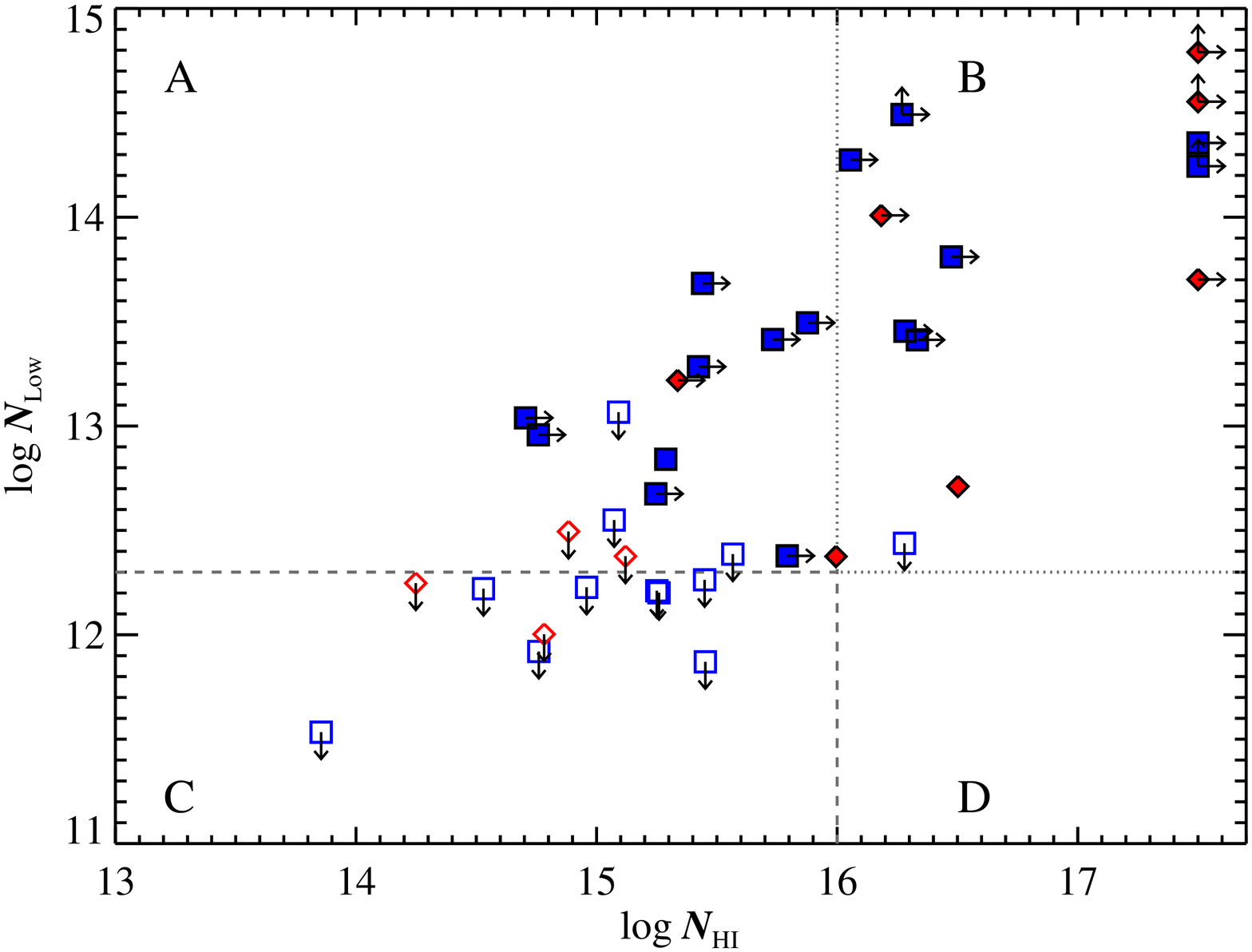}
\end{centering}
\caption{The \nlow\ value, generally $\N{Si^+}$ for values and
lower limits and $\N{Mg^+}$ for upper limits, against the measured
\nhi\ value for the \cosha\ sample.  
The dotted/dashed lines demarcate four regions of the plot:
(A)  $\mnhi \le 10^{16} \cm{-2}$ and $\mnlow \ge 10^{12.3} \cm{-2}$, where the gas highly ionized; 
(B) $\mnhi \ge 10^{16} \cm{-2}$ and $\mnlow \ge 10^{12.3} \cm{-2}$,
where the majority of positive detections for low-ions lie; and
(C) $\mnhi \le 10^{16} \cm{-2}$ and $\mnlow < 10^{12.3} \cm{-2}$ where
the majority of non-detections are located; and (D) $\mnhi \ge 10^{16} \cm{-2}$ and 
$\mnlow < 10^{12.3} \cm{-2}$, a region that would indicate metal poor gas. This region is unoccupied, suggesting the CGM of $L^*$ galaxies is significantly
enriched. Another key point shown by this figure is that low-ion column density is significantly
coupled to \ion{H}{1} column density, with detections essentially
requiring a non-negligible opacity at the \ion{H}{1} Lyman limit.
}
\label{fig:low_NHI}
\end{figure}

In Figure~\ref{fig:low_NHI}, we plot \nlow\ against \nhi\ for the full
sample.   Despite the preponderance of limits, there are several
results to glean from the measurements (illustrated, in part, by the
dashed and dotted lines on the figure).  First, there is a complete
absence of systems with $\mnhi \ge 10^{16} \cm{-2}$ and $\mnlow <
10^{12.3} \cm{-2}$.  In fact, only two of the 12 systems with these
\nhi\ values have $\mnlow
< 10^{13.3} \cm{-2}$.  This suggests a low incidence of
metal-poor gas in the CGM of $L^*$ galaxies \citep[e.g.][]{ribaudo11}.  
Second, with only one
exception, sightlines definitively exhibiting $\mnhi \le 10^{16}
\cm{-2}$ (ignoring lower limits) are all consistent with having
$\mnlow < 10^{12.3} \cm{-2}$.  (One notable counterexample from the literature is the sightline studied by Tripp et al.  2011\nocite{tripp11} which has
many components with $\mnhi \le 10^{16}
\cm{-2}$  and very extensive low-ion absorption.) While in principle this could result from lower metallicity at lower
\nhi, we suspect this result follows from the ionization state of the
gas possibly combined with a lower surface density of total hydrogen.
Third, all systems with $\mnlow \ge 10^{13} \cm{-2}$ are consistent
with $\mnhi > 10^{16} \cm{-2}$  and all systems with $\mnlow <
10^{13.3} \cm{-2}$ have $\mnhi < 10^{16} \cm{-2}$.
In at least rough terms, therefore, $\mnhi = 10^{16} \cm{-2}$
demarcates a relatively rapid transition from significant to nearly negligible low-ion
absorption.  By the same token, the presence of
significant low-ion absorption (e.g.\ a \ion{Mg}{2} doublet) implies
gas with a non-negligible opacity at the Lyman limit, at least for
sightlines intercepting the CGM of an $L^*$ galaxy.

Turning to the relation of low-ion absorption with galaxy properties,
the top two panels of Figure~\ref{fig:lowint_gal} plot \nlow\ against (a) SFR and (b)
stellar mass, with the symbol size inversely proportional to the impact
parameter.   In terms of SFR, there may be a general trend of lower \nlow\
values (average or median) with decreasing SFR but there is
substantial scatter at all values.  This scatter appears well
correlated with \Rperp\ for the SF galaxies but possibly
anti-correlated with \Rperp\ for the non-SF galaxies. 
In any case, we conclude that it is 
unlikely that SFR is a dominant factor driving
the strength of low-ion absorption, which runs contrary to the inferences drawn by \cite{menard09}. 

In contrast, the \nlow\ values appear well-correlated with stellar
mass, indicating that there is more circumgalactic gas in more massive galaxy halos.   For example, the galaxies with $M_* > 10^{11} \mmsun$ show the
highest \nlow\ values and a high incidence of positive detections
(6/7) even though the sample is almost exclusively red, non-SF
galaxies.   
Meanwhile, the non-SF galaxies with $M_* < 10^{11} \mmsun$ are
dominated by non-detections.  In fact, restricting the evaluation to
the non-SF galaxies, there is a significant trend with $M_*$.   
Even with the small sample size, the non-SF galaxies with $M_* <
10^{11} \mmsun$ have a distribution of \nlow\ values that differs from
the $M_* > 10^{11} \mmsun$ at 95\%\ confidence.  
This cannot be simply attributed to differences in impact parameter between
the two sub-samples.
Despite an apparent trend of decreasing \nlow\ with decreasing $M_*$
for the SF galaxies with $M_* < 10^{11} \mmsun$, the null hypothesis
of no correlation is not ruled out at very high confidence ($94\%$) 
by a generalized Kendall-tau test. 
If such a trend exists, it will require a larger sample of galaxies,
possibly controlled for impact parameter.

\subsection{Intermediate-Ions}
\label{sec:inter}

For the purposes of this work, we define an `intermediate ion' as the first species of an element which requires 
an excess of 1\,Ryd in energy to produce it. 
For example, the intermediate ions of C and Si are C$^{++}$ and
Si$^{++}$, 
which require 24.4\,eV and 16.3\,eV of energy respectively to ionize
the previous (low-ion) stage. 
Owing to their higher ionization potential, one predicts that these
ions trace more highly ionized gas, e.g.\ the inner portions of an
\ion{H}{2} region.

\begin{figure*}[t!]
\begin{centering}
\hspace{0.65in}
\includegraphics[width=0.67\linewidth,angle=0]{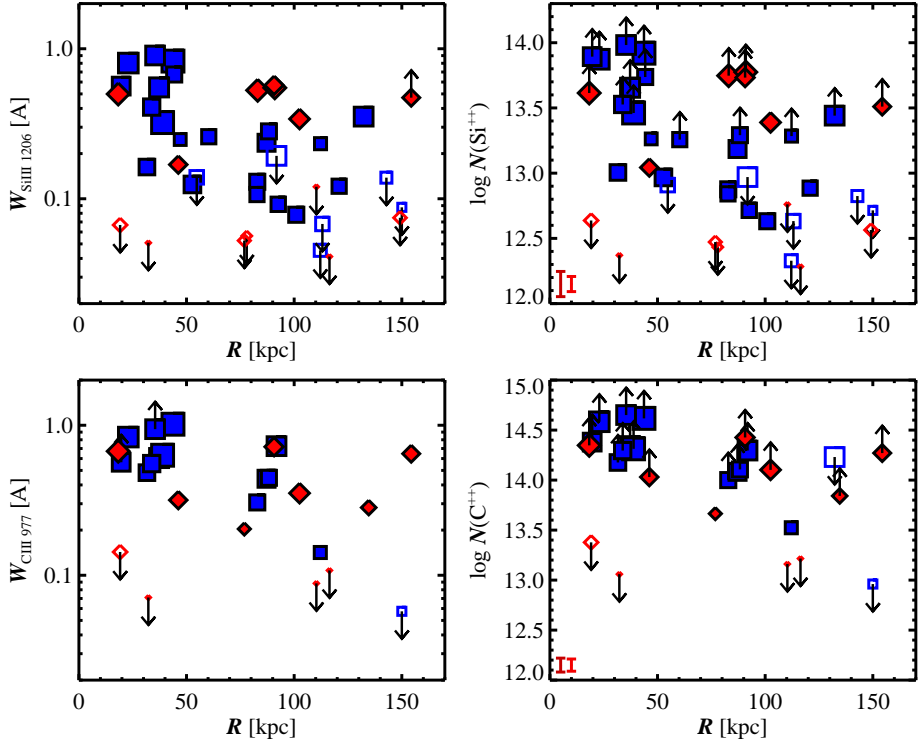}
\end{centering}

\caption{
Similar to Figure~\ref{fig:low_ions} but for the intermediate ions of
Si$^{++}$ (upper) and C$^{++}$ (lower).  There is a high incidence ($>80\%)$
of detections for these ions, espcially within the SF population.
There is also a very significant anti-correlation ($>99\%$ c.l.)
between intermediate-ion absorption strength and impact parameter.
Despite the lower incidence of positive detections among the non-SF
galaxies, the detected values
match or even exceed the typical values of SF
galaxies. 
}
\label{fig:inter}
\end{figure*}

In the far-UV, there are two particularly strong transitions from
these intermediate-ions, 
\ion{Si}{3}~$\lambda$1206 and
\ion{C}{3}~$\lambda$977. 
Measurements of their equivalent widths and column densities for
the \cosha\ sample
are presented in Figure~\ref{fig:inter}.  It is
immediately evident that the sightlines show a high covering fraction,
even exceeding the observed values for the low-ions.  For
\ion{C}{3}~977, the SF galaxies exhibit a nearly 100\%\ $C_f$
(Table~\ref{tab:cover_ew}) with the only non-detection  in 14 cases
having $\mRperp \approx 160$\,kpc.
Within $\mRperp = 50$\,kpc, the results are especially striking.  The
SF galaxies uniformly exhibit equivalent widths $W_{977} > 400$\,m\AA\
corresponding to lower limits on $\N{C^{++}}$ of $10^{14.2} \cm{-2}$
and higher.   Some of the values may even exceed the observed
\ion{H}{1} column densities.

\begin{figure*}[t!]
\begin{centering}
\hspace{0.55in}
\includegraphics[height=0.75\linewidth,angle=90]{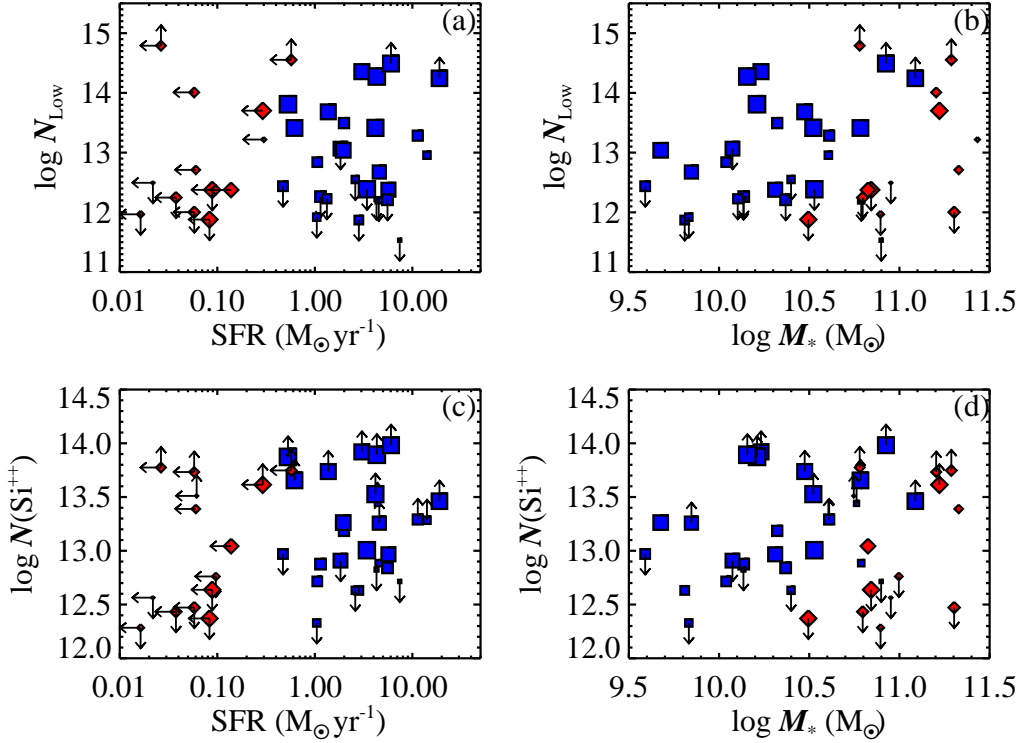}
\end{centering}
\caption{
Panels (a) and (b) show the low-ion column density against the galaxy
properties SFR and stellar mass.  Regarding SFR, there is little
correlation between the two measurements;  the only notable result is
a possible bimodality in the \nlow\ values for the non-SF population.
In terms of stellar mass, there are two key results: 
(1) an apparent trend of increasing \nlow\ with increasing $M_*$ and
(2) a nearly perfect separation between the measurements of the SF and
non-SF populations.
In all panels, the symbol size is inversely proportional to the impact
parameter.  The lower panels (c) and (d) show the Si$^{++}$ column
density against the galaxy properties.  Again, we note no obvious
trend with SFR but observe a strong dependence with $M_*$.
}
\label{fig:lowint_gal}
\end{figure*}

%%%%%%%%%%%%%%%
%%%% Intermediate-ions

\afterpage{\clearpage}

\begin{figure*}
\begin{centering}
\hspace{0.4in}
\includegraphics[height=0.75\linewidth,angle=90]{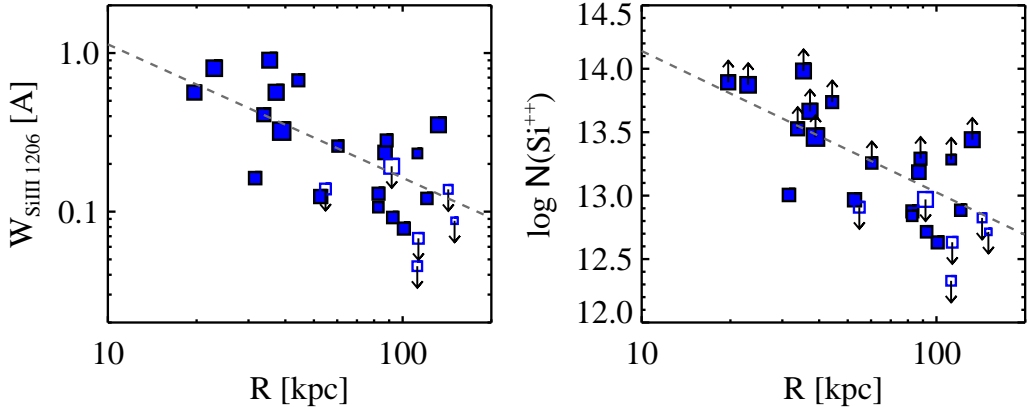}
\end{centering}
\caption{
Fitted equivalent widths and column densities for Si$^{++}$ as a
function of impact parameter, restricted to the SF galaxies.  If we
(incorrectly) treat all measurements as values, we measure power-law
exponents for the variation in \ew\ and $N$ with \Rperp:
$\alpha_W = \alphwv$ and $\alpha_N = \valphn$.
A proper treatment of limits would only steepen these slopes.
}
\label{fig:si3_fit}
\end{figure*}

The results for Si$^{++}$ are similar, albeit with a greater dynamic
range of equivalent widths and column densities, and somewhat
lower $C_f$ values. Si$^{++}$ ( f$_{\lambda,1206}$ = 1.660) has a larger oscillator strength than Si$^{+}$  (f$_{\lambda, 1260}$  = 1.007),  which may play a role in this difference. Nonetheless, the greater dynamic range reveals a key result:
the absorption strength of intermediate ions is highly correlated with
impact parameter.  Restricting to the positive detections of the SF
galaxies and treating all of these as 
values\footnote{A proper treatment of the limits, lower and upper,
  would only strengthen the statistical significance.}, we measure a
Spearman's rank correlation of $-0.65$ 
and $-0.69$ for $W_{1206}$ and $\N{Si^{++}}$ vs.\ \Rperp\, implying an
anti-correlation at $99.95\%$ and 99.99\%\ confidence.
Motivated by the strong anti-correlation, we modeled the radial
variation in $W_{1206}$ and $\N{Si^{++}}$ with single power-law
expressions:

\begin{equation}
W_{1206} = W_{(10\,\rm kpc)} \ltp \frac{R}{10\,{\rm kpc}} \rtp^{\alpha_W} 
\end{equation}
\begin{equation}
\N{Si^{++}} = N_{\rm (10\, kpc)} \ltp \frac{R}{10\,\rm kpc} \rtp^{\alpha_N} 
\end{equation}
In each case, we have restricted the analysis to the positive
detections and have taken lower limits at their measured value.
Minimizing $\chi^2$ with equal weighting for \\
\\
\\
\vspace{0.2in}
each measurement, we derive 
$\log W_{(10\, \rm kpc)} = \lgwv$,
$\alpha_W = \alphwv$ and 
$\lgnr = \vlgnr$,
$\alpha_N = -1.11 \pm 0.29$.
This simple model is overplotted on the data in
Figure~\ref{fig:si3_fit}.
The preponderance of lower limits to \nsit\ at $R<50$\,kpc implies a
even higher $\lgnr$ and steeper $\alpha_N$ values than
reported by this simple analysis.  We discuss the implications and
possible origins of this central result in $\S$~\ref{sec:discuss}.

\begin{figure}
\begin{centering}
\includegraphics[height=0.90\linewidth,angle=90]{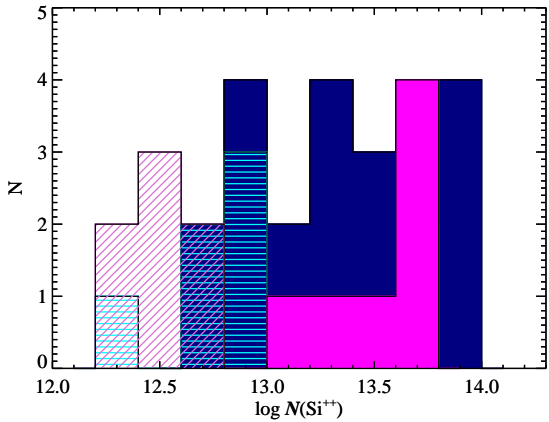}
\end{centering}
\caption{
Histogram of Si$^{++}$ column densities for the star-forming galaxies
(solid blue = detections and lower limits; cyan hatches are upper
limits) and non-SF galaxies (solid magenta = detections and lower limits;
pink hatches are upper limits).  Whereas the values for the SF
galaxies exhibit a relatively uniform distribution of values, the
non-SF galaxies have a bimodal distribution of positive detections and
upper limits.  This implies that the CGM of the non-SF galaxies has a
bimodal behaviour either locally (e.g.\ a patchy distribution around
each galaxy) and/or globally (the CGM exists around some but not all
non-SF galaxies).
}
\label{fig:Si3_hist}
\end{figure}

Regarding the non-SF galaxies, the incidence of positive detections
for the intermediate-ions is lower than that observed for the SF
galaxies, with $C_f \approx 50\%$ for Si$^{++}$ and C$^{++}$.  On the other hand,
the equivalent widths and column densities of the positive detections
are comparable or even higher than those for the SF galaxies.  This
gives a bimodal distribution in the non-SF galaxy subset which is
not evident in the SF galaxy sample (Figure~\ref{fig:Si3_hist}).
The implication is that the gas is either less smoothly distributed
within individual halos of non-SF galaxies and/or that there is a
bimodal separation in the CGM amongst the galaxies themselves.
A similar result holds for C$^{++}$ as well.  A less likely option to explain this result would be that all of the non-SF galaxies for which we have detected low and intermediate ion absorption have a fainter, SF galaxy in their vicinity that is associated with the gas. We have not completed a redshift survey of all faint sources in the COS-Halos quasar fields, and the SDSS images are not sensitive to galaxies fainter than ~0.1L$^{*}$ at z$\sim$ 0.2, so it is impossible to comment in detail about this option. However, we do note that a visual inspection of the SDSS images of the non-SF galaxy QSO fields yields very few additional  blue L$>$0.1L$^{*}$ candidates within two arcminutes. Even if there are other star-forming galaxies in their vicinity, the CGM gas is nonetheless 
physically associated with the elliptical and very likely bound to its dark matter halo. Establishing the origin of the CGM is beyond the scope of this empirically-focused paper.

To explore further the characteristics of the intermediate-ions, we
plot in the lower two panels of Figure~\ref{fig:lowint_gal} \nsit\ against the (c) SFR and (d)
stellar mass of the associated galaxies.  Similar to the low-ion
absorption, there is no obvious correlation with intermedate-ion
column density and SFR.  
%For the SF galaxies there are
%non-detections and lower limits at all SFR values.  Perhaps the only
%result worth noting is an apparent bimodality within the non-SF
%population where the detections all exhibit $\mnsit > 10^{13.3}
%\cm{-2}$ and all but one non-detection has $\mnsit < 10^{12.8}
%\cm{-2}$, i.e. an $\approx 0.5$\,dex offset.
%[This is repetitive with above]
Regarding $M_*$, however, one identifies notable trends.  Restricting
first to the SF population, the \nsit\ values appear correlated with
$M_*$;  the null hypothesis is ruled out at 99\%\ confidence by a
generalized Kendal-tau correlation test.  Similarly, the median column
density for $M_* > 10^{10.5} \mmsun$ is 
$\Ntild {Si^{++}}= 10^{13.4} \cm{-2}$  and 
$\Ntild {Si^{++}}= 10^{12.9} \cm{-2}$  for $M_* < 10^{10.5} \mmsun$.
Within the non-SF population, there is a
preponderance of low \nsit\ values for $M_* < 10^{11} \mmsun$ (6/8).
Indeed, this stands in stark contrast to the SF population.  
%[Do these non-SF, non-detections show OVI??]
The notion of a correlation, however, is tempered by several
positive detections at these masses.
Another notable result revealed by Figure~\ref{fig:lowint_gal} is that
the quiescient galaxies are well separated from the SF galxies within
the $M_*$ panels, both low and intermediate-ions.

\begin{figure*}
\begin{centering}
\hspace{0.4in}
\includegraphics[height=0.80\linewidth,angle=90]{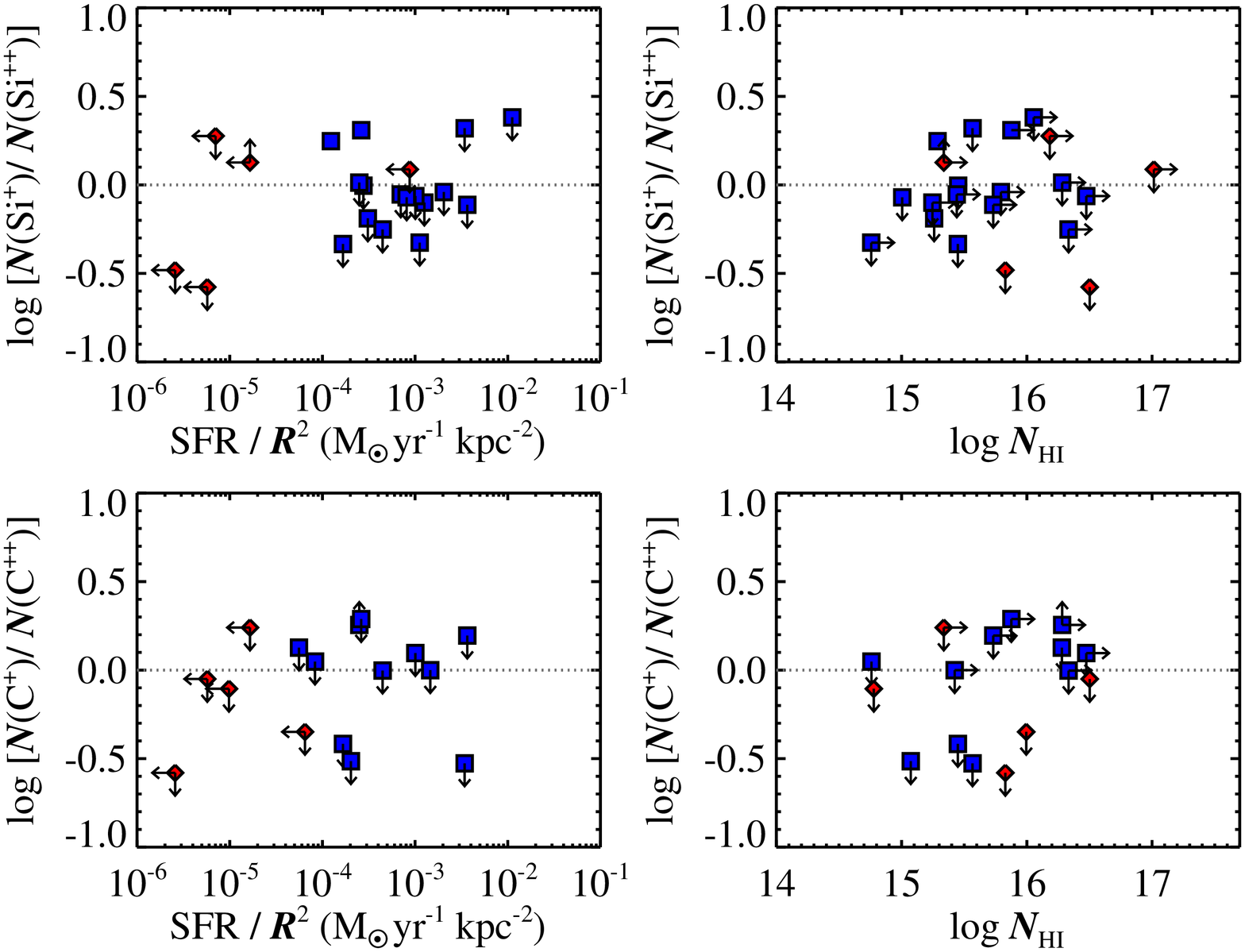}
\end{centering}
\caption{
Comparison of the low (Si$^+$, C$^+$) and intermediate (Si$^{++}$,
C$^{++}$) ionic column density ratios against a crude estimate of the
ionizing flux from star-forming galaxies (SFR/$R^2$) and the
\ion{H}{1} column densities of the gas \nhi.  The principal result is that
the ratios are low: $\N{Si^{+}} / \N{Si^{++}}$ and
$\N{C^+}/\N{C^{++}}$ are less than one for the vast majority
of systems and nearly all are consistent with falling below unity. 
This result indicates that the CGM gas of $L^*$ galaxies is highly ionized.
Note that 22(27) systems are not plotted for Si(C) because both
species are lower limits or upper limits and/or there was no coverage of CIII $\lambda977$ in the data.
}
\label{fig:low_vs_inter}
\end{figure*}

In concluding this section, we compare the column densities and component structure of
the low and intermediate-ions.  We supplement this analysis by comparing these profile fits with those of O$^{+5}$ \citep{tumlinson11}. Figure~\ref{fig:low_vs_inter} plots
the \nsiw/\nsit\ and \nciw/\ncit\ ratios against two quantities which
may be expected to correlate with the ionization state of the gas:
\ion{H}{1} column density and the SFR scaled inversely by the impact parameter
squared.  For the former, one expects stronger low-ion absorption as
\nhi\ increases and the gas becomes optically thick to ionizing
radiation. The quantity SFR/$\mRperp^2$, meanwhile, represents  an estimate of the ionizing flux from massive stars. This
assertion is subject, of course,  to uncertainties in the escape
fractions of galaxies, to error in using
\Rperp\ as the distance from source to gas, and to the relative flux of the galaxy
to the extragalactic UV background.  

The main point to emphasize from Figure~\ref{fig:low_vs_inter} is
that the \nsiw/\nsit\ ratios are uniformly low;  nearly every
system\footnote{We caution that 5 systems have lower limits to both
  \nsiw\ and \nsit\ and 13 systems have non-detections for both ions.}
is consistent with \nsiw/\nsit~$< 1$.  If the gas arises in a single
phase, this requires that the majority of CGM gas 
is significantly ionized.  One draws the same conclusion
from the observed \nciw/\ncit\ ratios.  Inspection of
Figure~\ref{fig:low_vs_inter} also reveals there are no obvious
trends with SFR/$\mRperp^2$ or \nhi.  Perhaps the only noteworthy
observation in this regard is that the two systems with \nsiw/\nsit\
exceeding unity have amongst the lowest SFR/$\mRperp^2$
values in the star-forming population.

\begin{figure*}
\begin{centering}
\hspace{0.4in}
\includegraphics[width=0.80\linewidth]{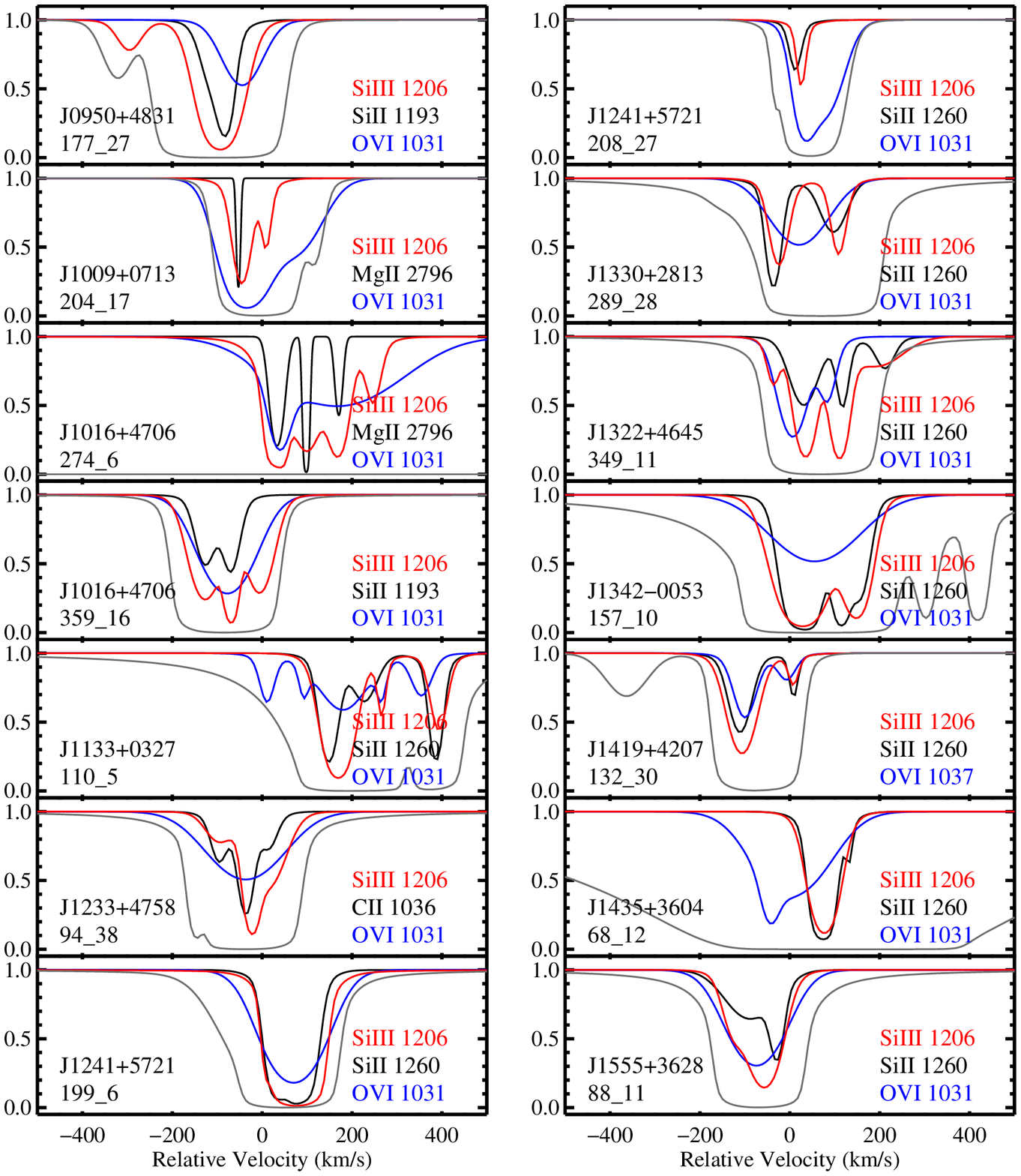}
\end{centering}

\caption{
Comparison of the normalized profile-fits for the low (black),  intermediate (red)
and high (O$^{+5}$, blue) ion transitions for selected systems where there is a strong
detection of each. We also show the profile fits to the HI L$\alpha$ line (saturated in each of these cases) in dark gray for reference. 
}
\label{fig:prof_compare}
\end{figure*}

If the low and intermediate ions arise from a single phase, then one
would expect the kinematic component structure of their absorption
profiles to be similar. 
We may test this prediction through an analysis of the profile fits
generated from the absorption-lines (Table~\ref{tab:final_columns}).
As we impose no restrictions on the component
structure between different ionization states of the same element, any
qualitative and quantitative similarities between the fits arise
naturally. Indeed, a visual inspection of the fitted absorption
features indicates strong similarities between low and intermediate
ionization states of Si, C, and N (figures~\ref{fig:stackend} and \ref{fig:prof_compare}).  
Errors in the COS wavelength solution
should only weaken such correlations. Generally, 
the intermediate ion absorption tends to be stronger than the low ion
absorption in the CGM of $L^*$ galaxies, but there is no evidence that the
low and intermediate ionic absorption arise from different gas phases
based on their component structure alone.  

We have also shown in Figure \ref{fig:prof_compare} the profile fits
to the O$^{+5}$ absorption line, which were derived from our COS dataset
and analyzed previously by \cite{tumlinson11}. Unlike the clear
correspondence between the black (low-ion) and red (intermediate-ion)
lines, a qualitative assessment of the overall agreement and component
structure of the blue (high-ion: O$^{+5}$) lines is complex. In nearly
all cases, the individual fitted components of the O$^{+5}$ absorption
lines are broader than those of the lower ionization states, but there
is generally good correspondence between the shape and number of
components.  There are impressive alignments (e.g. J1419$+$4207
132\_30; J1009$+$0713 204\_17),  but just as many, if not more,
complete misalignments (e.g. J1330$+$2813 289\_13; J1435$+$3604
68\_12). This comparison suggests the relationship between low and
high ionization states of gas along the same sightlines is not
straightforward, and commenting further here is beyond the scope of
this paper. Future work will perform a quantitative comparison of
these profiles and examine the implications for the origins and overlap of
this broad range of circumgalactic gas ionization states.

\subsection{Covering Fraction ($C_f$)}
\label{sec:Cf}

A  valuable measure of assessing the distribution of
absorption by the CGM is through estimations of the covering fraction
$C_f$.  Strictly speaking, the $C_f$ value describes the
fraction of random sightlines that pass within a given impact
parameter to a galaxy (or population of galaxies) that exhibit a
specified absorption strength for a given transition or ion.
For example, one can measure the fraction of sightlines exhibiting $W_{\rm r, 2796} >
0.3$\AA\ when passing within\footnote{In principle, one must also
  define $C_f$ over a velocity interval relative to the galaxy.  This
  is important for \ion{H}{1} absorption where one may be confused by
  the intergalactic medium.  Heavy elements are sufficiently rare, 
  however, that the results presented here are insensitive to this issue
  provided one chooses a sufficiently large window ($>300\mkms$). } 
50\,kpc of a star-forming $L^*$ galaxy at
$z=0$.  The covering fraction assesses how gas in the CGM is distributed (as
projected on the sky), gives a rough assessment of the strength of
absorption, and reduces complex distributions to a single number.
Furthermore, it presents a simplified target for theoretical models to
consider. 

Despite its apparent simplicity, estimating the $C_f$ values for the
equivalent width of a specific transition or the column density of a
given ion is non-trivial.  This is due to several factors.  First, we
have a limited sample of galaxies under study and cannot claim to have
a complete sample nor (necessarily) even a fully representative sample
of any given population.  Second, 
we have limited sampling of the CGM
on scales of $\mRperp < 30$\,kpc.  
Third, the projected area scales with radius whereas we designed the
\cosha\ project to uniformly sample impact parameter from $\mRperp
\approx 30-160$\,kpc.  A strict comparison of models to our dataset
must account for these issues.
Lastly, both the equivalent width and column
density measurements include a mix of measurements and upper/lower
limits.    Given these limitations, we proceed conservatively.

For the covering factors relating to the equivalent width measured for
a specific transition (e.g.\ $W_{\rm r, 2796}, W_{\rm r, 1334}$), we chose
thresholds that exceeded nearly all of the observed upper limits.
These values are listed in column~4 of Table~\ref{tab:cover_ew}.  In
cases where an upper limit exceeds this threshold, we included that
measurement as if it exceeds the theshold, thereby increasing $C_f$.
The error reported in $C_f$,  which follows a standard binomial Wilson
score, accounts for this uncertainty by calculating the 68\%\
confidence interval assuming that upper limits above the threshold do
not satisfy it.

The evaluation of $C_f$ for column density measurements are more
challenging because they often include a mix of upper and lower
limits.  Again, we aimed for a column density threshold that
exceeded the upper limits while not exceeding the lower limit values.
Similar to the $C_f$ for \ew\ values, we treated upper limits above
the threshold as detections.  We also considered lower limits below
the line as satisfying the threshold.  In this respect the resultant $C_f$
values may be considered maximal, aside from the uncertainty of
Poisson statistics. 

\begin{figure}
\begin{centering}
\hspace{-0.1in}
\includegraphics[width=0.99\linewidth,angle=0]{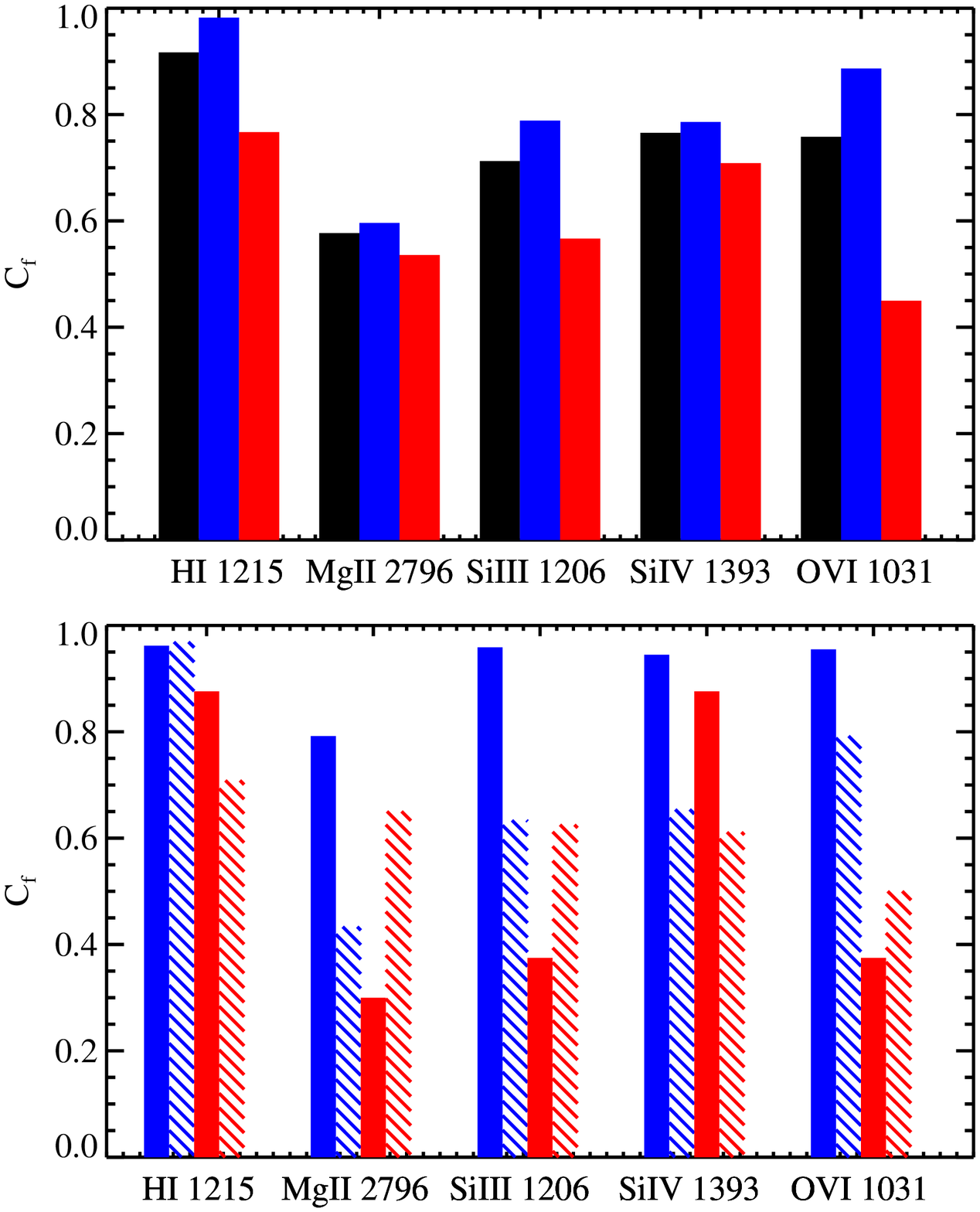}
\end{centering}
\caption{
(upper):
Covering fraction $C_f$ of ionic absorption (equivalent widths)
for the \cosha\ sample of galaxies, as measured to $\mRperp =
160$\,kpc (Table~\ref{tab:cover_ew}).  
The black/blue/red bars indicate the full/SF/non-SF samples respectively.  
The high incidence of \ion{H}{1} gas requires a cool CGM surrounding
$L^*$ galaxies \citep{tumlinson12}, and the significant $C_f$ values
for the various metal-line transitions implies a metal-enriched gas.
Within the metal-line transitions, one notes that $C_f$ rises with
increasing ionization state, especially for the SF population.
(lower):
Covering fraction $C_f$ of ionic absorption (equivalent widths)
separated by impact parameter: solid bars refer to $\mRperp <
75$\,kpc and hashed are for $\mRperp \ge 75$\,kpc.
Within the SF population (blue), there is a clear difference between the two
radial cuts, aside from \ion{H}{1} absorption.  This suggests 
declining metal-enrichment with increasing \Rperp\ which we will
explore in a future paper \citep{werk12b}.   Oddly, the opposite trend
is noted for the quiescent galaxies (red).  We are concerned, however,
that this is driven by small sample size of this sub-population.
}
\label{fig:covering}
\end{figure}

In Figure~\ref{fig:covering} we present a comparison of the $C_f$
values for transitions spanning a wide range of ionization state.
The results for \ion{H}{1} and O${+5}$ are drawn from the
\cite{tumlinson12} and \cite{tumlinson11} papers, respectively.  It is
evident that the \ion{H}{1} gas exhibits the highest covering
fraction, not only for the SF galaxies; 
$L^*$ galaxies of all spectral
type exhibit a CGM of cool, hydrogen gas \citep{thom12}.  
Examining the metal-line
transitions, which are arranged from lowest to highest ionization
potential, one observes a significant covering $(C_f > 0.5)$ for all ions
and populations except O${+5}$ for the non-SF galaxies

%%%%%%%%%%%%%%%%

%%%%%%%%%%%%%%
\citep{tumlinson11}.  
This demonstrates significant metal-enrichment throughout the CGM of
$L^*$ galaxies.  It further suggests a multi-phase medium because it is
nearly physically impossible to find strong
low-ion and O${+5}$ absorption in the same gas. 
One also observes a trend of increasing $C_f$
with increasing ionization state.  For the SF galaxies, there is a
monotonic increase in $C_f$ from 60\%\ in the low-ion species to
nearly 90\%\ for the highly ionized O${+5}$ gas.
This implies that the gas is signifcantly ionized.
The non-SF galaxies also show increasing $C_f$ until O${+5}$ where
one finds much less gas \citep{tumlinson11}.  
The CGM of quiescient $L^*$ galaxies is also
enriched, but that there is a qualitative difference between the
two populations at the highest ionization states.

We explore variations in $C_f$ with impact parameter in
the lower panel of Figure~\ref{fig:covering}. 
Within the SF population (blue), there is a clear difference between the two
radial cuts for metals such that the metal covering fraction is lower for higher impact parameters. While an declining metallicity gradient could be responsible, this trend is more likely the result of the decreasing total gas surface density with radius \citep{werk12b}.  Oddly, the opposite trend
is noted for the quiscient galaxies (red).  We are concerned, however,
that this may be masked by the small number statistics of this
sub-population. 

%This suggests declining metal-enrichment with increasing \Rperp\ which we will explore in a future paper \citep{werk12b}.
%[Consider showing fig\_ion\_profile.ps here]

\subsection{Comparison to Previous Work}
\label{sec:previous}

As a check on our analysis, we may wish to compare our results
against similar measurements from the literature.  
Such comparisons may reveal systematics in our line-measurements or
anomalies in the sample definition.
Unfortunately, very few
analyses have been performed on the low and intermediate ionization
states of metal-line absorption in the CGM of $L^*$ galaxies.  The
obvious exception is Mg$^+$, which has been extensively surveyed
over the past several years \citep[e.g.][]{bc09,chg+10}, and to a lesser-extent C$^+3$ (not covered by this work), which is found to extend out to 150 kpc \citep{clw01}. 

\begin{figure}
\begin{centering}
\hspace{-0.3in}
\includegraphics[width=0.90\linewidth,angle=90]{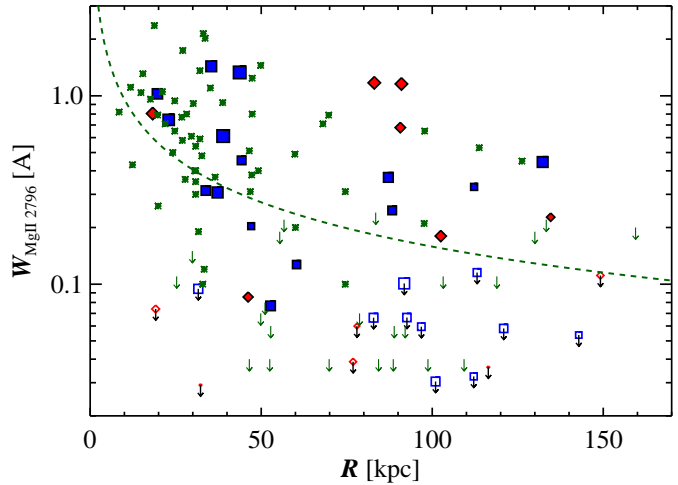}
\end{centering}

\caption{
The equivalent width of the MgII $\lambda$2796 line as a function of impact parameter for the COS-Halos data (red and blue points) and comparing to the data published by Chen et al. 2010 (green asterisks and arrows). We also show the fit to the data from Chen et al. 2010. 
}
\label{fig:mg2}
\end{figure}

In Figure~\ref{fig:mg2} we present a comparison of our $W_{\rm r, 2796}$
measurements against impact parameter with those reported by
\cite{chg+10}.  We also plot their favored model for the variation
of $W_{\rm r, 2796}$ with impact parameter, adopting $L=L^*$ in the
calculation.  Qualitatively, we find that the two datasets are in good
agreement.  The positive detections occupy similar regions of the
$W_{\rm r, 2796}, \mRperp$ parameter space and the upper limits occur 
primarily at $\mRperp > 50$\,kpc.  Both datasets show a very large
dispersion with respect to the $W_{\rm r, 2796}(R)$ model;  in fact the
line appears to best describe the division between detections and
non-detections at $\mRperp > 70$\,kpc.  We conclude that our sample,
which extends to many ions beyond Mg$^+$, has no especially anomolous
characteristics. 

%%%%%%%%%%%%%%%%%%%%%%%%%%%%%%%%%%
\section{Dominant characteristics of the cool CGM}
 \label{sec:discuss}

%\subsection{The Cool, Metal-enriched CGM }
%\begin{itemize}
%  \item Enriched CGM in a cool-phase.  Discuss distribution of
%    $b$-values?
%   \item Surface density gradient of low and intermediate ions
%   \item Bimodal nature of the non-SF galaxies
%   \item Si$^{++}$/Si$^+$ implies a highly ionized gas (provided a single
%     phase)
%   \item Covering fraction
%\end{itemize}

In the previous section, we presented results derived primarily from
measurements of the equivalent widths and column densities of low and
intermediate ionization states of metals arising in the CGM of
low-$z$, $L \approx L^*$ galaxies.  These data 
provide an empirical
assessment of several fundamental properties of the CGM. Table \ref{tab:ewcolm_stats} summarizes the absorption line strength statistics for our sample (and various subsamples) of galaxies, giving median and average values as a function of impact parameter. 
Before discussing these results, we
summarize key results from our previous and parallel analyses
on this \cosha\ sample of galaxies to provide context.  

First, nearly all of these
galaxies exhibit strong ($\mwlya > 0.5$\AA) \lya\ absorption with
\ion{H}{1} column densities \nhi\ exceeding $10^{14} \cm{-2}$ and often 
$10^{16} \cm{-2}$ \citep[Figure~\ref{fig:covering};][]{thom12,tumlinson12}.  Only 5 of the
\ntot\ sightlines analyzed in this paper have $\mnhi < 10^{14} \cm{-2}$.  
Second, line-profile analysis of the \lya\ absorption yields Doppler
parameters that require the \ion{H}{1} gas to be predominantly cool, i.e.\ $T <
10^5$\,K.  
These data establish the presence of a cool CGM surrounding $L \approx
L^*$ galaxies to at least 160\,kpc, consistent with previous works on
\ion{H}{1} \lya\ \citep[e.g.][]{lbt+95,wakker09,pwc+11}. 
Third, the SF population exhibits a very high incidence
($\gtrsim 90\%$ to $\mRperp = 160$\,kpc) of strong O${+5}$
absorption \citep[Figure~\ref{fig:covering};][]{tumlinson11}.  
The measured column densities require
a large mass of highly ionized gas in the CGM of these galaxies
($M_{\rm CGM}^{\rm high}>10^9 \mmsun$).
In contrast, the quiescent systems
show a significantly lower frequency of detected O${+5}$ revealing a dichotomy in the CGM that
traces galaxy properties. 

In this paper, we have analyzed far-UV spectra from {\emph {HST}}/COS and
near-UV spectra from Keck/HIRES to examine over 30 metal-line
transitions of low and intermediate ionization states with 
ionization potentials ranging from 
$\approx 5-100\,$eV.  Of the \ntot\ galaxies analyzed
(Table~\ref{tab:galprops}), we report positive detections 
for at least one metal-line transition from the CGM in 33 systems.  
Consistent with
the O${+5}$ results, this demonstrates a metal-enriched CGM for 
$L^*$ galaxies to at least $\mRperp = 160$\,kpc.  In contrast to the
O${+5}$ results, our detections frequently include
the quiescent galaxy population: 9/16 systems have positive
metal-line detections, including both low and intermediate-ions. By comparison, the star-forming population exhibits a metal-line detection rate of 23/27. Our results indicate that the absence of strong O${+5}$ in
quiescent galaxies is not the simple consequence of a very low
metal-enrichment for their halos.  
One must invoke other mechanisms to explain this dichotomy, presumably related to the physical conditions of the more diffuse or hotter gas. 

Another fundamental property of the CGM probed by our observations
is that it must be cool, i.e.\ $T < 10^5$\,K.  The positive detection
of low-ion transitions in \ndlow\ galaxies demands this
result.  Collisional ionization of gas at $T > 10^{4.6}$\,K would
yield negligible quantities of ions like Mg$^+$, C$^+$, and Si$^+$,
even if at solar abundance and with a large total gas column
\citep[e.g.][]{gs07}.  The
galaxies that only exhibit intermediate ions (C$^{++}$, Si$^{++}$) 
are constrained to $T < 2 \sci{5}$\,K to avoid extremely large gas
surface densities.
Such temperatures are consistent with the measured line-widths from
profile-fitting (Table~\ref{tab:final_columns}), 
provided the gas has a modest ($\approx 5-10\mkms$)
turbulence. 
Furthermore, these metals trace the
majority of observed \ion{H}{1} gas, as evidenced by the line-profiles
(Figure~\ref{fig:stackend}) and the observed trend between \nhi\ and
metal column density (Figure~\ref{fig:low_NHI}).    
We conclude that the lower ionization states measured in this paper
are physically arising in the same regions as the majority of
observed \ion{H}{1} gas.  
Together, these metals and \ion{H}{1} gas form the `cool' CGM of $L^*$
galaxies. 
Future work will combine the two 
datasets and use ionization modeling to estimate the gas metallicity
in this component \citep{werk12b}.

Despite the cool temperature $(T<10^5$\,K), we have argued that the gas is
predominantly ionized based on the
relative abundances of the intermediate and low-ions
(Figure~\ref{fig:low_vs_inter}).  
This conclusion depends, in part, on the assertion that the low-ion
gas arises in the same regions as the intermediate-ion gas. 
Direct inspection of the line profiles
(Figure~\ref{fig:stackend}) and comparisons of the profile fits
(Figure~\ref{fig:prof_compare}) support this conclusion. 
The dominance of the intermediate ions then implies that the hydrogen gas is at
least 50\%\ ionized and more likely $>90\%$.  One can limit the degree
of ionization through an additional comparison with the 
high-ion states (e.g.\ \ion{Si}{4}, \ion{C}{4}).  
Although these transitions have poorer data quality, they generally
indicate that the majority of cool gas is not very highly ionized.
Of course, this comparison ignores the nearly ubiquitous O${+5}$
gas present in SF galaxies, which must be very highly ionized. A direct implication, to be explored further in
future work, is that the cool CGM 
traces a distinct gas from that
bearing the O${+5}$ absorption.  

\begin{figure}
\begin{centering}
\hspace{-0.35in}
\includegraphics[width=0.85\linewidth,angle=90]{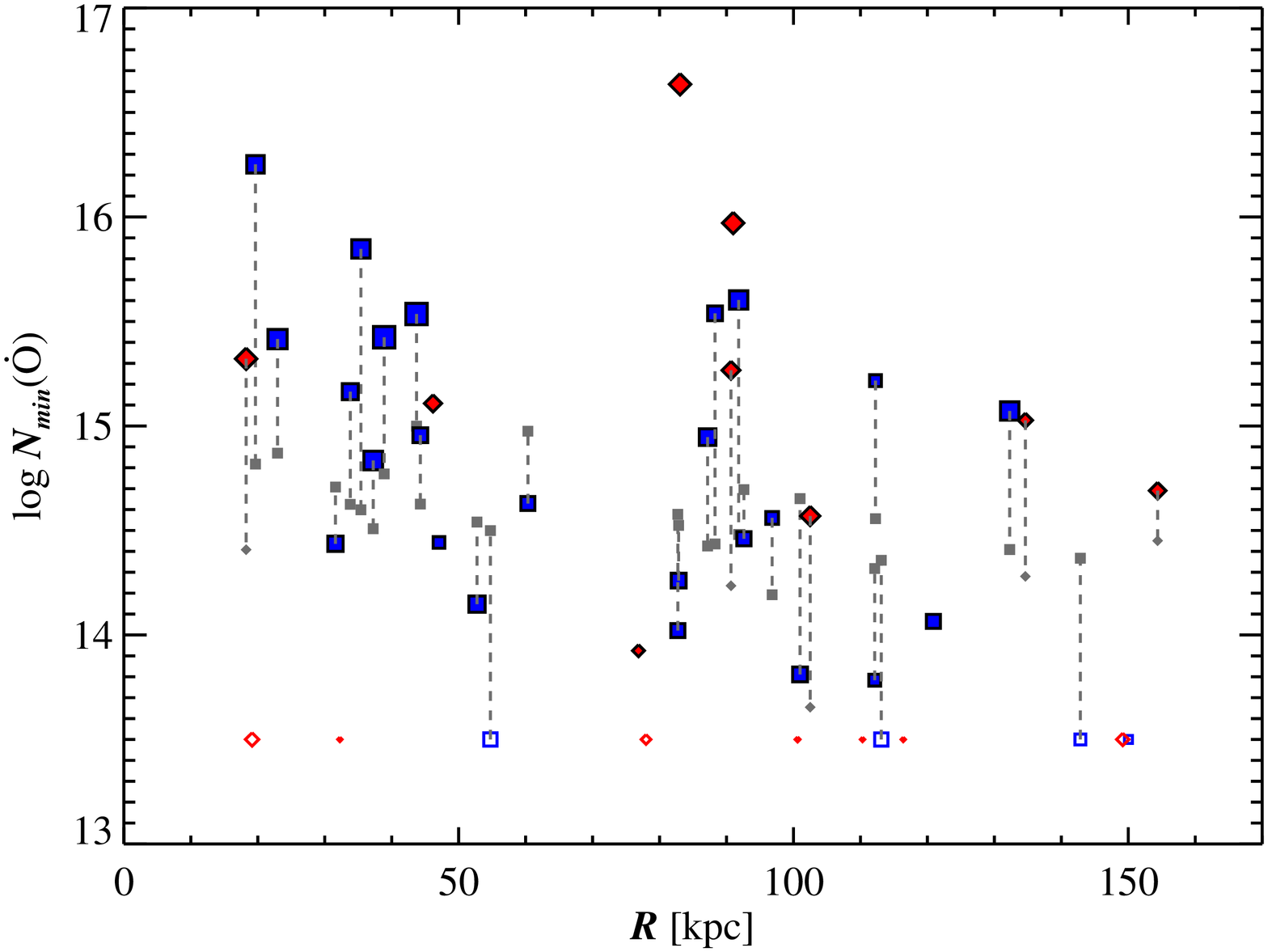}
\end{centering}
\caption{
\nmin, the minimum metal column density in the cool phase of the CGM (T $<$ 10$^{5}$ K)  described in $\S$ 5,  versus impact parameter. The gray points connected to this quantity by gray dashed lines show the values of Log N$_{\rm O^{+5}}$ published by \cite{tumlinson11} for comparison. In cases where we do not detect gas in this low-ionization state cool phase, we plot open symbols with \nmin =  10$^{13.5}$ cm$^{-2}$. Symbol sizes correspond to N$_{\rm HI}$, where bigger color symbols have larger N$_{\rm HI}$. 
}
\label{fig:metal}
\end{figure}

Our survey also demonstrates that the cool CGM of $L^*$ galaxies 
hosts a large mass reservoir of metals and gas.
Figure~\ref{fig:metal} presents the minimum metal column density, \nmin, 
for each system against impact parameter, with the former defined as
follows: For each system, we consider the positive detections of the
low and intermediate ions (values and
lower limits) and convert these to a pseudo-oxygen ionic column
density assuming
solar relative abundances, i.e.\ the \nmin\ value for $\N{Si^{++}} =
10^{13.5} \cm{-2}$ is \nmin $= 10^{14.68} \cm{-2}$.  These are {\emph {minimum}} 
values because: (i) many of the lines are saturated 
and (ii) these columns are derived from a single ionization state such that we assume the ion fraction of the given species is one.  
For each system, we take the maximum \nmin\ value recorded for all
detected ions, as presented in Figure~\ref{fig:metal}. 
%In the
%calculation of \nmin\ we have ignored higher ionization states (i.e.\
%O${+5}$, \ion{S}{6}).  
The measurements are dominated
by intermediate ionization states, primarily Si$^{++}$, C$^{++}$ and
N$^{+}$ with only two values derived from a low-ion\footnote{To be conservative, for this analysis we have ignored measurements derived from \ion{N}{3}~989 owing to concerns that it is blended with \ion{Si}{3}~989.  This has only a minor effect on the results, however.}.
The \nmin\ values are large,
frequently exceeding $10^{15} \cm{-2}$ with a median of $10^{14.6}
\cm{-2}$ for the full sample.  The value is actually comparable to the
median \nhi\ of the sample ($10^{15.4} \cm{-2}$). 
This result is not surprising if we simply consider that the gas is predominantly
ionized.  

Independent of the direct \ion{H}{1} measurements,
if the gas has a solar metallicity or lower then the median
\nmin\ value implies 
a total hydrogen column of $N_{\rm H} > 10^{18} \cm{-2}$.
We stress again that this is a conservative lower limit because it
is derived from a single ionization state
and many of the measured ionic column densities are
lower limits because of line-saturation.  
As important, the average \nmin\ value is nearly
$10\times$ larger than the median. The implied total
gas mass for the cool CGM, therefore, is at least 

\begin{equation}
M_{\rm CGM}^{\rm cool} > 10^{9}  \ltp \frac{\mRperp}{\rm 160\,kpc}
\rtp^2 \ltp \frac{N_{\rm H}}{10^{18} \cm{-2}} \rtp 
\ltp \frac{Z_\odot}{Z} \rtp\ \mmsun
\end{equation}
In this equation, we have taken the mass as a surface area times a median column density, such that M$_{\rm CGM} = 1.4  \pi \rm R^2 \tilde{\rm N_{\rm H}} m_{\rm p}$.  Here, $\tilde{\rm N_{\rm H}}$ is the median total hydrogen column (which we assume has a covering fraction of $\sim$100\%, consistent with the data),  m$_{\rm p}$ is the mass of a proton, and the factor of 1.4 corrects for helium and metals. For reference, $N_{\rm H} = 10^{18} \cm{-2}$, which is the median hydrogen column computed at solar metallicity for a single ionization state, corresponds to $M_{\rm H} = 10^{3.9} \mmsun\ \rm kpc^{-2}$. Future work will improve this very conservative lower limit
through detailed ionization modeling of the gas \citep{werk12b}, and we
expect the value to rise by at least one order of magnitude.

Figure~\ref{fig:metal} also plots the O${+5}$ column densities for
these systems, connected to the \nmin\ values by a dashed gray line.  It is difficult to identify any particular correlation between
the two sets of measurements.  The $\N{O^{+5}}$ values occupy a
relatively narrow range of values and tend to be lower than the \nmin\
estimates. In this
respect, the figure suggests that the lower ionization states dominate
the metal reservoir.  Unfortunately, this hypothesis is difficult to
test because the physical origin (e.g.\ its ionization mechanism)
of O${+5}$ is unknown and corrections for ionization may be very large.  Indeed, it is generally accepted that the ionization fraction of O${+5}$ does not exceed 0.2, so in this respect the values are likely to be scaled up by a factor of 5 if they are to be compared to the total oxygen (for consistency, we have left f = 1.0 as for the low and intermediate ions).

Another trend  emphasized in the previous section (e.g.\ $\S$~\ref{sec:inter}) and also revealed by
Figure~\ref{fig:metal} is that the metal column densities decrease
with increasing impact parameter.  This is true for both the detection
rate and the measured values.  We conclude that the metal surface
density of the CGM decreases with radius, both in the cool gas traced
by the transitions analyzed in this manuscript and the highly ionized
gas traced by O${+5}$.  Future work will study whether this trend
is driven by a decline in metallicity, total surface density, or both
\citep{werk12b}.

Despite the preponderance of positive detections, there are 11 systems
without significant absorption from low or intermediate ions;  each is
shown with an open symbol at an arbitrary
value of $10^{13.5} \cm{-2}$ in Figure~\ref{fig:metal}.
Nearly all of these systems also
exhibit $\mnhi < 10^{14.5} \cm{-2}$.  
The majority of
cases are associated with quiescient galaxies and none of these show 
O${+5}$ absorption.   These non-detections
arise either because of a `hole' in the CGM of these galaxies or
because the galaxy lacks a CGM altogether.  
In contrast, all but one the SF galaxies
exhibit a positive O${+5}$ detection and strong \ion{H}{1}
absorption.
This suggests the cool CGM of SF galaxies is ``patchy'', corresponding physically to a non-unity volume filling factor.  This is in contrast with the 
the highly ionized CGM traced by O${+5}$. 

Such a patchy characteristic is further supported by the observed
distribution of equivalent widths for low-ion gas at large impact
parameters ($\mRperp > 70$\,kpc; Figure~\ref{fig:low_ions}).
Considering both detections and limits, one observes a very wide
dispersion in the measurements (at least an order of magnitude) that is
uniformly distributed.  This occurs despite the fact that most
of the sightlines show strong \ion{H}{1} and O${+5}$ absorption
(Figure~\ref{fig:covering}).  We conclude that the CGM has a patchy
distribution of low-ion gas.
%[Why isn't there more variation in the line-profiles??]
%[Consider including the 2D `maps' here]

\begin{figure}
\begin{centering}
\includegraphics[width=0.95\linewidth,angle=0]{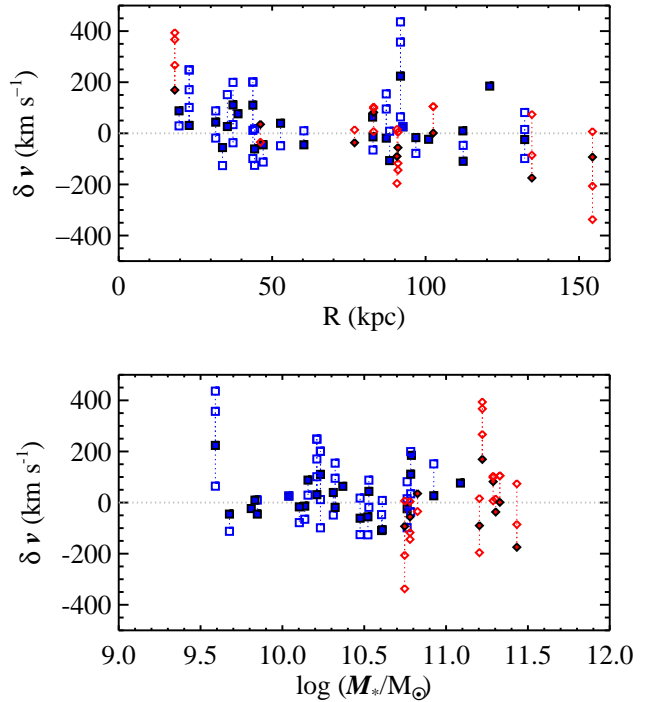}
\end{centering}

\caption{  Fitted component centroid velocities with respect to galaxy systemic redshift, $\delta$v, as a function of impact parameter and stellar mass for the low and intermediate ionization state metal lines examined in this work (one of: MgII, CII, CIII, SiIII).  Dotted lines span the full range in velocity over which the transition is defined. The dominant component is shown as a solid point (blue squares denote SF galaxies; red diamonds denote quiescent galaxies), while other fitted components are shown as open symbols. The majority of the gas is bound to the dark matter halo: v$_{\rm esc}$ at R$=$ 150 kpc for an L$^*$ galaxy is $\approx 200$ km $s^{-1}$. 
}
\label{fig:kin_prof}
\end{figure}

The kinematics of the gas offer additional insight into the nature of
the cool CGM.
In \cite{tumlinson12}, we compare the gas kinematics of the
\ion{H}{1} gas to estimates of the escape velocity and argued
that the majority of that material is not escaping.
Figure~\ref{fig:kin_prof} presents a similar analysis, based on the
line-profile fits for the metals in the cool CGM.  Two results should be
emphasized from the Figure.  First, the dominant gas components (solid points) have a
velocity dispersion about systemic of $\sigma \approx 85 \mkms$.
Wavelength error and uncertainty in the
galaxy redshifts contribute to this dispersion, but only on the order 
of $10-20\%$.  
Therefore, the gas has significant motions relative to the galaxy.
These motions, however, 
are consistent with the velocity dispersion
predicted for the dark matter halos hosting $L^*$ galaxies 
and we conclude that the majority of gas is confined to the system.
Second, the line-profiles tend to span only $50-100\mkms$; only a few
examples show motions in excess of $200\mkms$.  
Similar to \cite{tumlinson12}, we conclude the cool CGM is bound to
the galaxy.
%[Is there a possible trend with $M^*$?]   

Another aspect of the kinematics revealed in Figure~\ref{fig:kin_prof} is that
a majority of galaxies show metal-line profiles that span the
systemic velocity.  This contrasts with previous work on \ion{Mg}{2}
kinematics where the profiles have tended to lie
on only one side of systemic \citep{sks+02,kcs+07}. One notable difference between these previous efforts and our own is that both \cite{sks+02} and \cite{kcs+07} assembled a {\emph{post-facto}} galaxy sample, that is, they searched for the associated galaxy after finding metal-line absorption. 
%[Point out those were *all* searches for the galaxy after finding the
%gas] 
Our results also differ from models for accreting cool gas which
predict the material should have high angular momentum and therefore
show absorption with large velocity offset to one side of systemic
\citep{stewart11}. 

 \section{Summary and Conclusions}
\label{sec:sum}

In this paper, we have presented the equivalent width, column density measurements, and line profile fits for
low and intermediate ionization states of the CGM surrounding \ntot\ low-$z$, $L\approx L^*$ galaxies drawn from
the \cosha\ survey.  These measurements are derived from far-UV
transitions observed in HST/COS and Keck/HIRES spectra of background
quasars within an impact parameter $\mRperp < 160$\,kpc to the
targeted galaxies at z $\sim$0.2.  We have closely examined empirical relations between the CGM and galaxy properties. Additionally, we have analyzed adjacent ionization states of different species and compared the kinematic structure of their individual fitted components. Here, we summarize our main findings on the qualitative nature of the z $\sim0$ CGM
as described by our new observations:

\begin{enumerate}
\item Low and Intermediate ionization state metal absorption lines are a very common feature of the
  CGM for $L^*$ galaxies of all spectral types (33/44 galaxies show absorption from low/intermediate ionization state material).
This frequent presence of substantial lower ionization state material (e.g.\ Mg$^+$,
  Si$^+$, Si$^{++}$, C$^{++}$) requires a cool component for the CGM (T $<$ 10$^{5}$ K), consistent with our  results on the \ion{H}{1} gas \citep{tumlinson12, thom12}.
  \item   A comparison of the relative column densities of adjacent ionization states of low and intermediate ions indicates the gas is predominantly ionized.  
  \item The detection rates and column densities derived for the low and intermediate ionization state metal lines decrease with increasing impact parameter, a trend we interpret as a declining metal surface density profile for the CGM within its inner 160 kpc.  
\item There is large dispersion in low and intermediate ion absorption strengths, independent of galaxy properties. This variation implies a
  patchy distribution of dense, cool CGM gas.
\item The qualitative comparison of the absorption line profiles between low and intermediate ions indicates an overall correspondence of component structure and shape. We find no evidence that the low and intermediate ionization states of the cool CGM arise from distinct phases. However, an additional comparison of these profiles with those of O$^{+5}$, as analyzed by \cite{tumlinson11}, suggests a complex relationship between low/intermediate ionization states and more highly ionized gas that will be fully explored in future work. 
\item The gas kinematics derived from Voigt profile fits to the data suggest the CGM is largely bound to its host galaxy's dark matter halo.
  Furthermore, the frequent presence of multiple velocity components indicate the material  is clumpy.
  \item The metal column densities of the low and intermediate ionization state gas imply that the CGM is a massive gaseous reservoir, its baryonic content  estimated to far exceed $10^{9} \mmsun$ based on conservative assumptions.  This mass estimate was made independently of the mass estimate discussed by \cite{tumlinson11}, which was based on O$^{+5}$ measurements alone. The cool  (T $<$ 10$^{5}$ K) and warm (T $\approx$ 10$^{5 - 6}$ K) O$^{+5}$-traced CGM may represent completely distinct gas phases whose total masses will contribute separately to the CGM baryonic content.  An upcoming paper  (Werk et al., in prep) will report a full analysis of the ionized gas fraction and metallicity of the gas using photoionization modeling, and will provide a more reliable baryonic mass estimate for the cool CGM of z$\sim$0 L$^*$ galaxies. 
\end{enumerate}

 \section{Acknowledgements}
   
  Support for program GO11598 was provided by NASA through a grant
  from the Space Telescope Science Institute, which is operated by the
  Association of Universities for Research in Astronomy, Inc., under
  NASA contract NAS 5-26555.  Much of the data presented herein were
  obtained at the W.M. Keck Observatory, which is operated as a
  scientific partnership among the California Institute of Technology,
  the University of California and the National Aeronautics and Space
  Administration. The Observatory was made possible by the generous
  financial support of the W.M. Keck Foundation. The authors wish to
  recognize and acknowledge the very significant cultural role and
  reverence that the summit of Mauna Kea has always had within the
  indigenous Hawaiian community. We are most fortunate to have the
  opportunity to conduct observations from this mountain. 
  JXP acknowledges support from a Humboldt visitor fellowship to the Max
  Planck Institute for Astronomy where part of this work was
  performed. JKW thanks the referee of this work, John Stocke, for very helpful comments that improved the manuscript. 
\nocite{solabnd}
\nocite{blanton07}
    {{ Facilities:} \facility{Keck: HIRES} \facility{HST: COS}

%\bibliography{opticalspec}
%\bibliographystyle{apj}
%\bibliography{/u/xavier/paper/Bibli/allrefs}
\bibliographystyle{apj}
\bibliography{lowions}

\begin{thebibliography}{61}
\expandafter\ifx\csname natexlab\endcsname\relax\def\natexlab#1{#1}\fi

\bibitem[{{Aracil} {et~al.}(2006){Aracil}, {Tripp}, {Bowen}, {Prochaska},
  {Chen}, \& {Frye}}]{araciletal06}
{Aracil}, B., {Tripp}, T.~M., {Bowen}, D.~V., {Prochaska}, J.~X., {Chen},
  H.-W., \& {Frye}, B.~L. 2006, \mnras, 367, 139

\bibitem[{{Asplund} {et~al.}(2005){Asplund}, {Grevesse}, \& {Sauval}}]{solabnd}
{Asplund}, M., {Grevesse}, N., \& {Sauval}, A.~J. 2005, in ASP Conf. Ser. 336:
  Cosmic Abundances as Records of Stellar Evolution and Nucleosynthesis, ed.
  T.~G. {Barnes} \& F.~N. {Bash}, 25--+

\bibitem[{{Bahcall} \& {Spitzer}(1969)}]{bs69}
{Bahcall}, J.~N., \& {Spitzer}, L.~J. 1969, \apj, 156, L63

\bibitem[{{Barton} \& {Cooke}(2009)}]{bc09}
{Barton}, E.~J., \& {Cooke}, J. 2009, \aj, 138, 1817

\bibitem[{{Bergeron}(1986)}]{bergeron86}
{Bergeron}, J. 1986, \aap, 155, L8

\bibitem[{{Blanton} {et~al.}(2003){Blanton}, {Hogg}, {Bahcall}, {Brinkmann},
  {Britton}, {Connolly}, {Csabai}, {Fukugita}, {Loveday}, {Meiksin}, {Munn},
  {Nichol}, {Okamura}, {Quinn}, {Schneider}, {Shimasaku}, {Strauss}, {Tegmark},
  {Vogeley}, \& {Weinberg}}]{blantonetal03}
{Blanton}, M.~R., {et~al.} 2003, \apj, 592, 819

\bibitem[{{Blanton} \& {Roweis}(2007)}]{blanton07}
{Blanton}, M.~R., \& {Roweis}, S. 2007, \aj, 133, 734

\bibitem[{{Bowen} {et~al.}(1995){Bowen}, {Blades}, \& {Pettini}}]{bowen+95}
{Bowen}, D.~V., {Blades}, J.~C., \& {Pettini}, M. 1995, \apj, 448, 634

\bibitem[{{Bowen} \& {Chelouche}(2011)}]{bc11}
{Bowen}, D.~V., \& {Chelouche}, D. 2011, \apj, 727, 47

\bibitem[{{Bowen} {et~al.}(2002){Bowen}, {Pettini}, \& {Blades}}]{bowen+02}
{Bowen}, D.~V., {Pettini}, M., \& {Blades}, J.~C. 2002, \apj, 580, 169

\bibitem[{{Chen} {et~al.}(2010){Chen}, {Helsby}, {Gauthier}, {Shectman},
  {Thompson}, \& {Tinker}}]{chg+10}
{Chen}, H., {Helsby}, J.~E., {Gauthier}, J., {Shectman}, S.~A., {Thompson},
  I.~B., \& {Tinker}, J.~L. 2010, \apj, 714, 1521

\bibitem[{{Chen} {et~al.}(1998){Chen}, {Lanzetta}, {Webb}, \&
  {Barcons}}]{clw+98}
{Chen}, H., {Lanzetta}, K.~M., {Webb}, J.~K., \& {Barcons}, X. 1998, \apj, 498,
  77

\bibitem[{{Chen} {et~al.}(2001){Chen}, {Lanzetta}, \& {Webb}}]{clw01}
{Chen}, H.-W., {Lanzetta}, K.~M., \& {Webb}, J.~K. 2001, \apj, 556, 158

\bibitem[{{Cooksey} {et~al.}(2010){Cooksey}, {Thom}, {Prochaska}, \&
  {Chen}}]{ctp+10}
{Cooksey}, K.~L., {Thom}, C., {Prochaska}, J.~X., \& {Chen}, H. 2010, \apj,
  708, 868

\bibitem[{{Dav{\'e}} \& {Tripp}(2001)}]{dt01}
{Dav{\'e}}, R., \& {Tripp}, T.~M. 2001, \apj, 553, 528

\bibitem[{{Dunkley} {et~al.}(2009){Dunkley}, {Komatsu}, {Nolta}, {Spergel},
  {Larson}, {Hinshaw}, {Page}, {Bennett}, {Gold}, {Jarosik}, {Weiland},
  {Halpern}, {Hill}, {Kogut}, {Limon}, {Meyer}, {Tucker}, {Wollack}, \&
  {Wright}}]{wmap05}
{Dunkley}, J., {et~al.} 2009, \apjs, 180, 306

\bibitem[{{Feigelson} \& {Nelson}(1985)}]{fn85}
{Feigelson}, E.~D., \& {Nelson}, P.~I. 1985, \apj, 293, 192

\bibitem[{{Froning} \& {Green}(2009)}]{froning09}
{Froning}, C.~S., \& {Green}, J.~C. 2009, \apss, 320, 181

\bibitem[{{Ghavamian} {et~al.}(2009){Ghavamian}, {Aloisi}, {Lennon}, {Hartig},
  {Kriss}, {Oliveira}, {Massa}, {Keyes}, {Proffitt}, {Delker}, \&
  {Osterman}}]{cos_lsf}
{Ghavamian}, P., {et~al.} 2009, {Preliminary Characterization of the Post-
  Launch Line Spread Function of COS}, Tech. rep.

\bibitem[{{Gnat} \& {Sternberg}(2007)}]{gs07}
{Gnat}, O., \& {Sternberg}, A. 2007, \apjs, 168, 213

\bibitem[{{Green} {et~al.}(2012){Green}, {Froning}, {Osterman}, {Ebbets},
  {Heap}, {Leitherer}, {Linsky}, {Savage}, {Sembach}, {Shull}, {Siegmund},
  {Snow}, {Spencer}, {Stern}, {Stocke}, {Welsh}, {B{\'e}land}, {Burgh},
  {Danforth}, {France}, {Keeney}, {McPhate}, {Penton}, {Andrews},
  {Brownsberger}, {Morse}, \& {Wilkinson}}]{green+12}
{Green}, J.~C., {et~al.} 2012, \apj, 744, 60

\bibitem[{{Kacprzak} {et~al.}(2007){Kacprzak}, {Churchill}, {Steidel},
  {Ceverino}, {Klypin}, \& {Murphy}}]{kcs+07}
{Kacprzak}, G.~G., {Churchill}, C.~W., {Steidel}, C.~C., {Ceverino}, D.,
  {Klypin}, A.~A., \& {Murphy}, M.~T. 2007, ArXiv e-prints, 710

\bibitem[{{Kartaltepe} {et~al.}(2007){Kartaltepe}, {Sanders}, {Scoville},
  {Calzetti}, {Capak}, {Koekemoer}, {Mobasher}, {Murayama}, {Salvato},
  {Sasaki}, \& {Taniguchi}}]{kartaltepe07}
{Kartaltepe}, J.~S., {et~al.} 2007, \apjs, 172, 320

\bibitem[{{Koester} {et~al.}(2007){Koester}, {McKay}, {Annis}, {Wechsler},
  {Evrard}, {Bleem}, {Becker}, {Johnston}, {Sheldon}, {Nichol}, {Miller},
  {Scranton}, {Bahcall}, {Barentine}, {Brewington}, {Brinkmann}, {Harvanek},
  {Kleinman}, {Krzesinski}, {Long}, {Nitta}, {Schneider}, {Sneddin}, {Voges},
  \& {York}}]{maxbcg}
{Koester}, B.~P., {et~al.} 2007, \apj, 660, 239

\bibitem[{{Lanzetta} {et~al.}(1995){Lanzetta}, {Bowen}, {Tytler}, \&
  {Webb}}]{lbt+95}
{Lanzetta}, K.~M., {Bowen}, D.~V., {Tytler}, D., \& {Webb}, J.~K. 1995, \apj,
  442, 538

\bibitem[{{Lin} {et~al.}(2004){Lin}, {Koo}, {Willmer}, {Patton}, {Conselice},
  {Yan}, {Coil}, {Cooper}, {Davis}, {Faber}, {Gerke}, {Guhathakurta}, \&
  {Newman}}]{lin04}
{Lin}, L., {et~al.} 2004, \apjl, 617, L9

\bibitem[{{M{\'e}nard} \& {Chelouche}(2009)}]{menard09}
{M{\'e}nard}, B., \& {Chelouche}, D. 2009, \mnras, 393, 808

\bibitem[{{Morris} {et~al.}(1993){Morris}, {Weymann}, {Dressler}, {McCarthy},
  {Smith}, {Terrile}, {Giovanelli}, \& {Irwin}}]{mwd+93}
{Morris}, S.~L., {Weymann}, R.~J., {Dressler}, A., {McCarthy}, P.~J., {Smith},
  B.~A., {Terrile}, R.~J., {Giovanelli}, R., \& {Irwin}, M. 1993, \apj, 419,
  524

\bibitem[{{Morton}(2003)}]{morton03}
{Morton}, D.~C. 2003, \apjs, 149, 205

\bibitem[{{Mulchaey} \& {Chen}(2009)}]{mc09}
{Mulchaey}, J.~S., \& {Chen}, H.-W. 2009, \apjl, 698, L46

\bibitem[{{Okamoto} \& {Habe}(1999)}]{okamoto99}
{Okamoto}, T., \& {Habe}, A. 1999, \apj, 516, 591

\bibitem[{{Penton} {et~al.}(2000){Penton}, {Shull}, \& {Stocke}}]{pss00}
{Penton}, S.~V., {Shull}, J.~M., \& {Stocke}, J.~T. 2000, \apj, 544, 150

\bibitem[{{Penton} {et~al.}(2002){Penton}, {Stocke}, \& {Shull}}]{pss02}
{Penton}, S.~V., {Stocke}, J.~T., \& {Shull}, J.~M. 2002, \apj, 565, 720

\bibitem[{{Prochaska} {et~al.}(2004){Prochaska}, {Chen}, {Howk}, {Weiner}, \&
  {Mulchaey}}]{pks0405_uv}
{Prochaska}, J.~X., {Chen}, H.-W., {Howk}, J.~C., {Weiner}, B.~J., \&
  {Mulchaey}, J. 2004, \apj, 617, 718

\bibitem[{{Prochaska} {et~al.}(2011{\natexlab{a}}){Prochaska}, {Weiner},
  {Chen}, {Cooksey}, \& {Mulchaey}}]{pwc+11}
{Prochaska}, J.~X., {Weiner}, B., {Chen}, H.-W., {Cooksey}, K.~L., \&
  {Mulchaey}, J.~S. 2011{\natexlab{a}}, \apjs, 193, 28

\bibitem[{{Prochaska} {et~al.}(2011{\natexlab{b}}){Prochaska}, {Weiner},
  {Chen}, {Mulchaey}, \& {Cooksey}}]{pwc2+11}
{Prochaska}, J.~X., {Weiner}, B., {Chen}, H.-W., {Mulchaey}, J., \& {Cooksey},
  K. 2011{\natexlab{b}}, \apj, 740, 91

\bibitem[{{Ribaudo} {et~al.}(2011){Ribaudo}, {Lehner}, {Howk}, {Werk}, {Tripp},
  {Prochaska}, {Meiring}, \& {Tumlinson}}]{ribaudo11}
{Ribaudo}, J., {Lehner}, N., {Howk}, J.~C., {Werk}, J.~K., {Tripp}, T.~M.,
  {Prochaska}, J.~X., {Meiring}, J.~D., \& {Tumlinson}, J. 2011, \apj, 743, 207

\bibitem[{{Rudie} {et~al.}(2012){Rudie}, {Steidel}, {Trainor}, {Rakic},
  {Bogosavljevi{\'c}}, {Pettini}, {Reddy}, {Shapley}, {Erb}, \&
  {Law}}]{rudie12}
{Rudie}, G.~C., {et~al.} 2012, \apj, 750, 67

\bibitem[{{Savage} \& {Sembach}(1996)}]{ss96}
{Savage}, B.~D., \& {Sembach}, K.~R. 1996, \araa, 34, 279

\bibitem[{{Savage} {et~al.}(1998){Savage}, {Tripp}, \& {Lu}}]{stl98}
{Savage}, B.~D., {Tripp}, T.~M., \& {Lu}, L. 1998, \aj, 115, 436

\bibitem[{{Shull} {et~al.}(1996){Shull}, {Stocke}, \& {Penton}}]{ssp96}
{Shull}, J.~M., {Stocke}, J.~T., \& {Penton}, S. 1996, \aj, 111, 72

\bibitem[{{Steidel} {et~al.}(2002){Steidel}, {Kollmeier}, {Shapley},
  {Churchill}, {Dickinson}, \& {Pettini}}]{sks+02}
{Steidel}, C.~C., {Kollmeier}, J.~A., {Shapley}, A.~E., {Churchill}, C.~W.,
  {Dickinson}, M., \& {Pettini}, M. 2002, \apj, 570, 526

\bibitem[{{Stewart} {et~al.}(2011){Stewart}, {Kaufmann}, {Bullock}, {Barton},
  {Maller}, {Diemand}, \& {Wadsley}}]{stewart11}
{Stewart}, K.~R., {Kaufmann}, T., {Bullock}, J.~S., {Barton}, E.~J., {Maller},
  A.~H., {Diemand}, J., \& {Wadsley}, J. 2011, \apj, 738, 39

\bibitem[{{Stocke} {et~al.}(2006){Stocke}, {Penton}, {Danforth}, {Shull},
  {Tumlinson}, \& {McLin}}]{stockeetal06}
{Stocke}, J.~T., {Penton}, S.~V., {Danforth}, C.~W., {Shull}, J.~M.,
  {Tumlinson}, J., \& {McLin}, K.~M. 2006, \apj, 641, 217

\bibitem[{{Stocke} {et~al.}(1995){Stocke}, {Shull}, {Penton}, {Donahue}, \&
  {Carilli}}]{stocke+95}
{Stocke}, J.~T., {Shull}, J.~M., {Penton}, S., {Donahue}, M., \& {Carilli}, C.
  1995, \apj, 451, 24

\bibitem[{{Thom} \& {Chen}(2008{\natexlab{a}})}]{tc08b}
{Thom}, C., \& {Chen}, H. 2008{\natexlab{a}}, \apjs, 179, 37

\bibitem[{{Thom} \& {Chen}(2008{\natexlab{b}})}]{tc08a}
---. 2008{\natexlab{b}}, \apj, 683, 22

\bibitem[{{Thom} {et~al.}(2012){Thom}, {Tumlinson}, {Werk}, {Prochaska}, \&
  {Tripp}}]{thom12}
{Thom}, C., {Tumlinson}, J., {Werk}, J., {Prochaska}, J.~X., \& {Tripp}, T.
  2012, ArXiv e-prints

\bibitem[{{Tilton} {et~al.}(2012){Tilton}, {Danforth}, {Shull}, \&
  {Ross}}]{tilton+12}
{Tilton}, E.~M., {Danforth}, C.~W., {Shull}, J.~M., \& {Ross}, T.~L. 2012,
  ArXiv e-prints

\bibitem[{{Tripp} {et~al.}(2001){Tripp}, {Giroux}, {Stocke}, {Tumlinson}, \&
  {Oegerle}}]{trippetal01}
{Tripp}, T.~M., {Giroux}, M.~L., {Stocke}, J.~T., {Tumlinson}, J., \&
  {Oegerle}, W.~R. 2001, \apj, 563, 724

\bibitem[{{Tripp} {et~al.}(1998){Tripp}, {Lu}, \& {Savage}}]{tripp+98}
{Tripp}, T.~M., {Lu}, L., \& {Savage}, B.~D. 1998, \apj, 508, 200

\bibitem[{{Tripp} {et~al.}(2011){Tripp}, {Meiring}, {Prochaska}, {Willmer},
  {Howk}, {Werk}, {Jenkins}, {Bowen}, {Lehner}, {Sembach}, {Thom}, \&
  {Tumlinson}}]{tripp11}
{Tripp}, T.~M., {et~al.} 2011, Science, 334, 952

\bibitem[{{Tripp} \& {Savage}(2000)}]{ts00}
{Tripp}, T.~M., \& {Savage}, B.~D. 2000, \apj, 542, 42

\bibitem[{{Tripp} {et~al.}(2008){Tripp}, {Sembach}, {Bowen}, {Savage},
  {Jenkins}, {Lehner}, \& {Richter}}]{tripp08}
{Tripp}, T.~M., {Sembach}, K.~R., {Bowen}, D.~V., {Savage}, B.~D., {Jenkins},
  E.~B., {Lehner}, N., \& {Richter}, P. 2008, \apjs, 177, 39

\bibitem[{{Tumlinson} \& {Fang}(2005)}]{tf05}
{Tumlinson}, J., \& {Fang}, T. 2005, \apjl, 623, L97

\bibitem[{{Tumlinson} {et~al.}(2012){Tumlinson}, {Thom}, {Werk}, {Prochaska},
  \& {Tripp}}]{tumlinson12}
{Tumlinson}, J., {Thom}, C., {Werk}, J., {Prochaska}, J.~X., \& {Tripp}, T.
  2012, ArXiv e-prints

\bibitem[{{Tumlinson} {et~al.}(2011){Tumlinson}, {Thom}, {Werk}, {Prochaska},
  {Tripp}, {Weinberg}, {Peeples}, {O'Meara}, {Oppenheimer}, {Meiring}, {Katz},
  {Dave}, {Brady Ford}, \& {Sembach}}]{tumlinson11}
{Tumlinson}, J., {et~al.} 2011, ArXiv e-prints

\bibitem[{{Wakker} \& {Savage}(2009)}]{wakker09}
{Wakker}, B.~P., \& {Savage}, B.~D. 2009, \apjs, 182, 378

\bibitem[{{Werk} {et~al.}(2013){Werk}, {Prochaska}, {Thom}, {Tumlinson}, \&
  {Tripp}}]{werk12b}
{Werk}, J., {Prochaska}, J.~X., {Thom}, C., {Tumlinson}, J., \& {Tripp}, T.
  2013, ArXiv e-prints

\bibitem[{{Werk} {et~al.}(2012){Werk}, {Prochaska}, {Thom}, {Tumlinson},
  {Tripp}, {O'Meara}, \& {Meiring}}]{werk12}
{Werk}, J.~K., {Prochaska}, J.~X., {Thom}, C., {Tumlinson}, J., {Tripp}, T.~M.,
  {O'Meara}, J.~M., \& {Meiring}, J.~D. 2012, \apjs, 198, 3

\bibitem[{{Zepf} \& {Koo}(1989)}]{zepf89}
{Zepf}, S.~E., \& {Koo}, D.~C. 1989, \apj, 337, 34

\end{thebibliography}
\clearpage

\begin{deluxetable}{llcccccrrcc}
\tablewidth{0pc}
\tablecaption{Galaxy Properties\label{tab:galprops}}
\tabletypesize{\scriptsize}
\tablehead{\colhead{Galaxy} & \colhead{z} & \colhead{R} &\colhead{log(M$_{*}/\mmsun$)}  &\colhead{$L$} &\colhead{$u-r$} &
\colhead{SFR} & \colhead{Abund} \\
&  &  (kpc) & & ($L^*$) & & ($\mmsun \, \rm yr^{-1})$}
\startdata
J0226+0015\_268\_22&0.2274& 77&10.80& 0.51&$> 2.08$&$< 0.09$&****\\
J0401-0540\_67\_24&0.2197& 82&10.14& 0.37&$ 1.25$&$ 1.14$&8.55\\
J0803+4332\_306\_20&0.2535& 76&11.30& 1.63&$> 2.81$&$< 0.21$&****\\
J0910+1014\_34\_46&0.1427&112&10.61& 0.60&$ 1.51$&$14.12$&8.52\\
J0910+1014\_242\_34&0.2641&134&11.43& 2.16&$> 2.44$&$< 0.30$&****\\
J0914+2823\_41\_27&0.2443&100& 9.81& 0.37&$ 1.24$&$ 2.83$&8.62\\
J0925+4004\_196\_22&0.2475& 83&11.29& 1.48&$> 2.58$&$< 0.57$&****\\
J0928+6025\_110\_35&0.1540& 90&10.78& 0.52&$ 2.55$&$< 0.04$&****\\
\enddata
\tablecomments{{} Derived Properties of target and bonus galaxies (see Werk et al. 2012 for details)  (1) SDSS Field Identifier
 and Galaxy Identifier, where the first number is the position angle in degrees from the QSO
and the second number is the projected separation in arcseconds (impact parameter) from the QSO
(2) Spectroscopic Redshift as determined from LRIS and MagE data. 
(3) Projected separation between the galaxy and QSO in kpc, calculated in the restframe of the galaxy.
(4)  Stellar Mass from $kcorrect$ (Blanton et al. 2007) 
(5)  Galaxy Luminosity in terms of L$^*$, where $L^*$ is given by an R-band absolute magnitude of $-21.12$ (Blanton et al. 2003) 
(6) SDSS $u - r$ galaxy colors
(7) The Galaxy SFR, determined from a combination of emission-line spectroscopy and broadband photometry 
 (9) Oxygen Abundance, 12 + Log (O/H), determined for galaxies whose MagE and LRIS spectra showed strong emission lines,  where the solar abundance is 8.69 (Asplund et al. 2005).} 
\end{deluxetable}
 
\clearpage

\begin{deluxetable}{llccccrrcc}
\tablewidth{0pc}
\tablecaption{Keck/HIRES Observations\label{tab:hires}}
\tabletypesize{\scriptsize}
\tablehead{\colhead{Quasar} & \colhead{UT} & \colhead{$t_{\rm exp} (s)$} &\colhead{S/N$^a$} }
\startdata
J0033-0055&2008-10-06&3600& 9.6\\
J0200-0845&2008-10-06&3000& 6.5\\
J0226+0015&2008-10-06&3600& 6.9\\
J0401-0540&2008-10-06&3600& 9.3\\
J0803+4332&2012-04-12&2400&11.7\\
J0910+1014&2012-04-13&2400& 6.9\\
J0914+2823&2012-04-13&2400& 8.1\\
J0925+4004&2012-04-12&2400& 5.4\\
J0928+6025&2010-03-26&3600&12.0\\
J0943+0531&2010-03-26&5400&11.8\\
J0950+4831&2012-04-12&2400&12.7\\
J1009+0713&2010-03-26&3600& 9.3\\
J1016+4706&2012-04-13&2400& 8.6\\
J1112+3539&2012-04-12&2400& 7.2\\
J1133+0327&2012-04-13&2400& 5.2\\
J1157-0022&2012-04-13&2400& 8.2\\
J1233+4758&2012-04-12&2400&13.4\\
J1233-0031&2012-04-13&2400& 9.9\\
J1241+5721&2012-04-12&2400& 6.2\\
J1245+3356&2012-04-12&2400&13.1\\
J1322+4645&2010-03-26&3200& 6.8\\
J1330+2813&2012-04-13&2400& 4.5\\
J1342-0053&2012-04-12&2400&14.5\\
J1419+4207&2010-03-26&3200&11.4\\
J1435+3604&2012-04-13&2400&13.4\\
J1437+5045&2012-04-13&2400& 6.4\\
J1445+3428&2012-04-12&2400& 5.6\\
J1514+3619&2010-03-26&3200& 6.9\\
J1550+4001&2012-04-13&2400&12.7\\
J1553+3548&2012-04-13&2400&15.3\\
J1555+3628&2012-04-12&2400& 4.4\\
J1616+4154&2012-04-12&2400&14.3\\
J1619+3342&2010-09-02&3000&16.7\\
J2257+1340&2008-10-06&3600&13.5\\
J2345-0059&2008-10-06&2400&12.7\\
\enddata
\tablenotetext{a}{Signal-to-noise per $1.3 \kms$ pixel at $\lambda \approx 3400$\AA.} 
\end{deluxetable}
 
\clearpage

%%%%%%%%%%%%%%Line Table
\clearpage
\LongTables
 
\begin{deluxetable}{lccccccccc}
\tablewidth{0pc}
\tablecaption{Metal-Line Measurements: Baseline Integration\label{tab:lowlines}}
\tabletypesize{\scriptsize}
\tablehead{\colhead{Ion} & \colhead{$\lambda_{\rm rest}$}  &\colhead{$[v_{\rm min}, v_{\rm max}]$}&\colhead{W$_{rest}$} & \colhead{$\log N$} &
\colhead{Flg}  & \colhead{$\log N_{\rm Adopted}$} \\
&(\AA) & ($\kms$) & (m\AA) & (cm$^{-2}$)}
\startdata
\cutinhead{J0226+0015\_268\_22  $z=0.2274$  $\Rperp =  78$\,kpc}
H\, I&949& [$0,250$] &$<73$&$<14.85$&5&$14.25\pm0.03$\\
H\, I&1215& [$0,280$] &$486\pm21$&$14.25\pm0.03$&1\\
C\, II&1036& [$-12,332$] &$<34$&$<13.44$&5&$<13.44$\\
C\, II&1334& [$-12,332$] &$<96$&$<13.69$&5\\
N\, I&1200& [$-12,332$] &$<56$&$<14.01$&5&$<14.01$\\
N\, II&1083& [$-12,332$] &$<41$&$<13.59$&5&$<13.59$\\
N\, V&1238& [$-12,332$] &$<64$&$<13.48$&5&$<13.48$\\
O\, I&1302& [$-12,332$] &$<119$&$<14.29$&5&$<14.29$\\
\enddata
\tablecomments{{}   First Line: SDSS Field Identifier and  
galaxy identifier, where the first number is the position angle in degrees from the QSO
and the second number is the projected separation in arcseconds (impact parameter) from the QSO; 
Galaxy Redshift, spectroscopically determined \citep{werk12}; and impact parameter in kpc. 
Columns: (1) Metal Ion name ; (2) Transition Wavelength in $\AA$; (3) Velocity interval over which the transition was measured in km s$^{-1}$ (4) Equivalent Width in m$\AA$; (5) Log Column Density in cm$^{-2}$;   (6) Transition Flag: 1 $=$ good, 3 $=$ minorly blended, 5 $=$ non-detection (2$\sigma$ upper limit), 9 $=$ saturated, and 11 $=$ blended and saturated; (9) The adopted ionic column density, a weighted mean for ions with multiple detected transitions}  
\end{deluxetable}
 
\clearpage

%%%%%%%%%%Stacks

%%%%%%%%%%%%%%%%%
%%%% species stacks
\afterpage{\clearpage}

 \begin{figure*}[h!]
\includegraphics[width=0.99\linewidth]{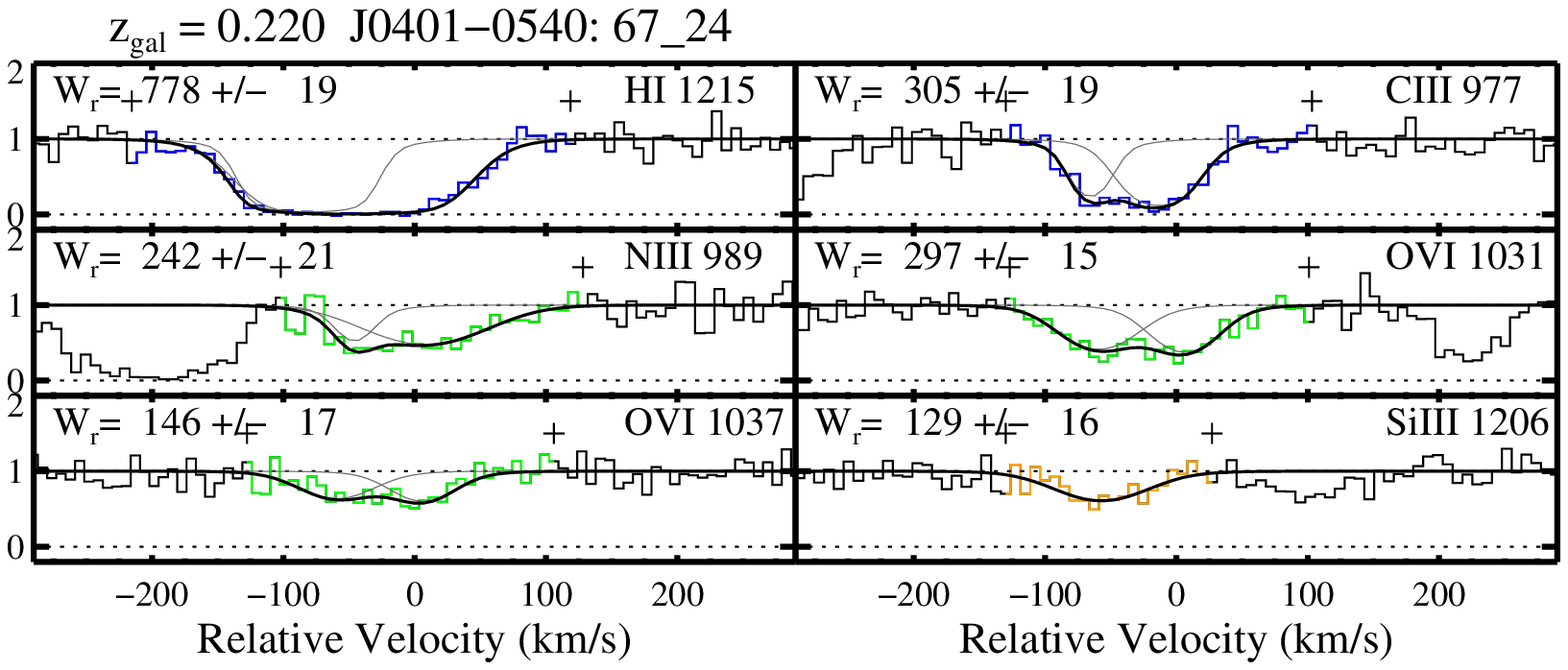}
\vspace{-0.2in}
\end{figure*}
\begin{figure*}[h!]
\includegraphics[width=0.99\linewidth]{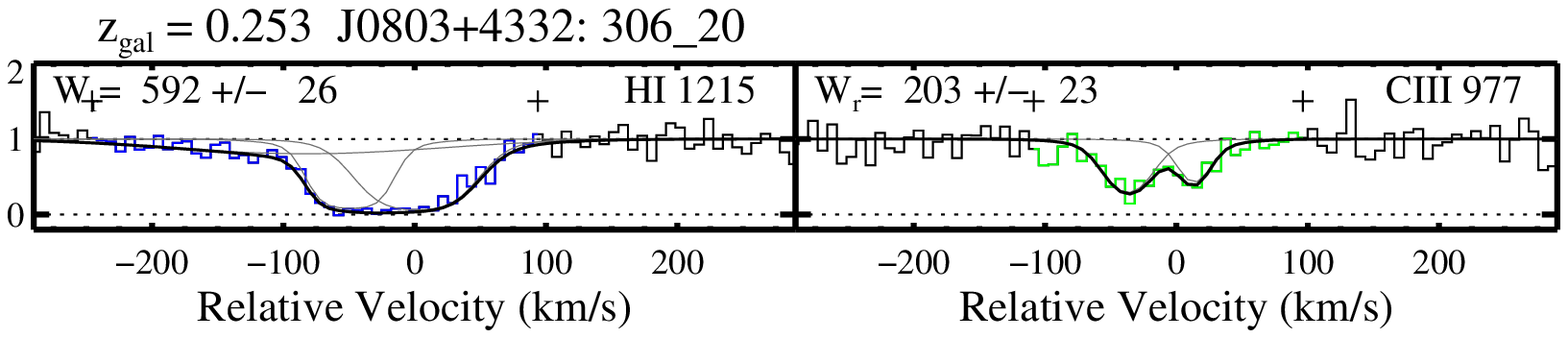}
\vspace{-0.2in}
\end{figure*}
\begin{figure*}[h!]
\includegraphics[width=0.99\linewidth]{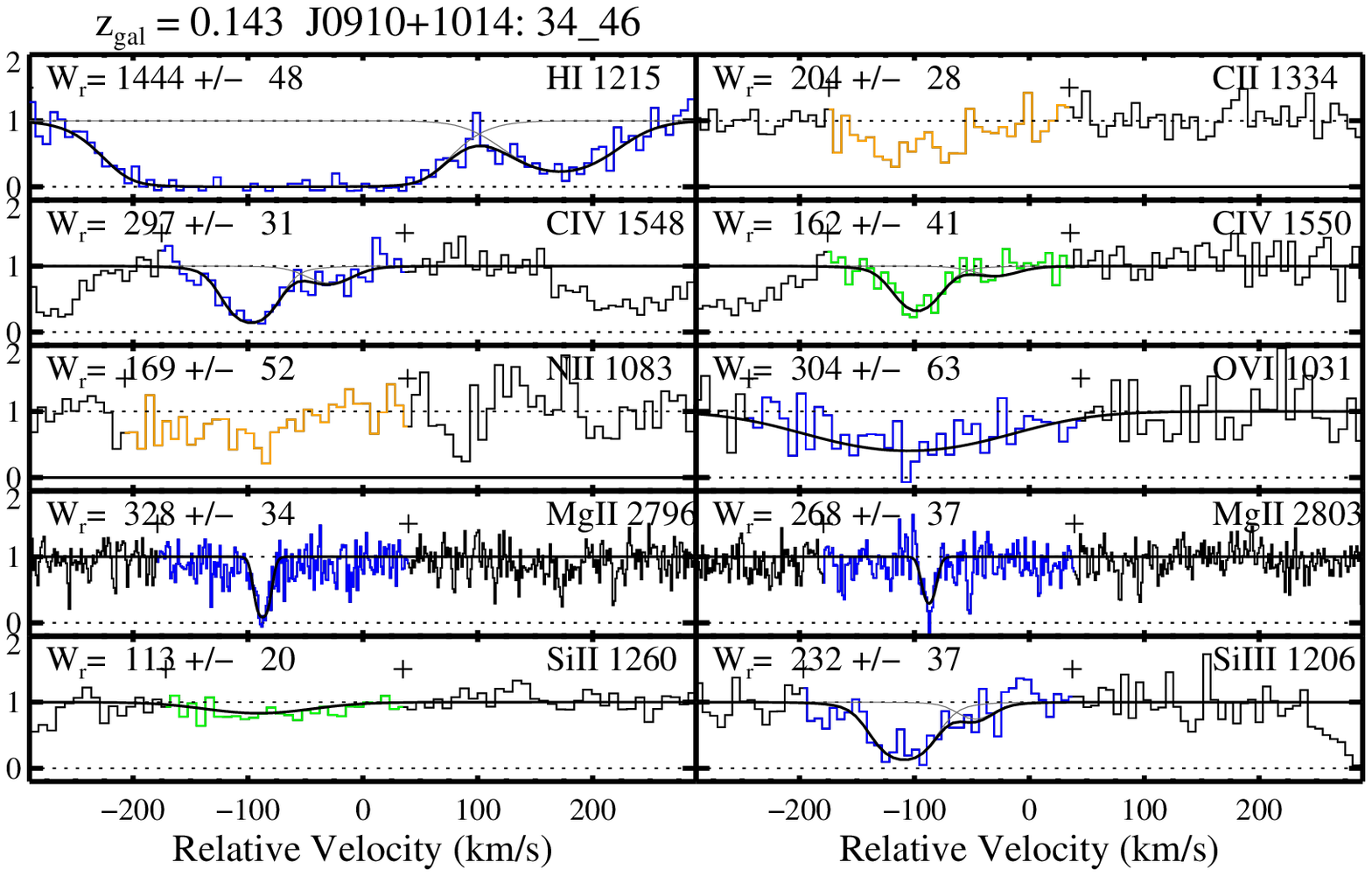}
\vspace{-0.2in}
\end{figure*}
\begin{figure*}[h!]
\includegraphics[width=0.99\linewidth]{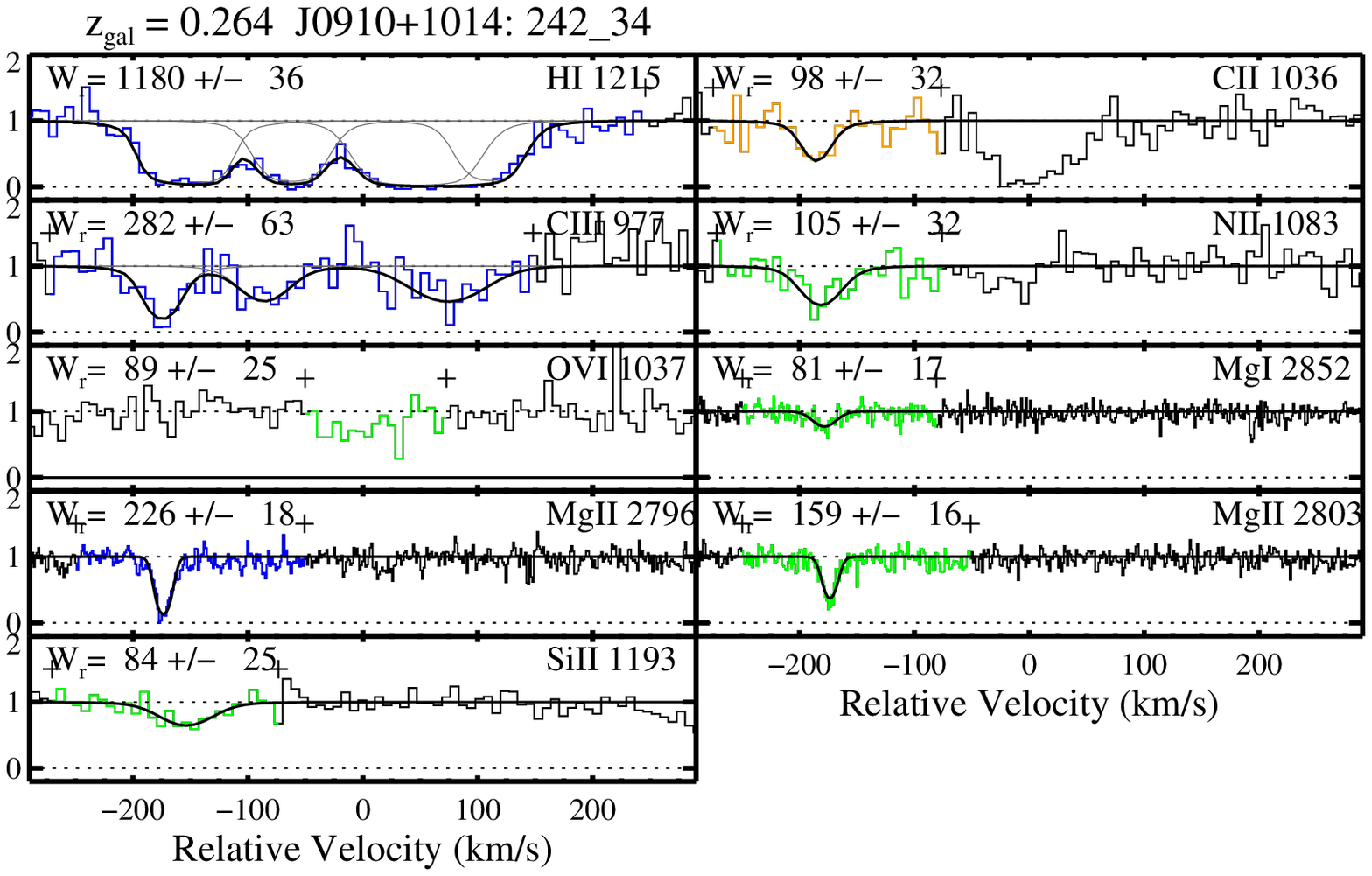}
\vspace{-0.2in}
\end{figure*}
\begin{figure*}[h!]
\includegraphics[width=0.99\linewidth]{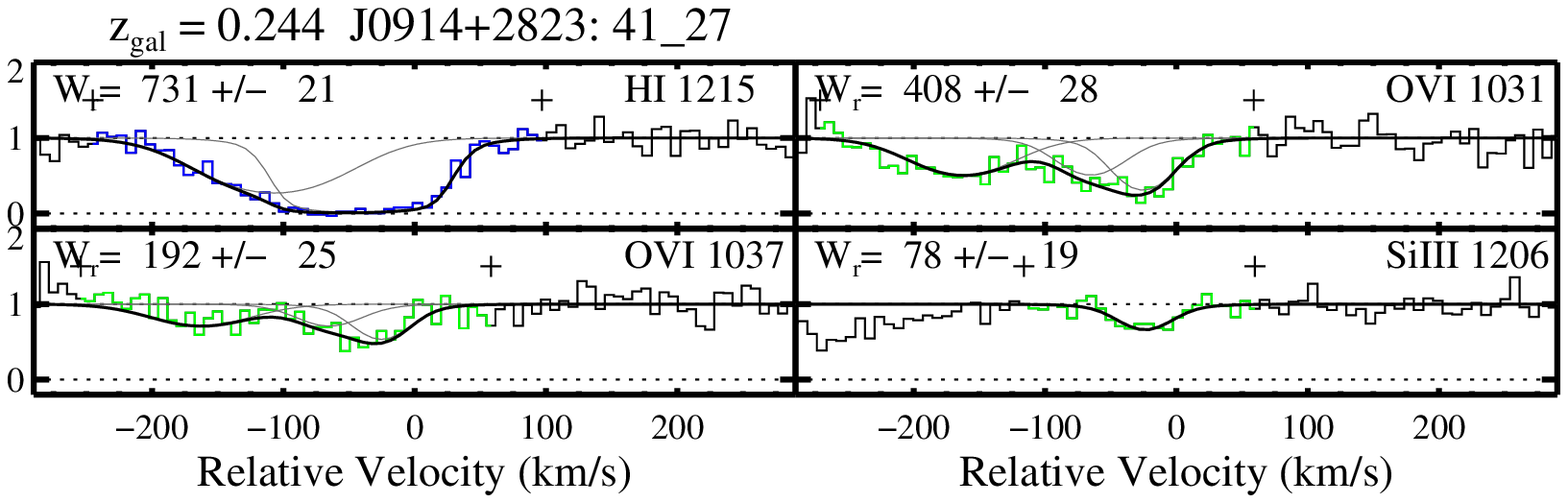}
\vspace{-0.2in}
\end{figure*}
\begin{figure*}[h!]
\includegraphics[width=0.99\linewidth]{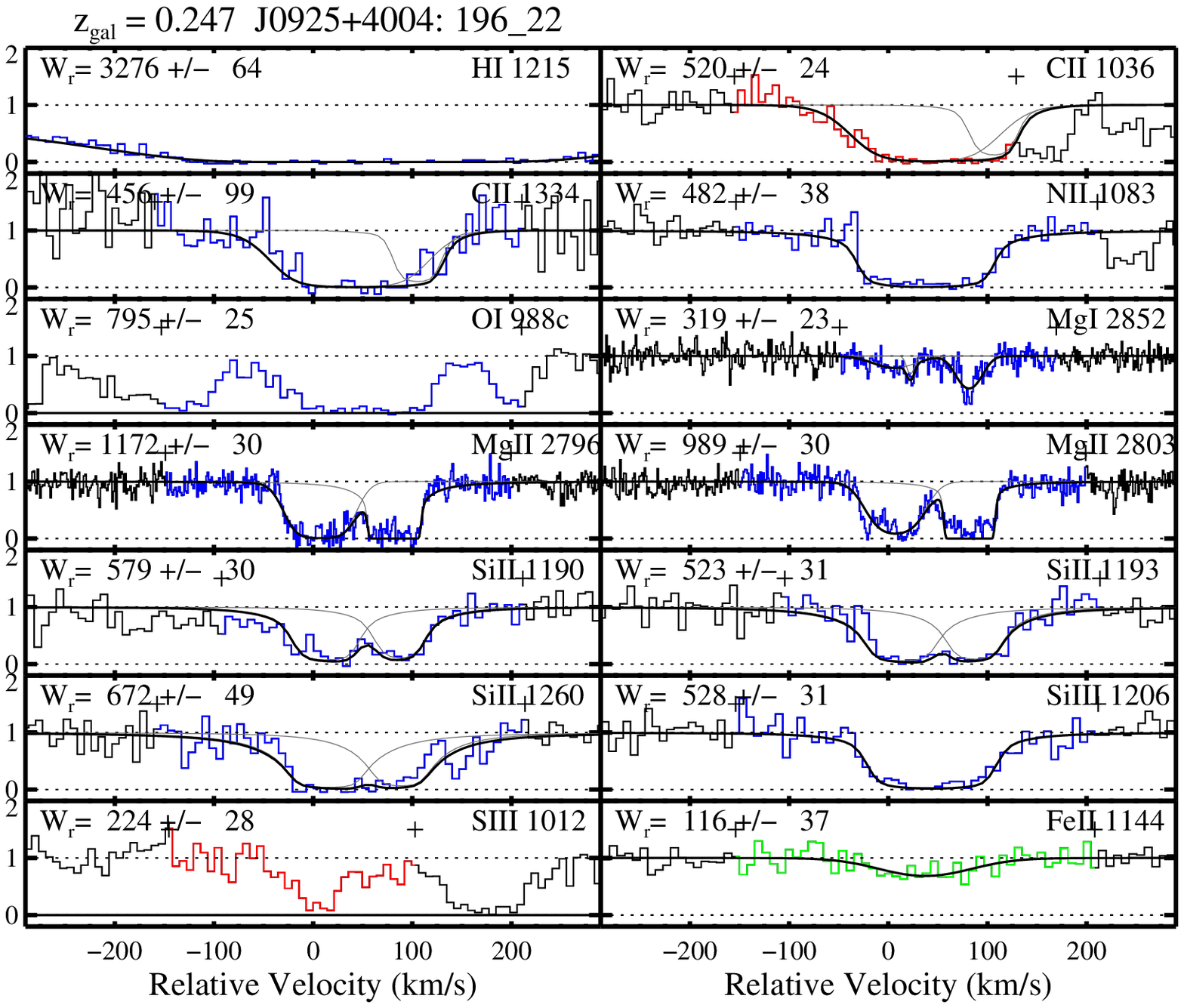}
\vspace{-0.2in}
\end{figure*}
\begin{figure*}[h!]
\includegraphics[width=0.99\linewidth]{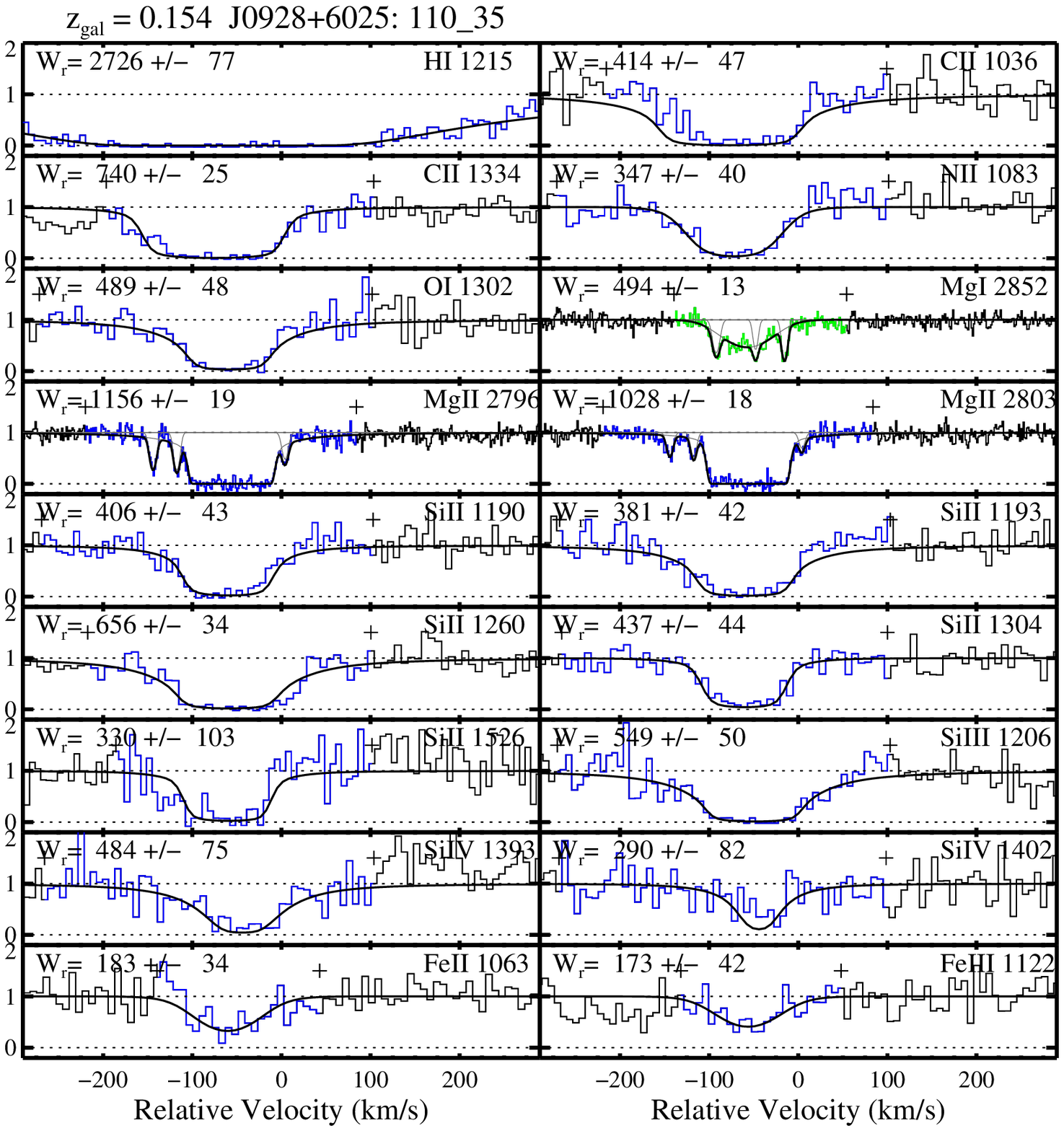}
\vspace{-0.2in}
\end{figure*}
\begin{figure*}[h!]
\includegraphics[width=0.99\linewidth]{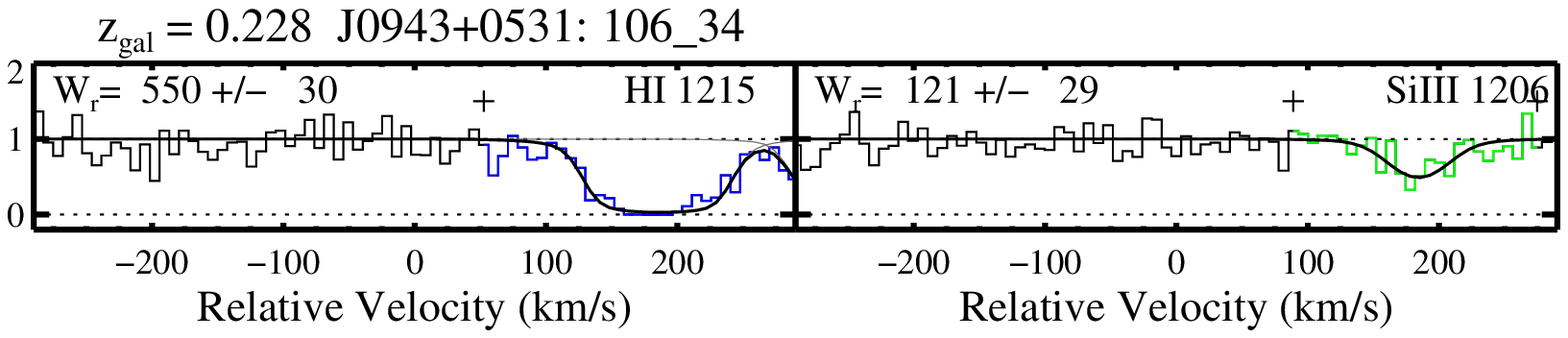}
\vspace{-0.2in}
\end{figure*}
\begin{figure*}[h!]
\includegraphics[width=0.99\linewidth]{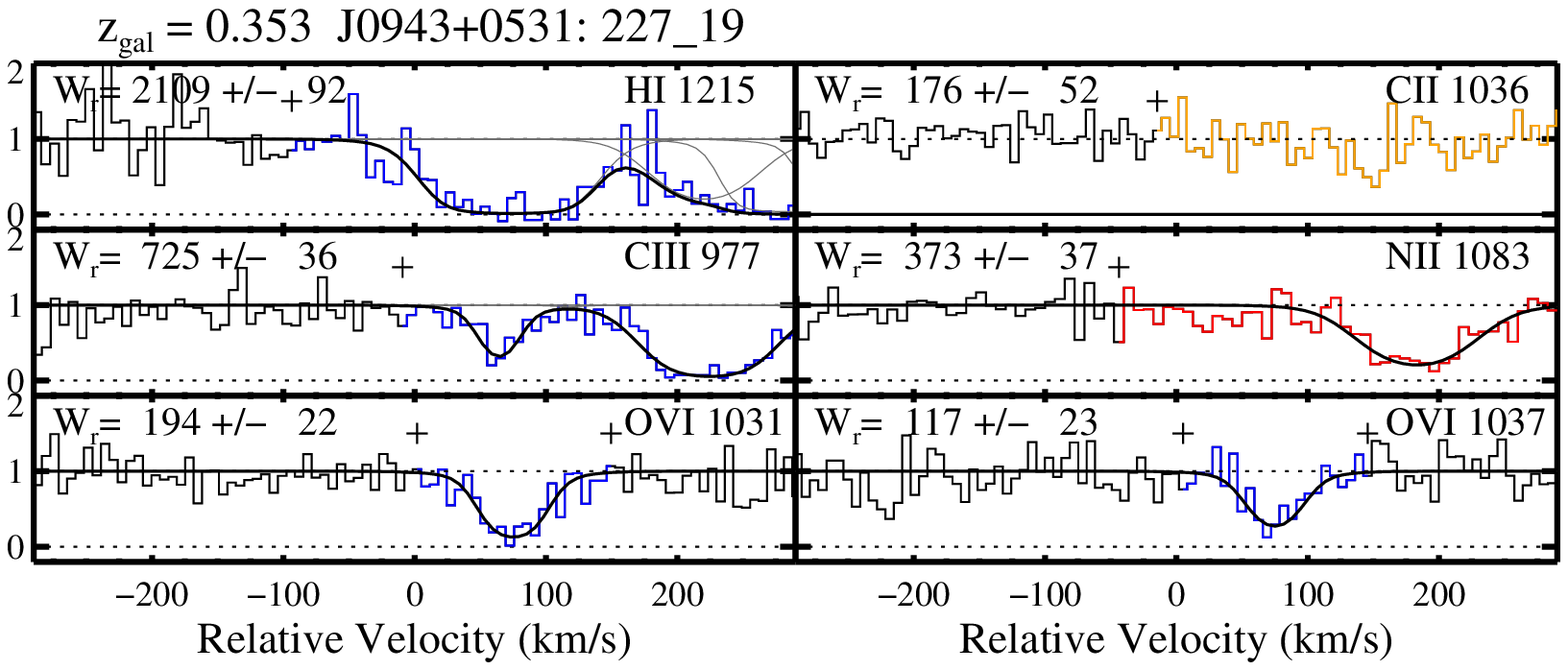}
\vspace{-0.2in}
\end{figure*}
\begin{figure*}[h!]
\includegraphics[width=0.99\linewidth]{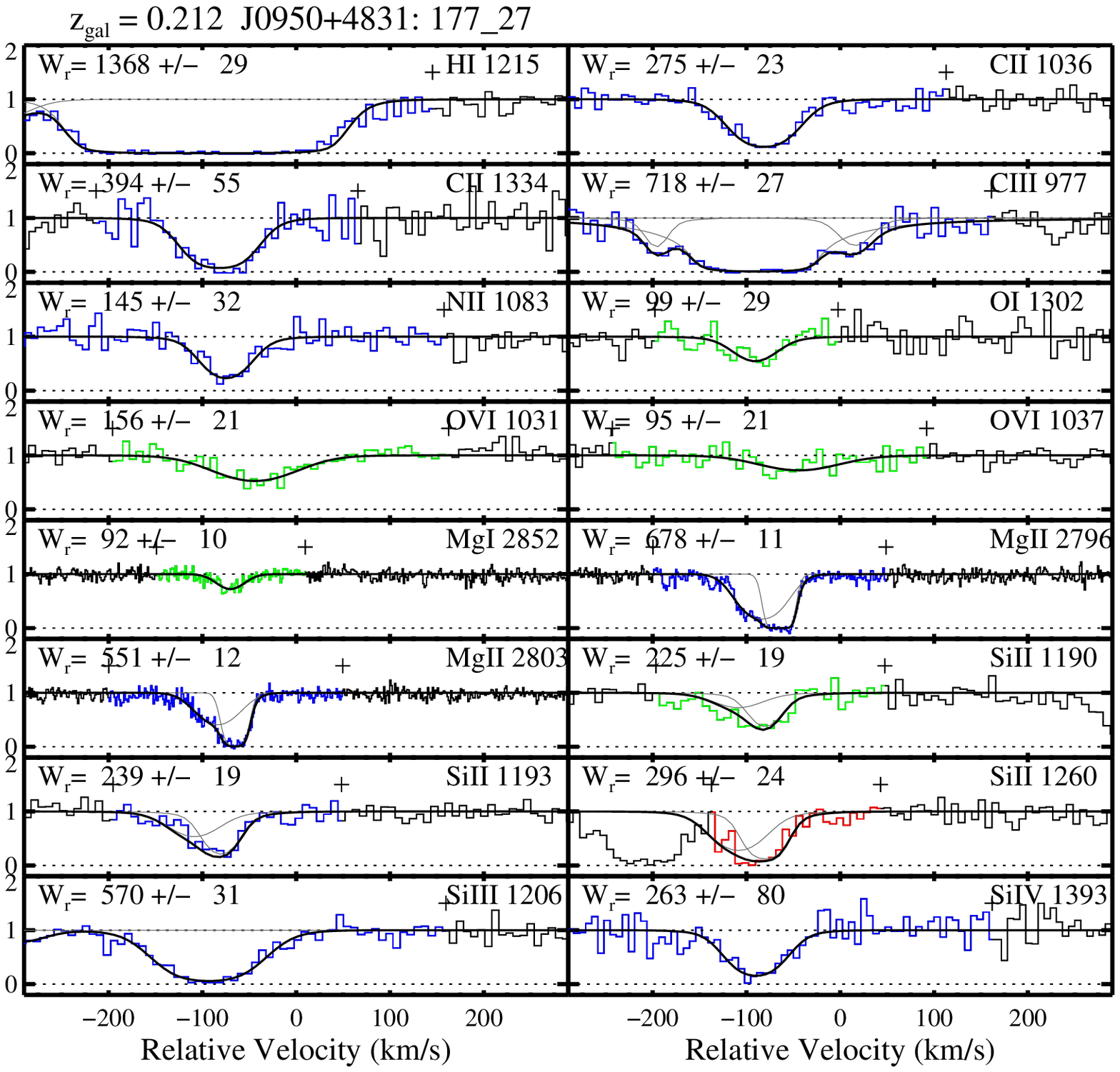}
\vspace{-0.2in}
\end{figure*}
\begin{figure*}[h!]
\includegraphics[width=0.99\linewidth]{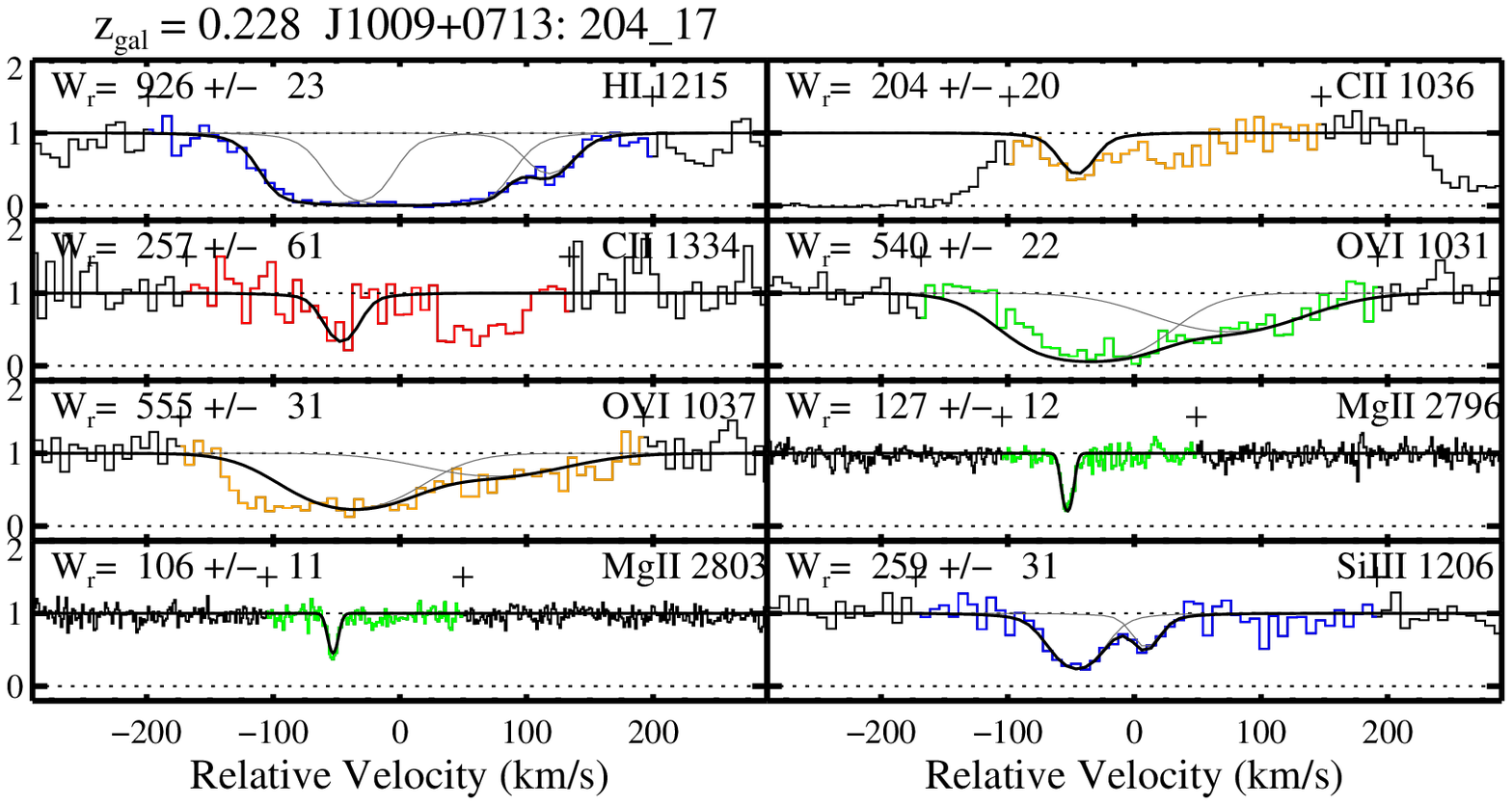}
\vspace{-0.2in}
\end{figure*}
\begin{figure*}[h!]
\includegraphics[width=0.99\linewidth]{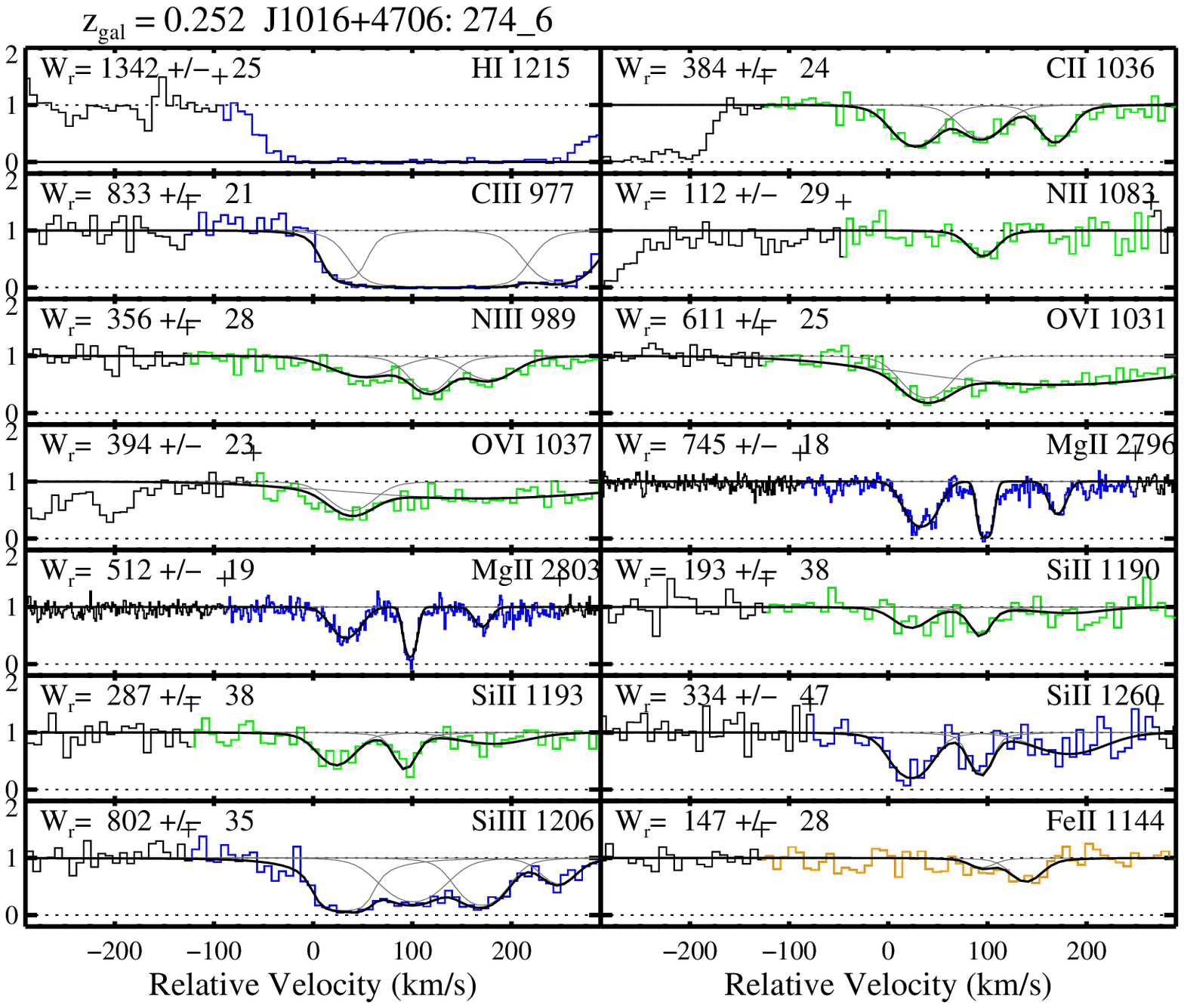}
\vspace{-0.2in}
\end{figure*}
\begin{figure*}[h!]
\includegraphics[width=0.99\linewidth]{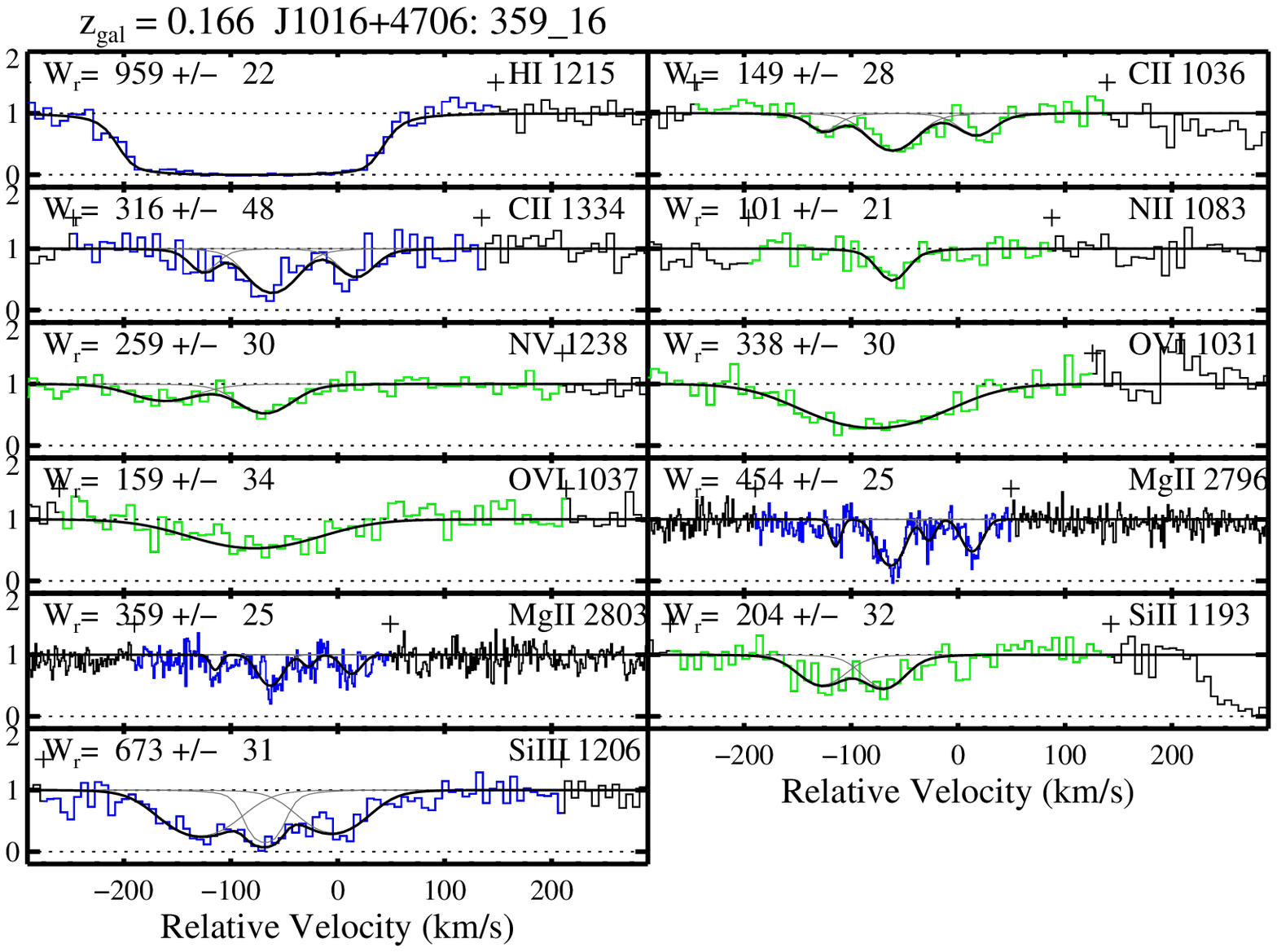}
\vspace{-0.2in}
\end{figure*}
\begin{figure*}[h!]
\includegraphics[width=0.99\linewidth]{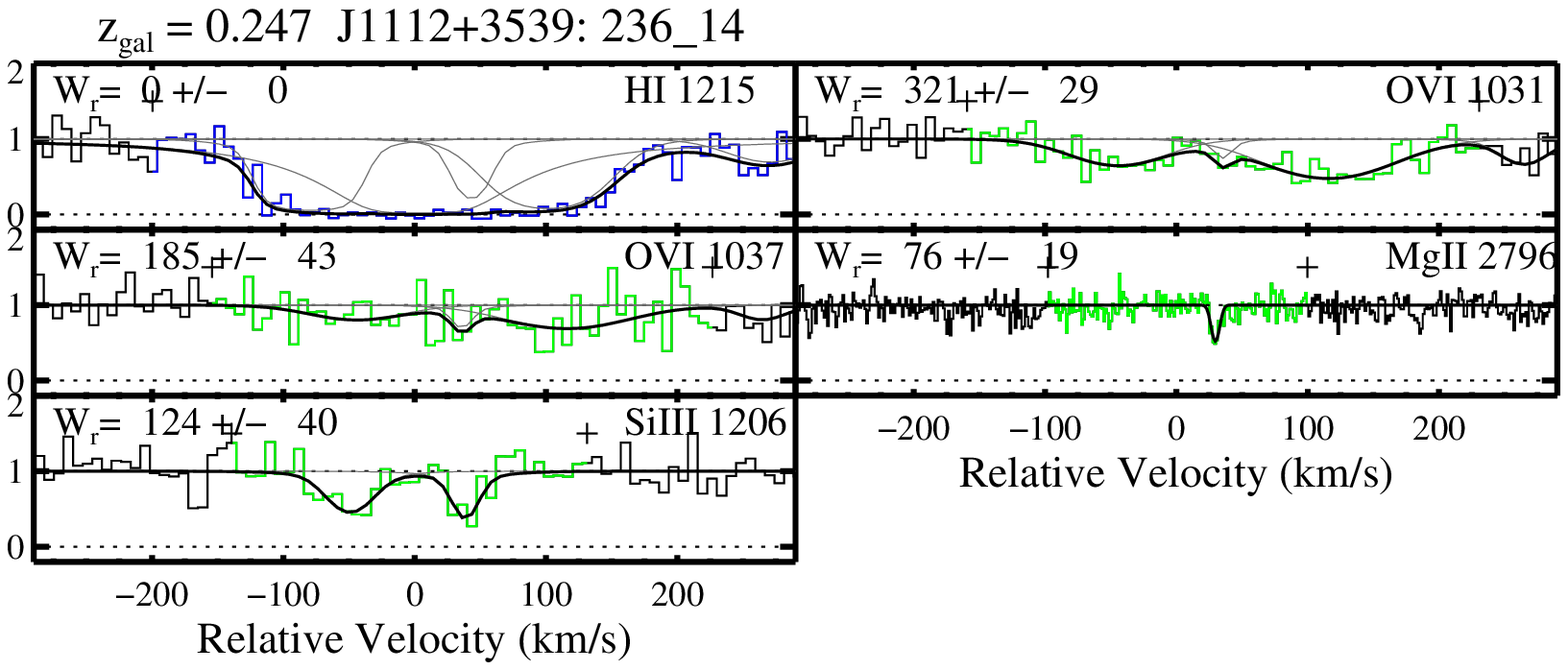}
\vspace{-0.2in}
\end{figure*}
\begin{figure*}[h!]
\includegraphics[width=0.99\linewidth]{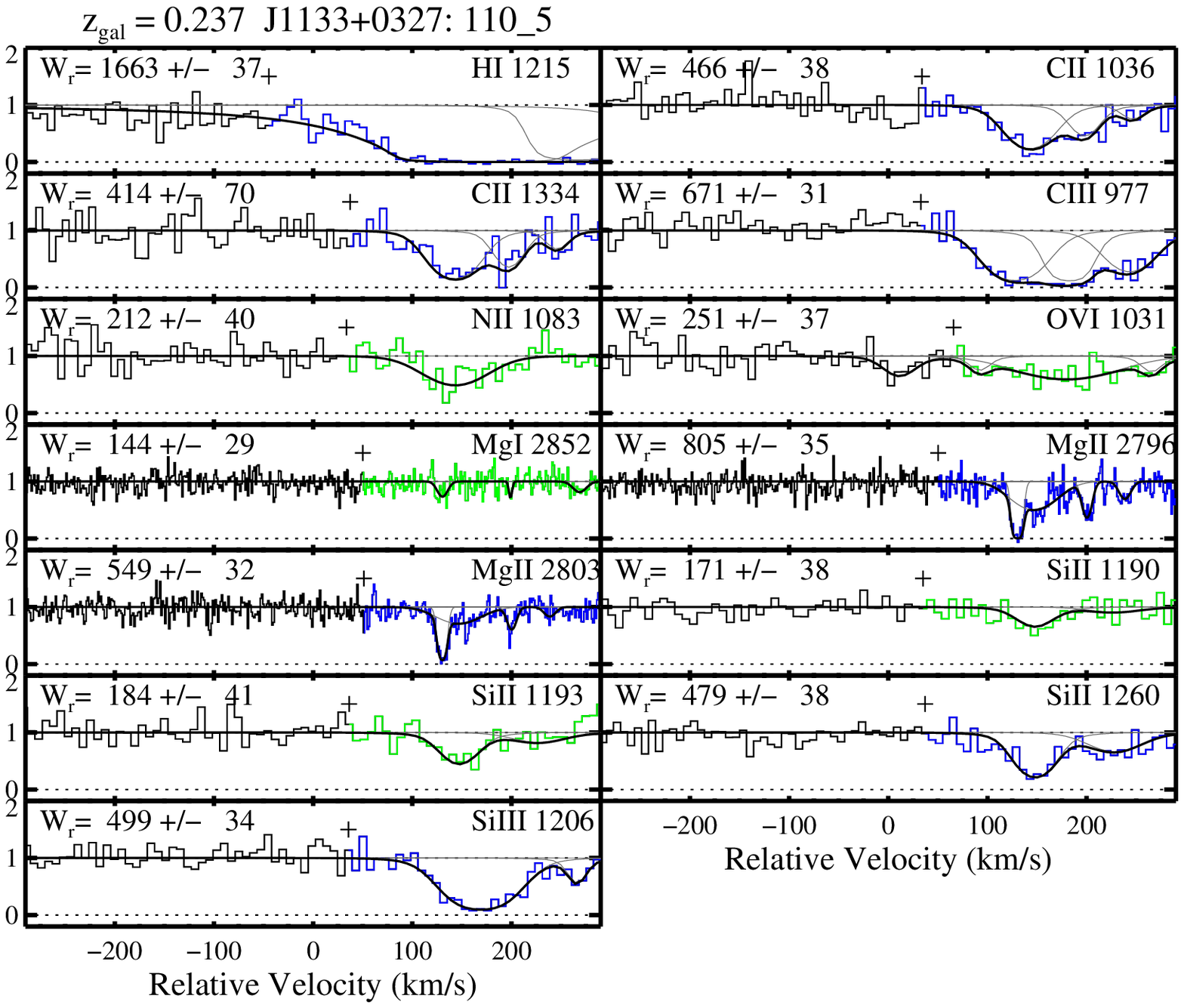}
\vspace{-0.2in}
\end{figure*}
\afterpage{\clearpage}
\begin{figure*}[h!]
\includegraphics[width=0.99\linewidth]{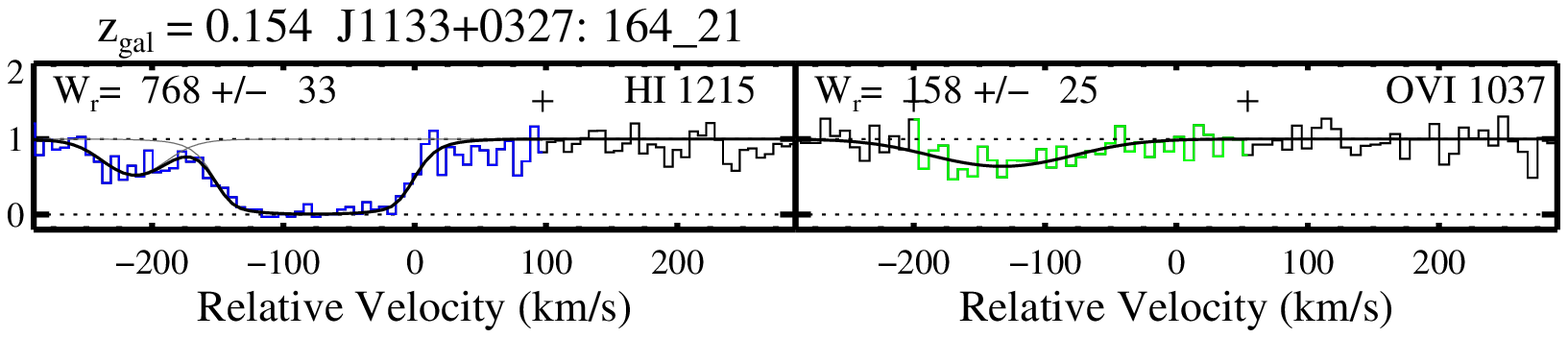}
\vspace{-0.2in}
\end{figure*}
\begin{figure*}[h!]
\includegraphics[width=0.99\linewidth]{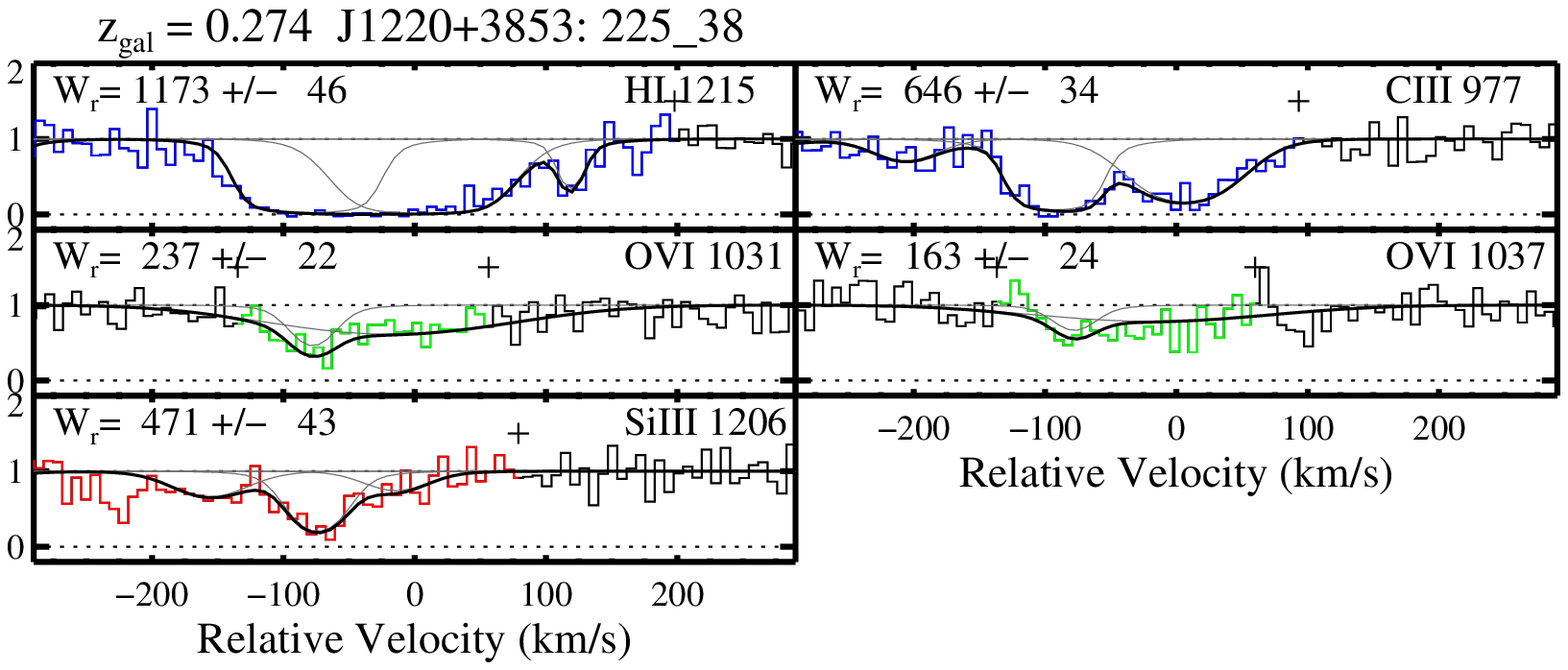}
\vspace{-0.2in}
\end{figure*}
\begin{figure*}[h!]
\includegraphics[width=0.99\linewidth]{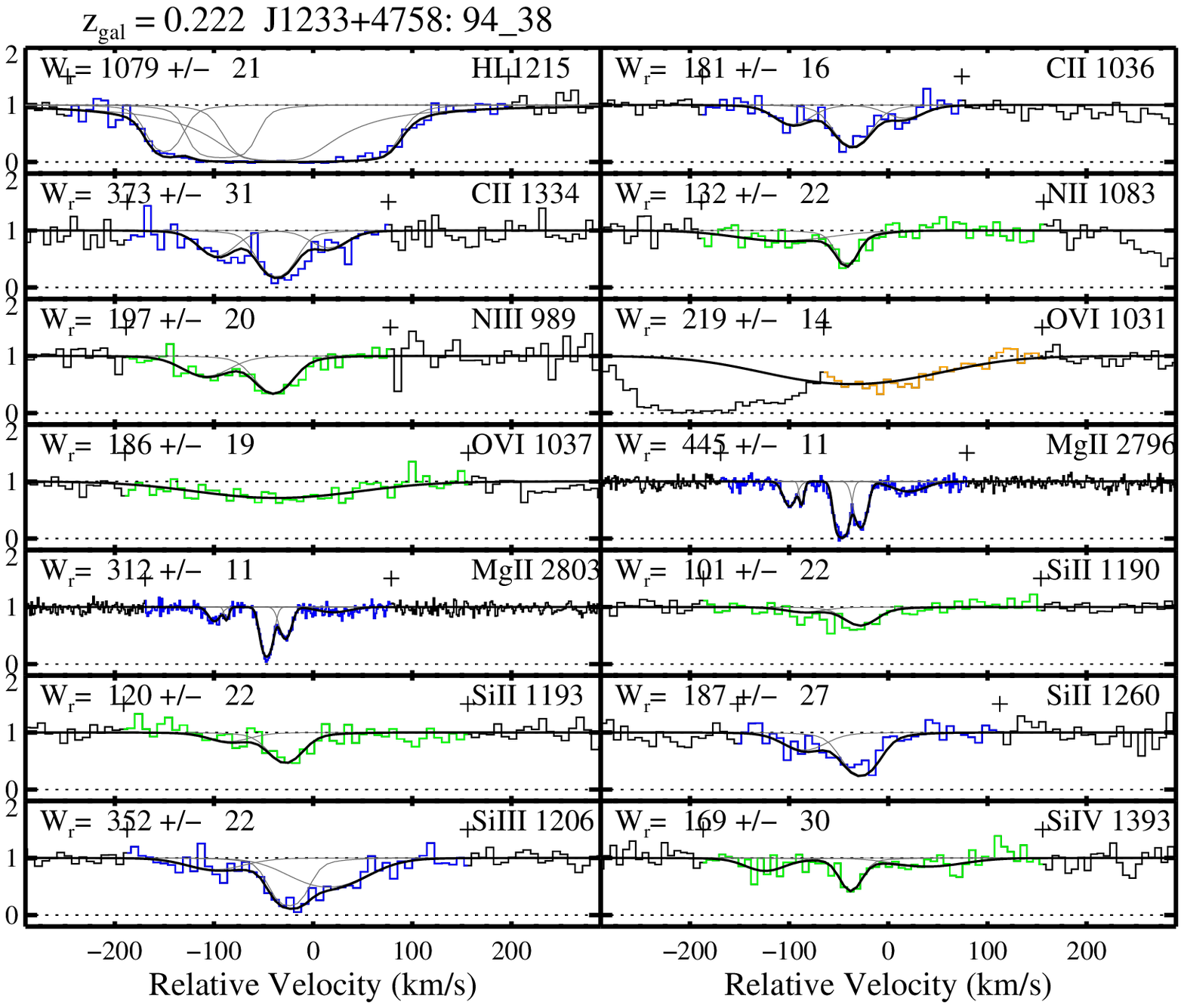}
\vspace{-0.2in}
\end{figure*}
\begin{figure*}[h!]
\includegraphics[width=0.99\linewidth]{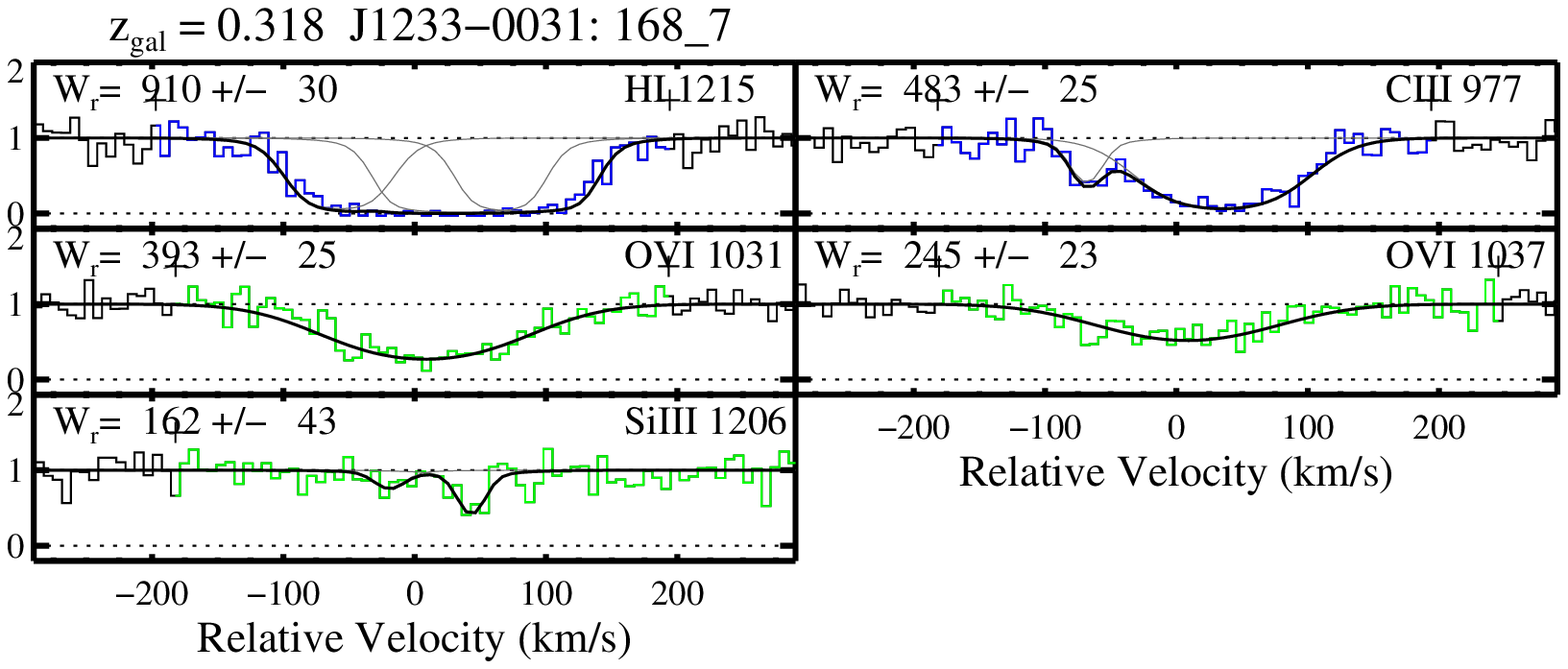}
\vspace{-0.2in}
\end{figure*}
\begin{figure*}[h!]
\includegraphics[width=0.99\linewidth]{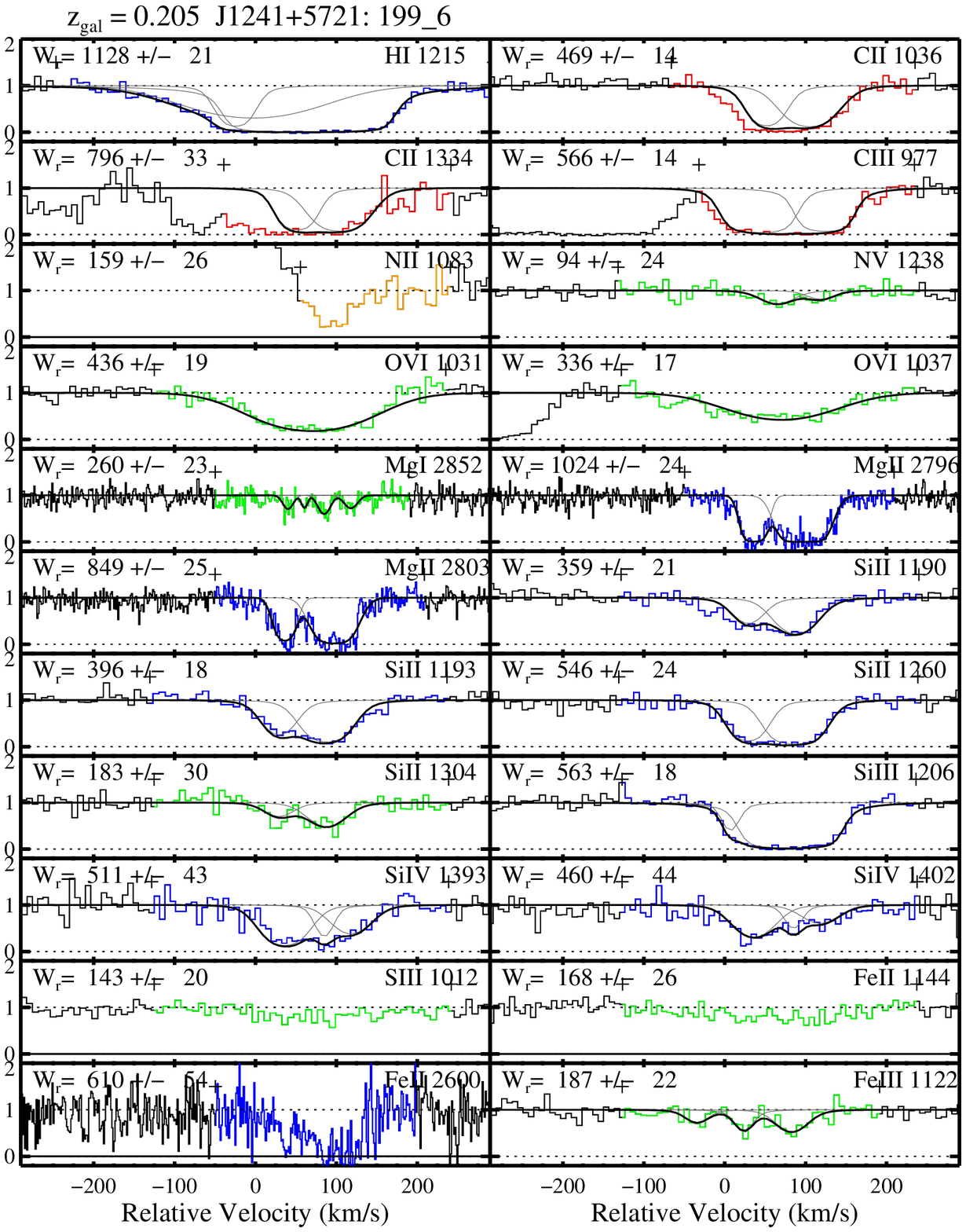}
\vspace{-0.2in}
\end{figure*}
\begin{figure*}[h!]
\includegraphics[width=0.99\linewidth]{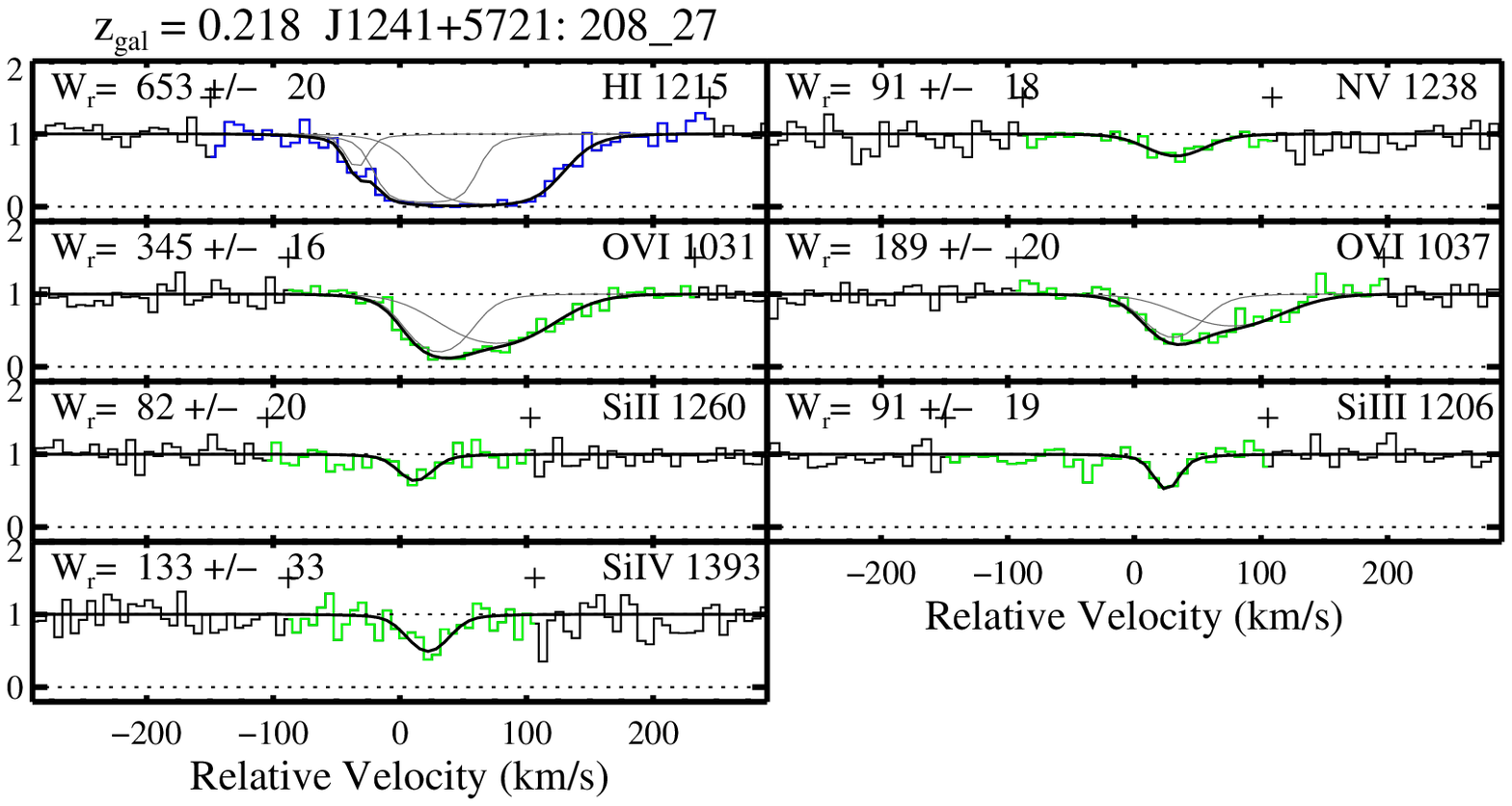}
\vspace{-0.2in}
\end{figure*}
\begin{figure*}[h!]
\includegraphics[width=0.99\linewidth]{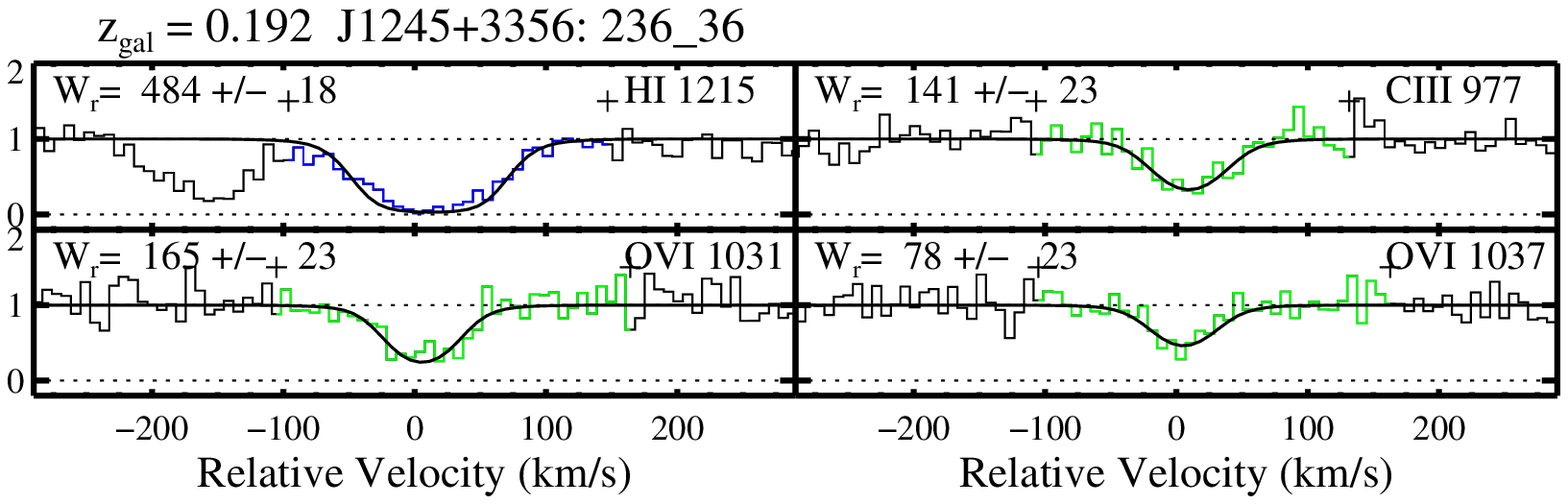}
\vspace{-0.2in}
\end{figure*}
\begin{figure*}[h!]
\includegraphics[width=0.99\linewidth]{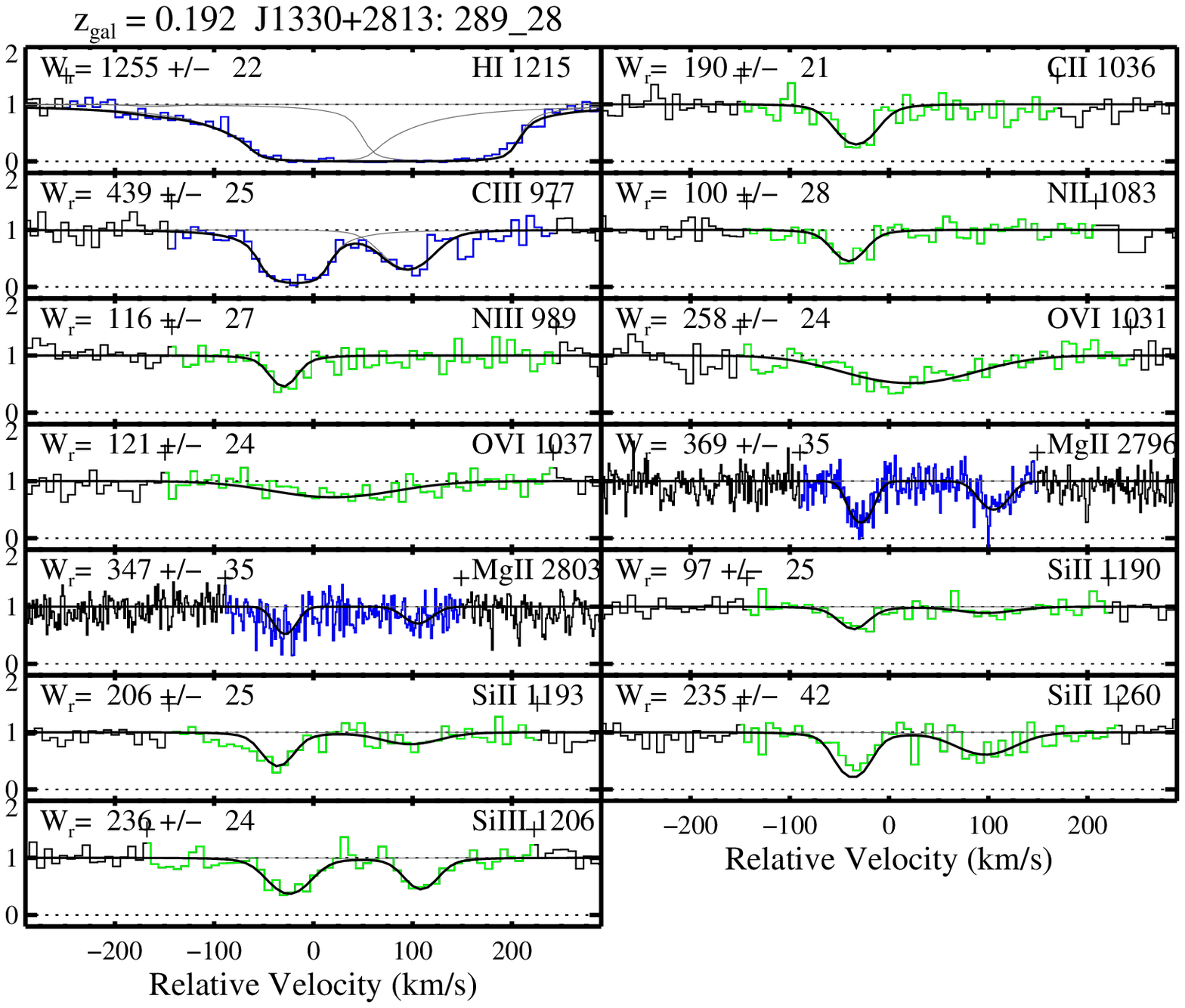}
\vspace{-0.2in}
\end{figure*}
\begin{figure*}[h!]
\includegraphics[width=0.99\linewidth]{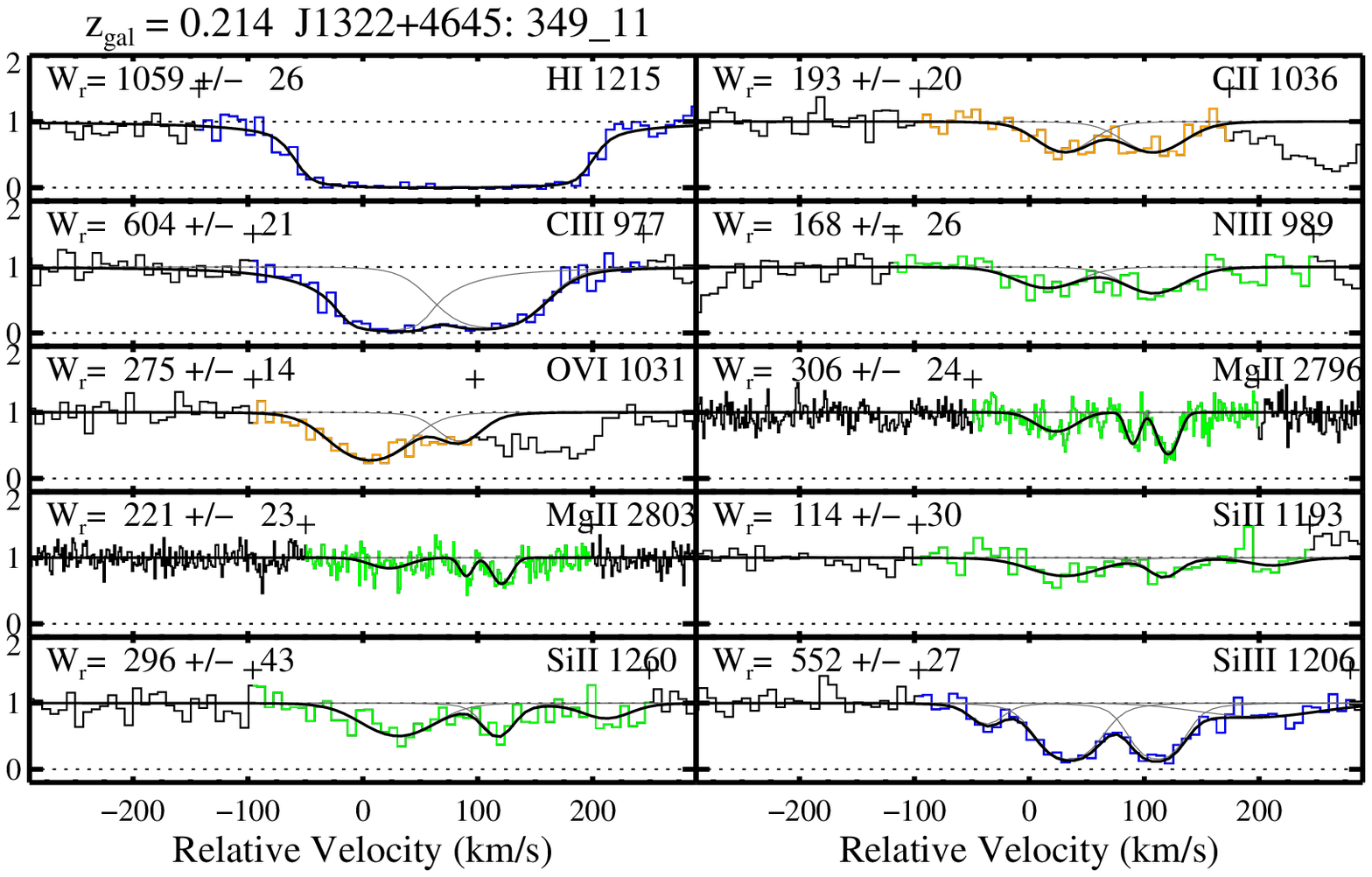}
\vspace{-0.2in}
\end{figure*}
\begin{figure*}[h!]
\includegraphics[width=0.99\linewidth]{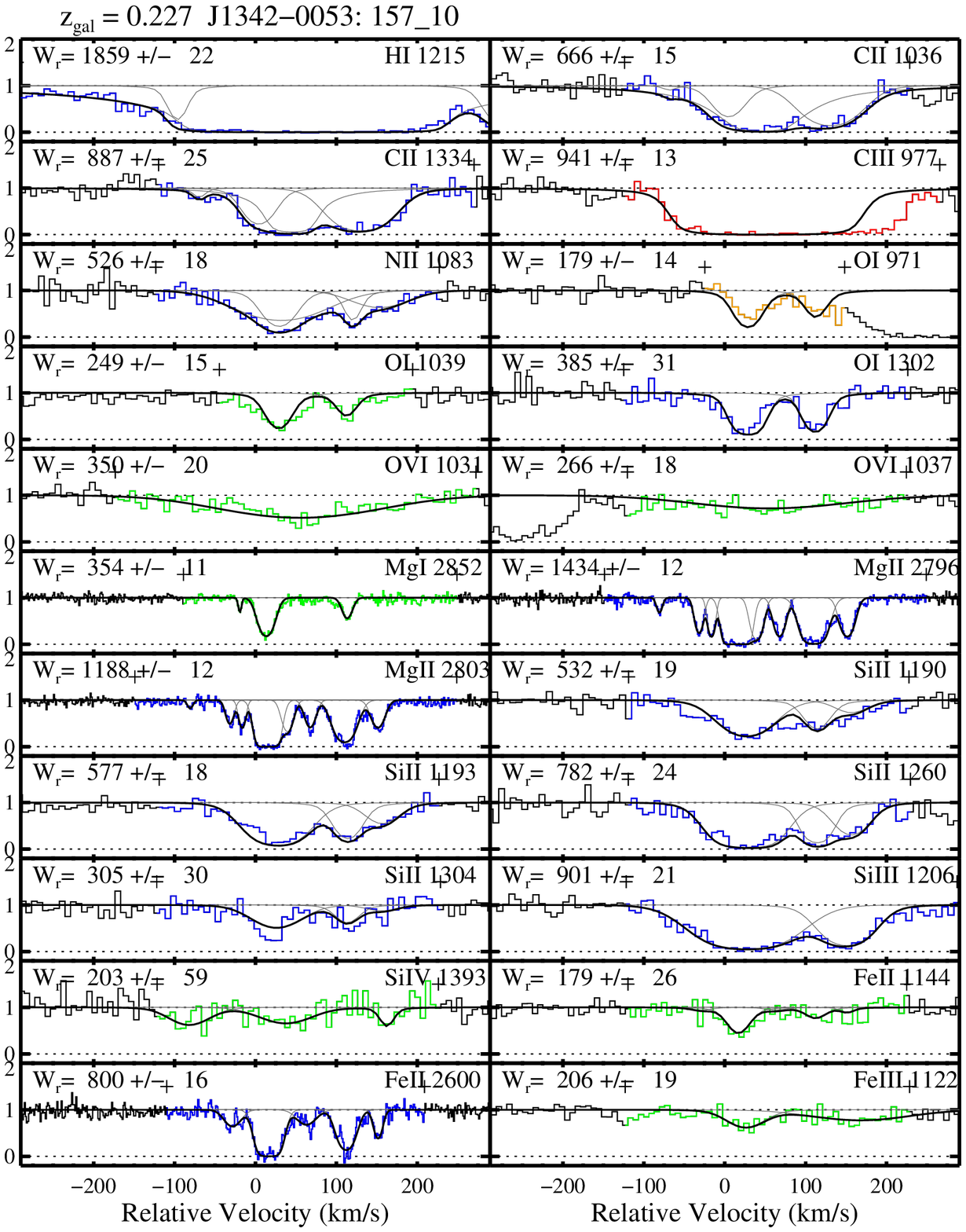}
\vspace{-0.2in}
\end{figure*}
\begin{figure*}[h!]
\includegraphics[width=0.99\linewidth]{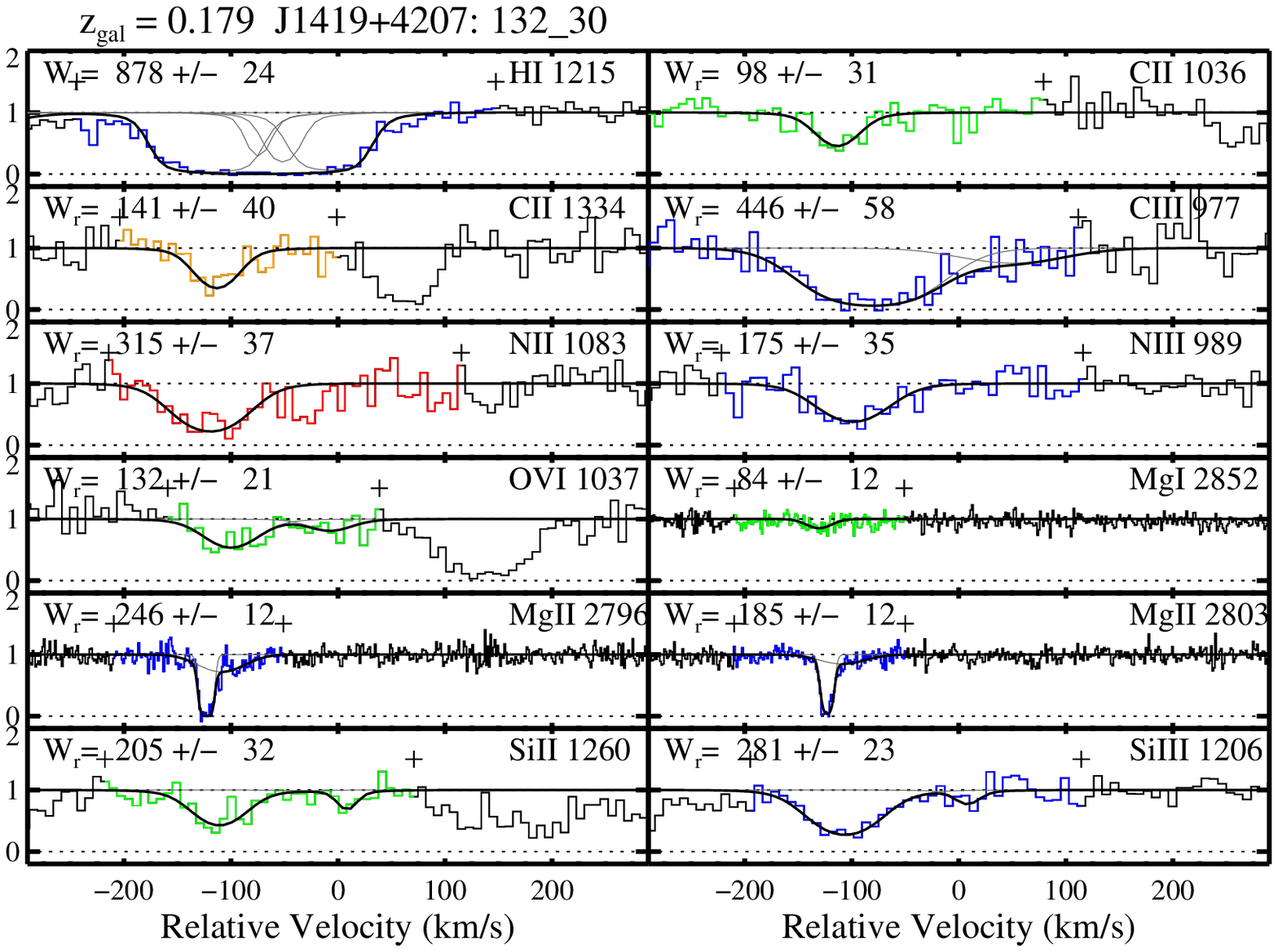}
\vspace{-0.2in}
\end{figure*}
\begin{figure*}[h!]
\includegraphics[width=0.99\linewidth]{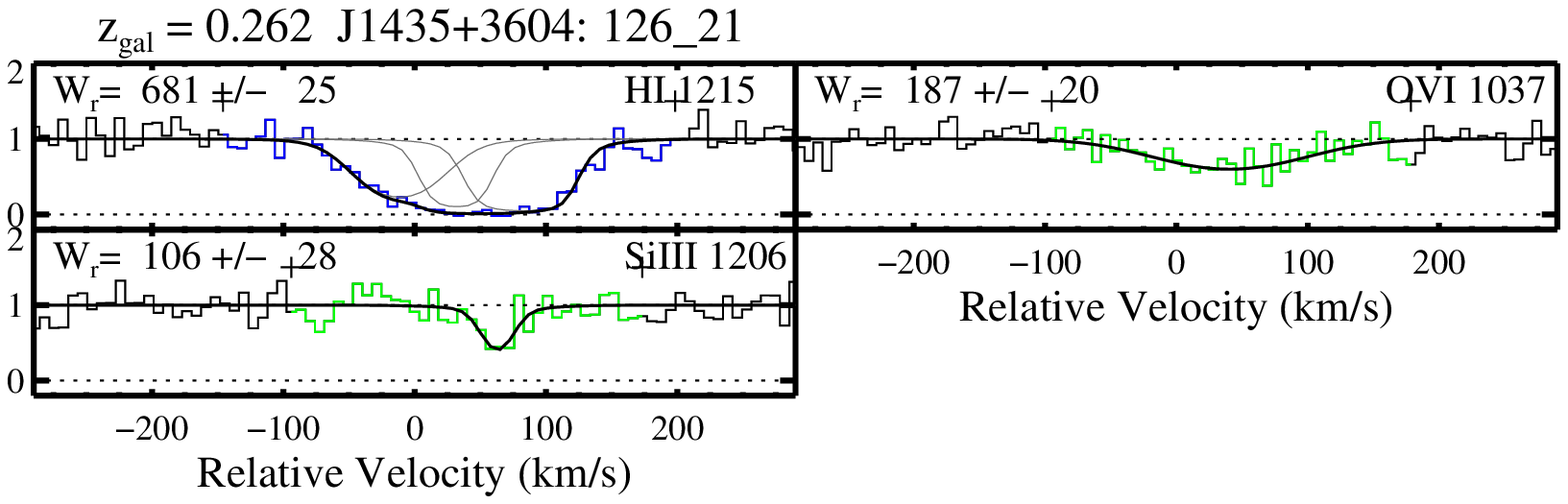}
\vspace{-0.2in}
\end{figure*}
\begin{figure*}[h!]
\includegraphics[width=0.99\linewidth]{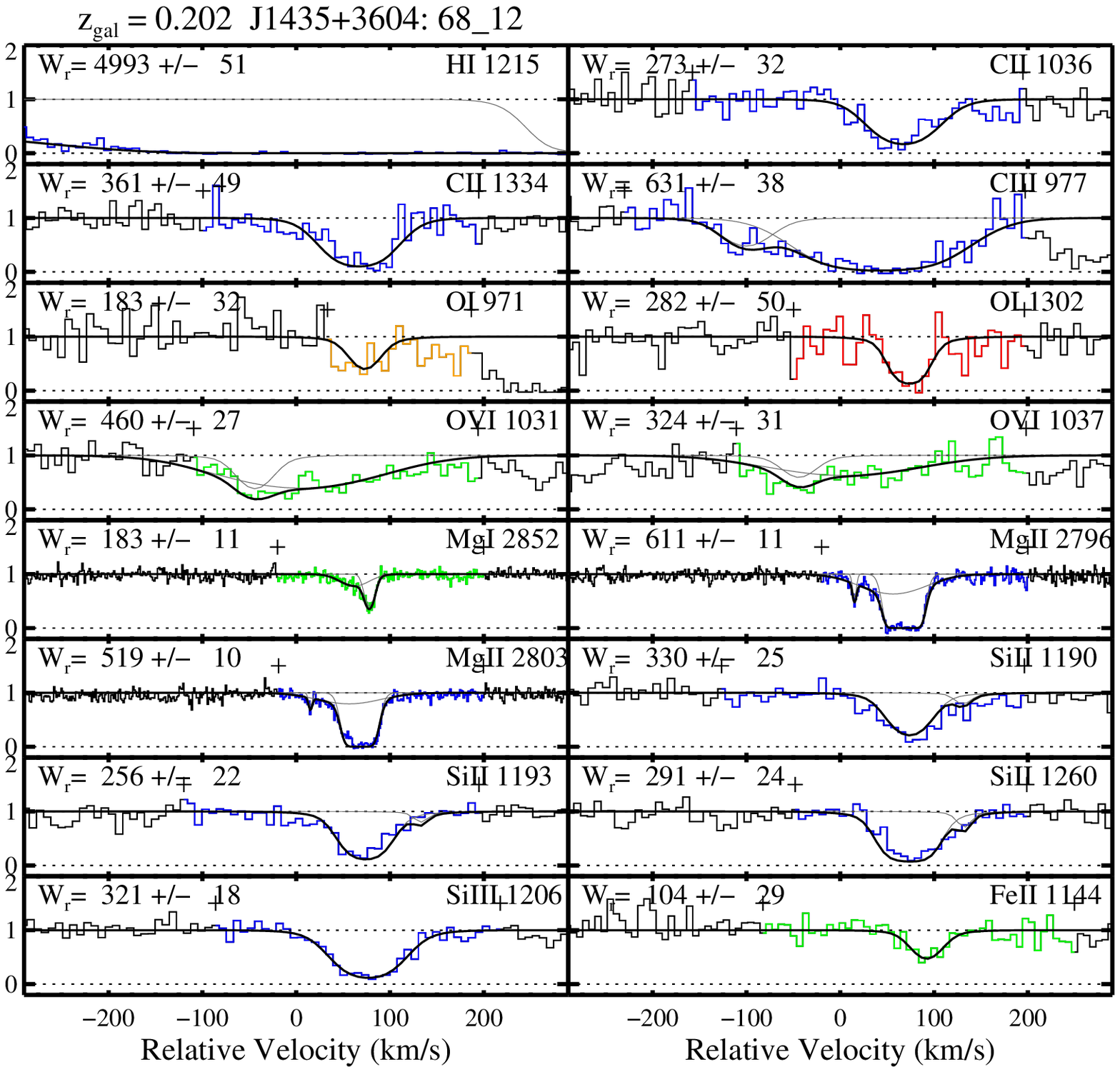}
\vspace{-0.2in}
\end{figure*}
\begin{figure*}[h!]
\includegraphics[width=0.99\linewidth]{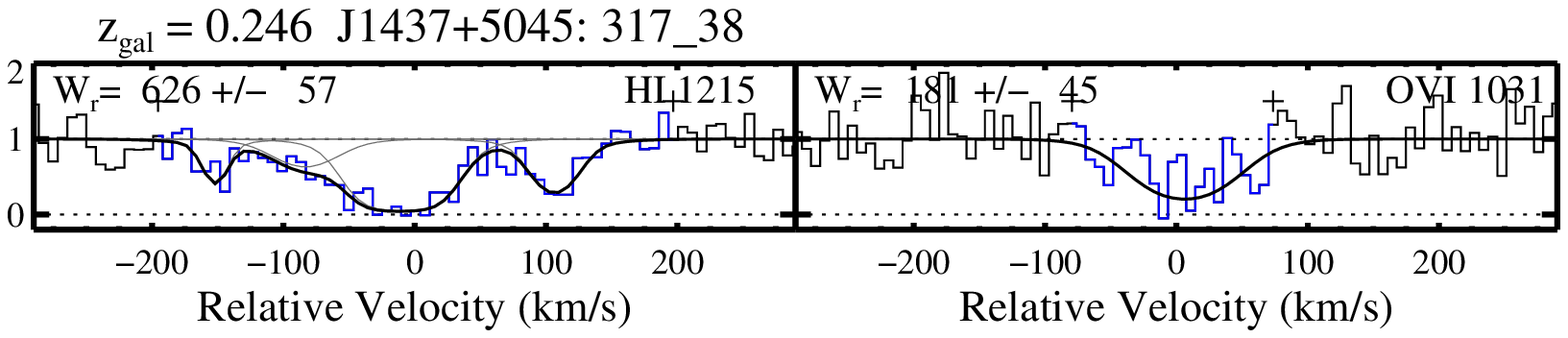}
\vspace{-0.2in}
\end{figure*}
\begin{figure*}[h!]
\includegraphics[width=0.99\linewidth]{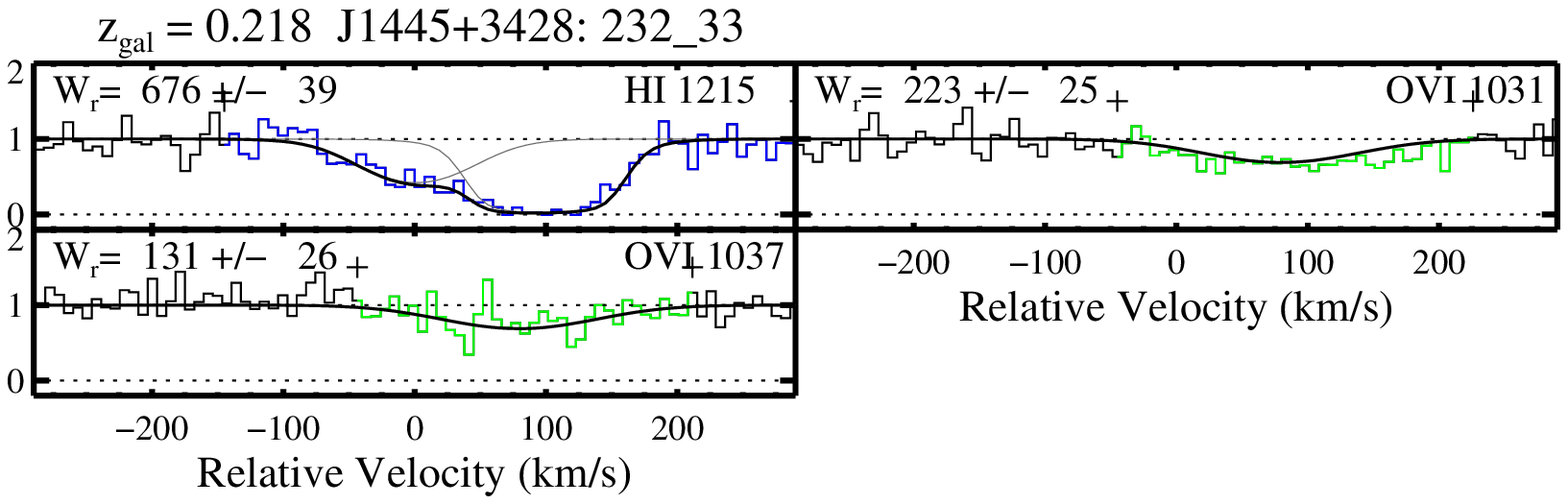}
\vspace{-0.2in}
\end{figure*}
\begin{figure*}[h!]
\includegraphics[width=0.99\linewidth]{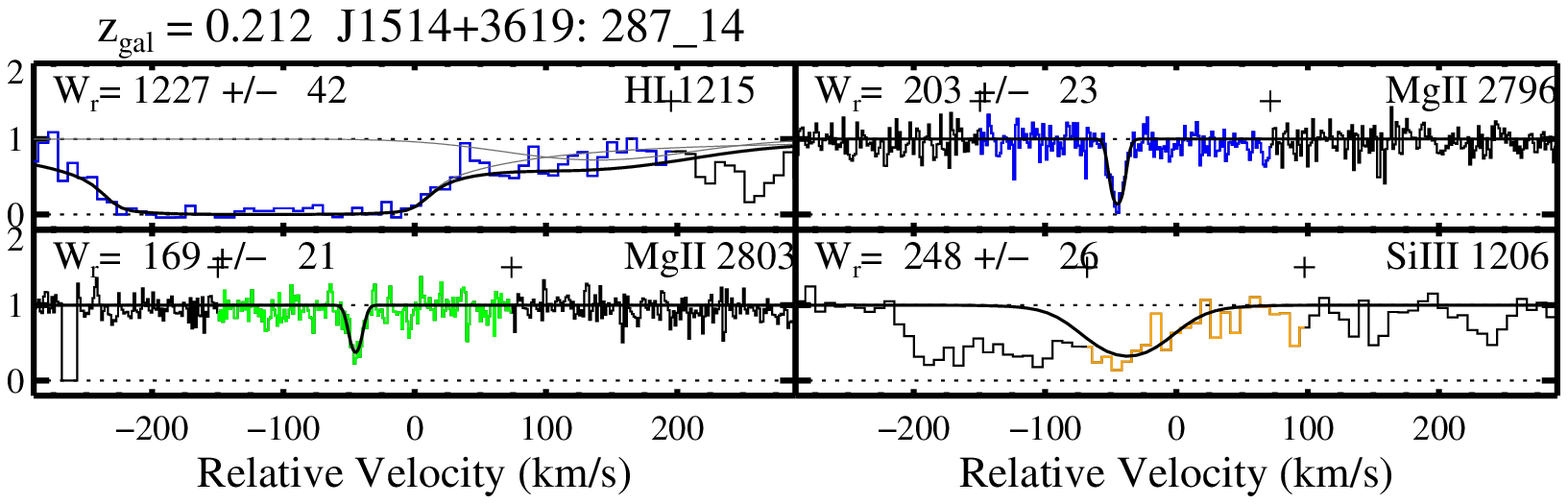}
\vspace{-0.2in}
\end{figure*}
\afterpage{\clearpage}
\begin{figure*}[h!]
\includegraphics[width=0.99\linewidth]{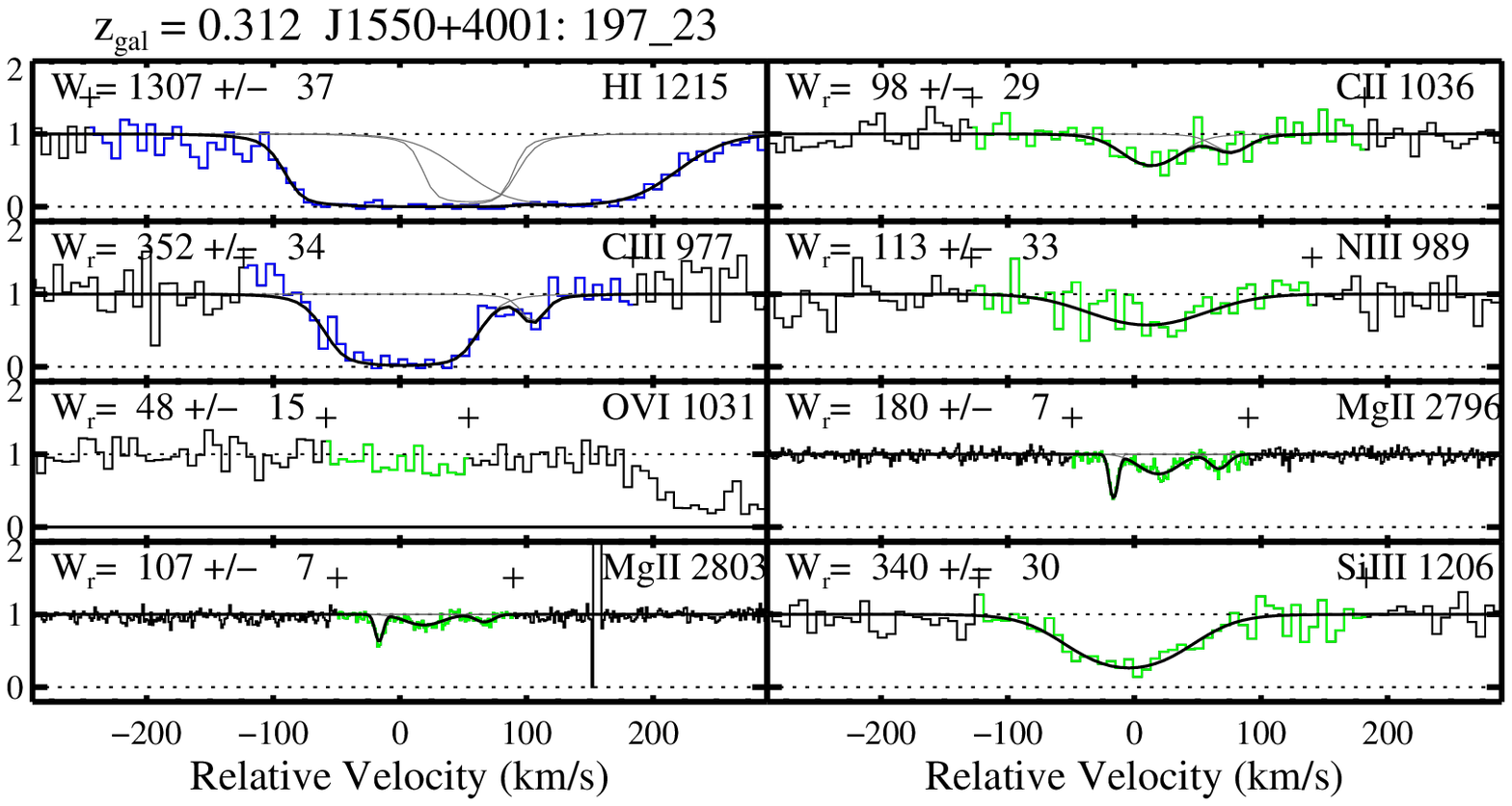}
\vspace{-0.2in}
\end{figure*}
\begin{figure*}[h!]
\includegraphics[width=0.99\linewidth]{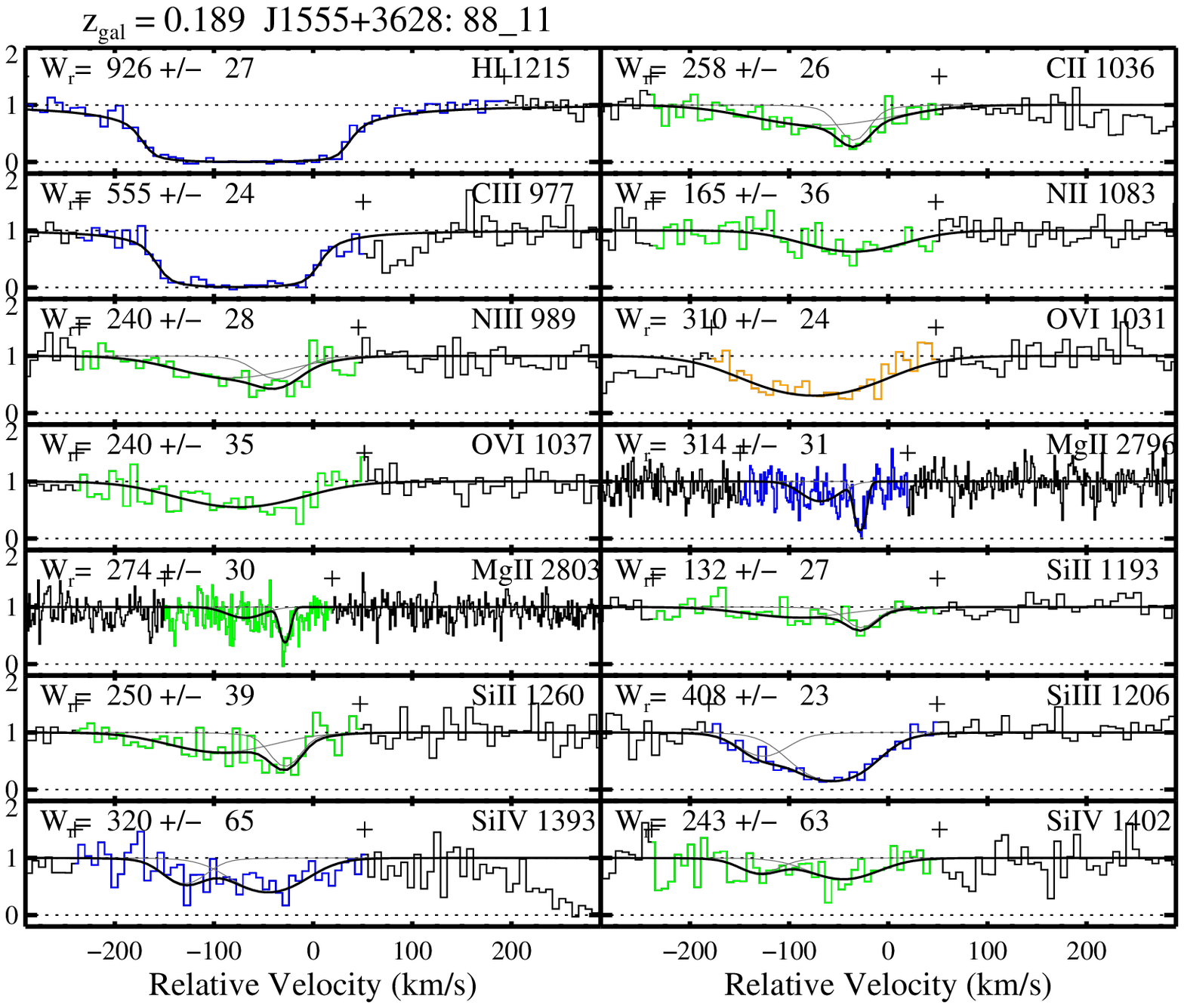}
\vspace{-0.2in}
\end{figure*}
\begin{figure*}[h!]
\includegraphics[width=0.99\linewidth]{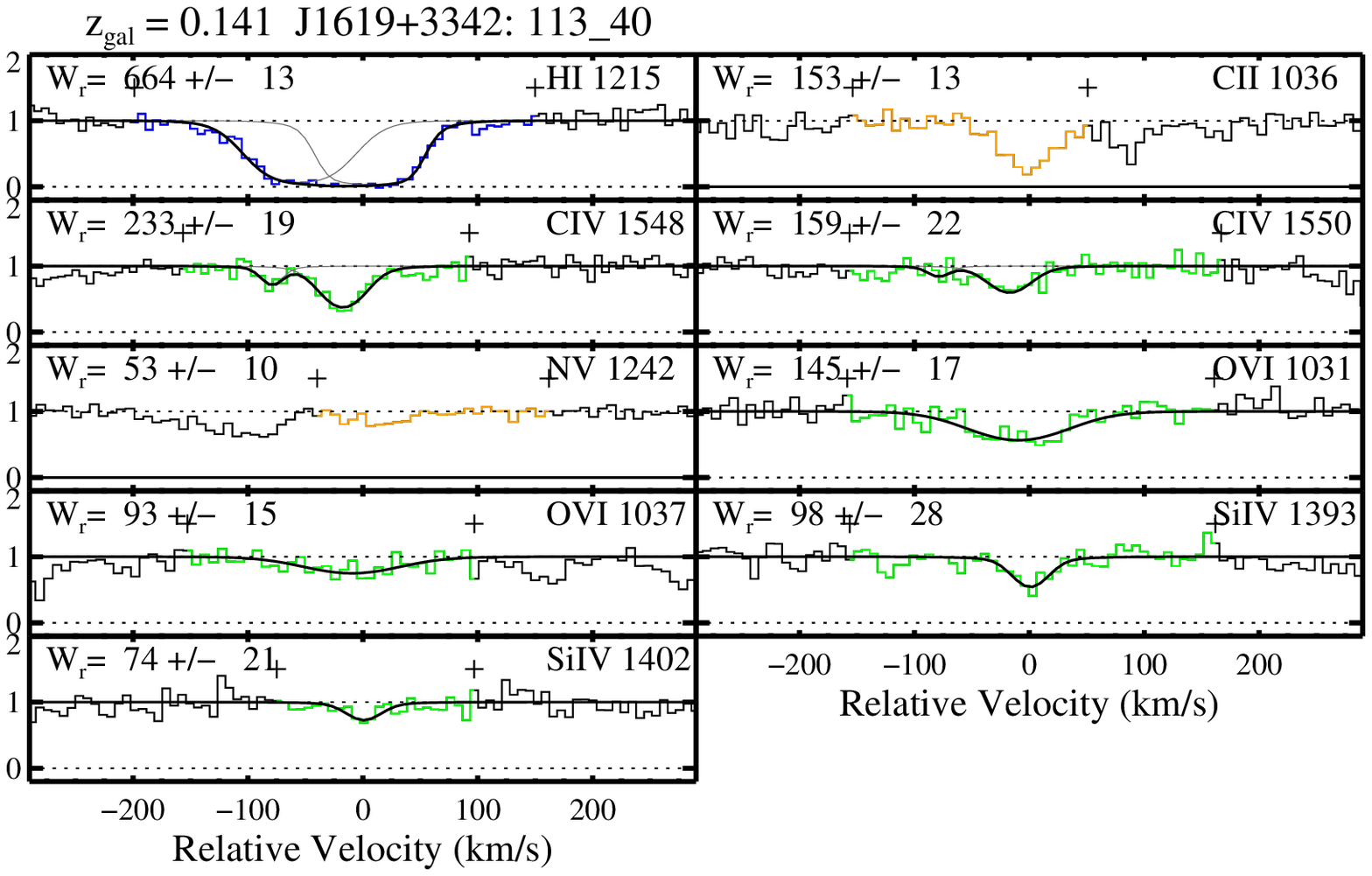}
\vspace{-0.2in}
\end{figure*}
\begin{figure*}[h!]
\includegraphics[width=0.99\linewidth]{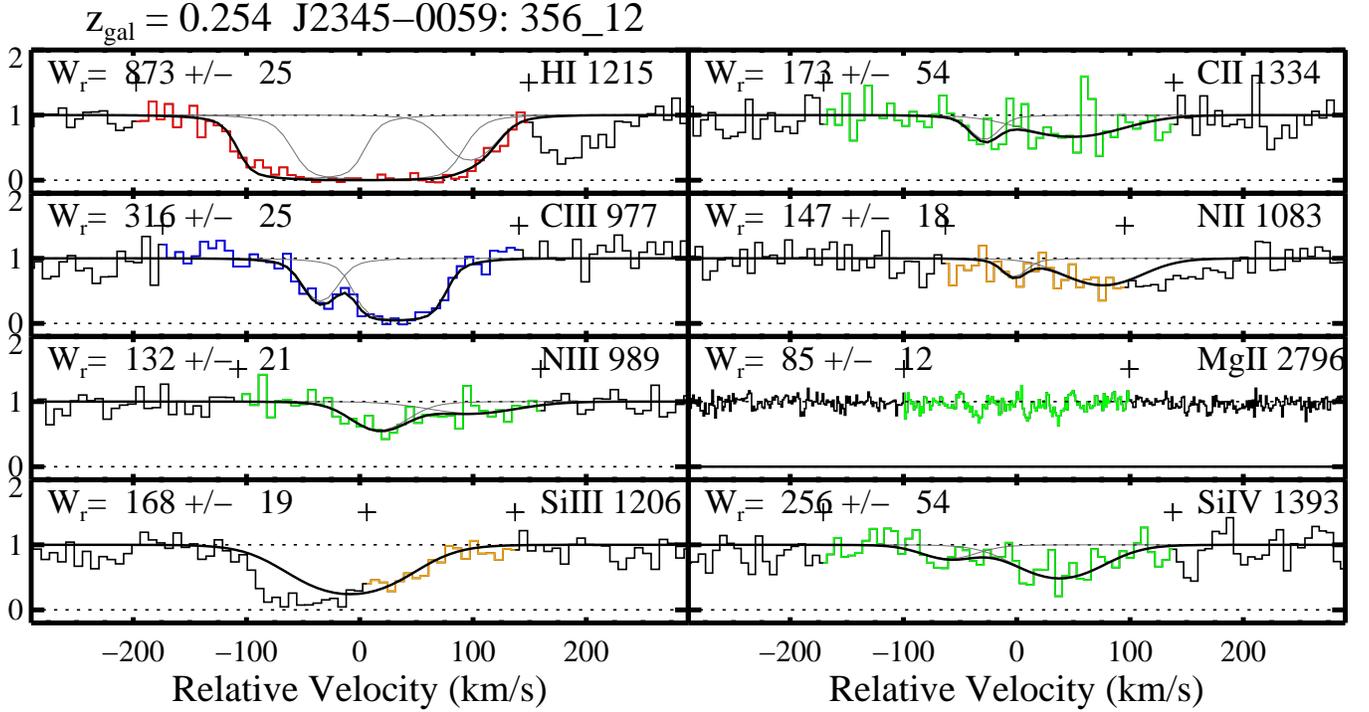}
\caption{Ionic species stack plots drawn from the COS and Keck spectra, centered on the absorption lines included in the analysis. Multi-component Voight profile fits to individual absorption lines are shown, where applicable. The X-axis ranges from $\pm$300 km/s from the associated galaxy redshift. Absorption lines are highlighted in color. Green signifies a good, uncorrupted detection; blue indicates the line is saturated; orange indicates the line is blended with some intervening absorption (often from the MW); and red shows lines that are both saturated and blended. The plus symbols mark the velocity range over which we define the absorption.}
\label{fig:stackend}
\end{figure*}
\afterpage{\clearpage}
%%%%%%%%%%%%%%%%

\begin{deluxetable}{lccccccc}
\tablewidth{0pc}
\tablecaption{Metal Line-Profile Fits and Final Column Densities\label{tab:final_columns}}
\tabletypesize{\scriptsize}
\tablehead{\colhead{Ion} & \colhead{$v^{a}$} & \colhead{$\sigma(v)$}&\colhead{$b$} & \colhead{$\sigma(b)$} & \colhead{$\log N$} & \colhead{$\sigma(N)$} & 
\colhead{$\log N_{\rm AODM}$}  \\
& ($\kms$) & ($\kms$) & ($\kms$) & ($\kms$) }
\startdata
\cutinhead{J0226+0015\_268\_22  $z=0.2274$}
\cutinhead{J0401-0540\_67\_24  $z=0.2197$}
C\, III&$ -65.2$&$  5.8$&$  12.7$&$  8.1$&$13.72$&$0.43$&$>14.00$ \\
&$ -13.9$&$  6.3$&$  25.1$&$  7.5$&$13.97$&$0.15$&$$ \\
N\, III&$ -45.7$&$  5.7$&$  14.7$&$ 13.6$&$14.01$&$0.38$&$14.57\pm0.05$ \\
&$   8.1$&$ 17.3$&$  54.4$&$ 16.8$&$14.45$&$0.14$&$$ \\
O\, VI&$ -59.3$&$  6.8$&$  33.3$&$  7.7$&$14.25$&$0.10$&$14.52\pm0.02$ \\
&$   6.0$&$  5.5$&$  27.9$&$  6.0$&$14.25$&$0.08$&$$ \\
Si\, III&$ -57.3$&$  5.2$&$  42.0$&$  7.7$&$12.91$&$0.06$&$12.88\pm0.06$ \\
\cutinhead{J0803+4332\_306\_20  $z=0.2535$}
C\, III&$ -36.8$&$  3.6$&$  17.1$&$  6.2$&$13.55$&$0.13$&$13.66\pm0.06$ \\
\enddata
\tablecomments{{}   First Line: SDSS Field Identifier and  
galaxy identifier, where the first number is the position angle in degrees from the QSO
and the second number is the projected separation in arcseconds (impact parameter) from the QSO; 
Galaxy Redshift, spectroscopically determined \citep{werk12}. \\ 
Columns: (1) Metal ion name ; (2) Component velocity in km s$^{-1}$ and  (3) associated error; (4) Component Doppler width and (5) associated error; (6) Log column density of the fitted component and (7) associated error;   (8) The AODM column density, for reference. A few of the component Doppler parameter errors of especially weak absorption lines are given as -0.0, a red flag that the fits should not be trusted in those particular instances.  }  
\end{deluxetable}
 
\clearpage

\begin{deluxetable}{llcccccc}
\tablewidth{0pc}
\tablecaption{Covering Fraction (EW)\label{tab:cover_ew}}
\tabletypesize{\scriptsize}
\tablehead{\colhead{Ion} & \colhead{Sample} & \colhead{$N_{\rm TOT}$} &
\colhead{$\mew_{\rm limit}$(m\AA)}
& \colhead{$\mCf [0,75]$}
& \colhead{$\mCf [75,160]$}
& \colhead{$\mCf [0,160]$}
}
\startdata
\cutinhead{Low-ion Transitions}
MgII\, 2796&All&39&100&$ 68^{+11}_{-11}$&$ 52^{+10}_{-22}$&$ 59^{+ 8}_{-15}$\\
MgII\, 2796&sSFR~$<10^{-11} \, {\rm yr^{-1}}$&13&100&$ 30^{+20}_{-20}$&$ 65^{+15}_{-26}$&$ 54^{+13}_{-20}$\\
MgII\, 2796&sSFR~$\ge 10^{-11} \, {\rm yr^{-1}}$&26&100&$ 81^{+11}_{-11}$&$ 43^{+13}_{-25}$&$ 61^{+ 9}_{-17}$\\
MgII\, 2796&$M_* < 10^{10.5}\mmsun$&17&100&$ 72^{+15}_{-15}$&$ 35^{+15}_{-31}$&$ 53^{+12}_{-23}$\\
MgII\, 2796&$M_* \ge 10^{10.5}\mmsun$&22&100&$ 61^{+16}_{-16}$&$ 63^{+12}_{-19}$&$ 63^{+10}_{-15}$\\
SiII\, 1260&All&39&150&$ 57^{+13}_{-13}$&$ 40^{+10}_{-24}$&$ 46^{+ 8}_{-18}$\\
SiII\, 1260&sSFR~$<10^{-11} \, {\rm yr^{-1}}$&14&150&$ 30^{+20}_{-20}$&$ 50^{+15}_{-32}$&$ 43^{+13}_{-25}$\\
SiII\, 1260&sSFR~$\ge 10^{-11} \, {\rm yr^{-1}}$&25&150&$ 68^{+14}_{-14}$&$ 34^{+12}_{-23}$&$ 48^{+10}_{-17}$\\
\enddata
\tablecomments{A table of covering fractions relating to the equivalent width measured for selected transitions, given for various ranges in impact parameter (in kpc). These values are shown as percentages in the last three columns along with their associated errors. The error reported in $C_f$,  which follows a standard binomial Wilson score, accounts for this uncertainty by calculating the 68\% confidence interval assuming that upper limits above the threshold do not satisfy it. We divide the sample into several subsamples (column 2) based on sSFR and mass. The total number of galaxies whose COS spectra cover the transition in question in each subsample is given in column 2. The values for HI and OVI were derived from the measurements presented in \cite{tumlinson12} and \cite{tumlinson11}, respectively.}
\end{deluxetable}
 
\clearpage

\begin{deluxetable}{llcccccc}
\tablewidth{0pc}
\tablecaption{Covering Fraction (Column Density)\label{tab:cover_colm}}
\tabletypesize{\scriptsize}
\tablehead{\colhead{Ion} & \colhead{Sample} & \colhead{$N_{\rm TOT}$} &
\colhead{$\log N_{\rm limit}$}
& \colhead{$\mCf [0,75]$}
& \colhead{$\mCf [75,160]$}
& \colhead{$\mCf [0,160]$}
}
\startdata
\cutinhead{Low-ion Transitions}
MgII&All&40&12.50&$ 68^{+11}_{-18}$&$ 42^{+10}_{-14}$&$ 52^{+ 8}_{-13}$\\
MgII&sSFR~$<10^{-11} \, {\rm yr^{-1}}$&13&12.50&$ 30^{+20}_{-20}$&$ 55^{+16}_{-16}$&$ 46^{+13}_{-13}$\\
MgII&sSFR~$\ge 10^{-11} \, {\rm yr^{-1}}$&27&12.50&$ 81^{+11}_{-20}$&$ 34^{+12}_{-17}$&$ 55^{+ 9}_{-17}$\\
MgII&$M_* < 10^{10.5}\mmsun$&18&12.50&$ 72^{+15}_{-27}$&$ 23^{+12}_{-19}$&$ 45^{+11}_{-21}$\\
MgII&$M_* \ge 10^{10.5}\mmsun$&22&12.50&$ 61^{+16}_{-16}$&$ 57^{+13}_{-13}$&$ 59^{+10}_{-10}$\\
SiII&All&42&13.00&$ 74^{+11}_{-36}$&$ 50^{+10}_{-31}$&$ 59^{+ 7}_{-31}$\\
SiII&sSFR~$<10^{-11} \, {\rm yr^{-1}}$&15&13.00&$ 30^{+20}_{-20}$&$ 54^{+14}_{-31}$&$ 47^{+12}_{-24}$\\
SiII&sSFR~$\ge 10^{-11} \, {\rm yr^{-1}}$&27&13.00&$ 88^{+ 8}_{-44}$&$ 47^{+12}_{-35}$&$ 66^{+ 9}_{-38}$\\
\enddata
\tablecomments{A table of covering fractions relating to the column density measured for selected transitions, given for various ranges in impact parameter (in kpc). These values are shown as percentages in the last three columns along with their associated errors. Here, the error reported in $C_f$,  which follows a standard binomial Wilson score, treats upper limits above the threshold as detections and considers lower limits below the threshold as satisfying it. In this case, $C_f$ values are maximal, aside from poisson error. We divide the sample into several subsamples (column 2) based on sSFR and mass. The total number of galaxies whose COS spectra cover the transition in question in each subsample is given in column 2. The values for HI and OVI were derived from the measurements presented in \cite{tumlinson12} and \cite{tumlinson11}, respectively.}
\end{deluxetable}
 
\clearpage
 
 \begin{landscape} 
\clearpage

\begin{deluxetable}{llcccccccccccc}
\tablewidth{0pc}
\tablecaption{Absorption Strength Statistics\label{tab:ewcolm_stats}}
\tabletypesize{\scriptsize}
\tablehead{\colhead{Ion} & \colhead{Sample} &
\colhead{$\tilde W$$^{a}$} & 
\colhead{$\tilde W$$^{a}$} & 
\colhead{$\tilde W$$^{a}$} & 
\colhead{$\avg{W}$$^{b}$} & 
\colhead{$\avg{W}$$^{b}$} & 
\colhead{$\avg{W}$$^{b}$} & 
\colhead{$\tilde{\log N}$$^{c}$} & 
\colhead{$\tilde{\log N}$$^{c}$} & 
\colhead{$\tilde{\log N}$$^{c}$} & 
\colhead{$\avg{{\rm log} N}$$^{d}$} & 
\colhead{$\avg{{\rm log} N}$$^{d}$} & 
\colhead{$\avg{{\rm log} N}$$^{d}$} \\ 
&&
[0,75] &
[75,160] &
[0,160] &
[0,75] &
[75,160] &
[0,160] &
[0,75] &
[75,160] &
[0,160] &
[0,75] &
[75,160] &
[0,160]
}
\startdata
\cutinhead{Low-ion Transitions}
MgII\, 2796&All&314&101&127&$482\pm463$&$245\pm333$&$343\pm403$&13.1&12.4&12.7&$13.1\pm0.67$&$12.6\pm0.70$&$12.8\pm0.72$\\
MgII\, 2796&sSFR~$<10^{-11} \, {\rm yr^{-1}}$& 86&180&111&$249\pm372$&$407\pm472$&$358\pm435$&12.4&12.7&12.5&$12.6\pm0.73$&$12.9\pm0.83$&$12.8\pm0.79$\\
MgII\, 2796&sSFR~$\ge 10^{-11} \, {\rm yr^{-1}}$&454& 67&203&$560\pm478$&$142\pm143$&$335\pm395$&13.3&12.3&13.0&$13.3\pm0.58$&$12.5\pm0.58$&$12.8\pm0.70$\\
MgII\, 2796&$M_* < 10^{10.5}\mmsun$&454& 66&101&$499\pm486$&$ 99\pm105$&$287\pm389$&13.1&12.3&12.4&$13.1\pm0.71$&$12.3\pm0.41$&$12.7\pm0.68$\\
MgII\, 2796&$M_* \ge 10^{10.5}\mmsun$&314&227&246&$466\pm472$&$339\pm395$&$385\pm418$&13.3&13.0&13.0&$13.1\pm0.68$&$12.8\pm0.80$&$12.9\pm0.75$\\
SiII\, 1260&All&292&115&142&$300\pm231$&$179\pm169$&$223\pm200$&13.4&13.0&13.2&$13.5\pm0.59$&$13.1\pm0.58$&$13.3\pm0.60$\\
SiII\, 1260&sSFR~$<10^{-11} \, {\rm yr^{-1}}$& 70&200&133&$169\pm208$&$261\pm239$&$235\pm226$&12.7&13.0&13.0&$13.0\pm0.49$&$13.3\pm0.78$&$13.2\pm0.72$\\
SiII\, 1260&sSFR~$\ge 10^{-11} \, {\rm yr^{-1}}$&297&114&145&$353\pm228$&$125\pm 67$&$216\pm187$&13.4&13.0&13.2&$13.6\pm0.54$&$13.0\pm0.33$&$13.3\pm0.55$\\
\enddata
\tablecomments{Median and average values for equivalent width and column density of selected metal-line transitions as a function of impact parameter in kpc. In all estimations, lower and upper limits are taken at their respective values.}
\tablenotetext{a}{$\tilde W$ is the median equivalent width for the given transition, in m\AA .}
\tablenotetext{b}{$\avg{W}$ gives the average and RMS of the equivalent width values, in m\AA.}
\tablenotetext{c}{$\tilde{{\rm log} N}$ is the median column density measured from all observed transitions of the given ion.}
\tablenotetext{d}{$\avg{{\rm log} N}$ gives the average and RMS of the logarithmic column densities.}
\end{deluxetable}

\clearpage
 
 \end{landscape} 
\end{document}